\RequirePackage[left,columnwise]{lineno}
\documentclass[aps,prc,twocolumn,floatfix,superscriptaddress,nofootinbib]{revtex4}

\usepackage{lipsum}
\usepackage{graphicx}
\usepackage{epstopdf}
\usepackage{subfigure}
\usepackage{epsfig}
\usepackage{amsmath,amssymb,amsfonts,eufrak,mathrsfs}
\usepackage{bm}
\usepackage{color}
\usepackage{lineno}

\newcommand{\bea}{\begin{eqnarray}}
\newcommand{\eea}{\end{eqnarray}}

\usepackage[utf8]{inputenc}

\begin{document}


\title{The JETSCAPE framework: p+p results}


\author{A.~Kumar}
\email{kumar.amit@wayne.edu}
\affiliation{Department of Physics and Astronomy, Wayne State University, Detroit MI 48201.}

\author{Y.~Tachibana}
\affiliation{Department of Physics and Astronomy, Wayne State University, Detroit MI 48201.}

\author{D.~Pablos}
\affiliation{Department of Physics and Astronomy, Wayne State University, Detroit MI 48201.}
\affiliation{Department of Physics and Astronomy, McGill University, Montr\'{e}al QC H3A-2T8.}

\author{C.~Sirimanna}
\affiliation{Department of Physics and Astronomy, Wayne State University, Detroit MI 48201.}

\author{R.~J.~Fries}
\email{rjfries@comp.tamu.edu}
\affiliation{Cyclotron Institute, Texas A\&M University, College Station TX 77843.}
\affiliation{Department of Physics and Astronomy, Texas A\&M University, College Station TX 77843.}



\author{A.~Angerami}
\affiliation{Lawrence Livermore National Laboratory, Livermore CA 94550.}

\author{S.~A.~Bass}
\affiliation{Department of Physics, Duke University, Durham NC 27708.}

\author{S.~Cao}
\affiliation{Department of Physics and Astronomy, Wayne State University, Detroit MI 48201.}

\author{Y.~Chen}
\affiliation{Laboratory for Nuclear Science, Massachusetts Institute of Technology, Cambridge MA 02139.}

\author{J.~Coleman}
\affiliation{Department of Statistics, Duke University, Durham NC 27708.}

\author{L.~Cunqueiro}
\affiliation{Department of Physics and Astronomy, University of Tennessee, Knoxville TN 37996.}

\author{T.~Dai}
\affiliation{Department of Physics, Duke University, Durham, NC 27708.}

\author{L.~Du}
\affiliation{Department of Physics, The Ohio State University, Columbus OH 43210.}

\author{H.~Elfner}
\affiliation{GSI Helmholtzzentrum f\"{u}r Schwerionenforschung, 64291 Darmstadt, Germany.}
\affiliation{Institute for Theoretical Physics, Goethe University, 60438 Frankfurt am Main, Germany.}
\affiliation{Frankfurt Institute for Advanced Studies, 60438 Frankfurt am Main, Germany.}

\author{D.~Everett}
\affiliation{Department of Physics, The Ohio State University, Columbus OH 43210.}

\author{W.~Fan}
\affiliation{Department of Physics, Duke University, Durham NC 27708.}


\author{C.~Gale}
\affiliation{Department of Physics and Astronomy, McGill University, Montr\'{e}al QC H3A-2T8.}

\author{Y.~He}
\affiliation{Key Laboratory of Quark and Lepton Physics (MOE) and Institute of Particle Physics, Central China Normal University, Wuhan 430079, China.}

\author{U.~Heinz}
\affiliation{Department of Physics, The Ohio State University, Columbus OH 43210.}

\author{B.~V.~Jacak}
\affiliation{Department of Physics, University of California, Berkeley CA 94270.}
\affiliation{Nuclear Science Division, Lawrence Berkeley National Laboratory, Berkeley CA 94270.}

\author{P.~M.~Jacobs}
\affiliation{Department of Physics, University of California, Berkeley CA 94270.}
\affiliation{Nuclear Science Division, Lawrence Berkeley National Laboratory, Berkeley CA 94270.}

\author{S.~Jeon}
\affiliation{Department of Physics and Astronomy, McGill University, Montr\'{e}al QC H3A-2T8.}

\author{K.~Kauder}
\affiliation{Department of Physics and Astronomy, Wayne State University, Detroit MI 48201.}
\affiliation{Department of Physics, Brookhaven National Laboratory, Upton NY 11973.}

\author{W.~Ke}
\affiliation{Department of Physics, Duke University, Durham, NC 27708.}

\author{E.~Khalaj}
\affiliation{Department of Computer Science, Wayne State University, Detroit MI 48202.}

\author{M.~Kordell~II}
\affiliation{Cyclotron Institute, Texas A\&M University, College Station TX 77843.}


\author{T.~Luo}
\affiliation{Key Laboratory of Quark and Lepton Physics (MOE) and Institute of Particle Physics, Central China Normal University, Wuhan 430079, China.}

\author{A.~Majumder}
\affiliation{Department of Physics and Astronomy, Wayne State University, Detroit MI 48201.}

\author{M.~McNelis}
\affiliation{Department of Physics, The Ohio State University, Columbus OH 43210.}

\author{J.~Mulligan}
\affiliation{Department of Physics, University of California, Berkeley CA 94270.}
\affiliation{Nuclear Science Division, Lawrence Berkeley National Laboratory, Berkeley CA 94270.}

\author{C.~Nattrass}
\affiliation{Department of Physics and Astronomy, University of Tennessee, Knoxville TN 37996.}

\author{D.~Oliinychenko}
\affiliation{Nuclear Science Division, Lawrence Berkeley National Laboratory, Berkeley CA 94270.}


\author{L.-G.~Pang}
\affiliation{Department of Physics, University of California, Berkeley CA 94270.}
\affiliation{Nuclear Science Division, Lawrence Berkeley National Laboratory, Berkeley CA 94270.}

\author{C. Park}
\affiliation{Department of Physics and Astronomy, McGill University, Montr\'{e}al QC H3A-2T8.}

\author{J.-F. Paquet}
\affiliation{Department of Physics, Duke University, Durham NC 27708.}

\author{J.~H.~Putschke}
\affiliation{Department of Physics and Astronomy, Wayne State University, Detroit MI 48201.}

\author{G.~Roland}
\affiliation{Department of Physics, Massachusetts Institute of Technology, Cambridge MA 02139.}

\author{B.~Schenke}
\affiliation{Department of Physics, Brookhaven National Laboratory, Upton NY 11973.}

\author{L.~Schwiebert}
\affiliation{Department of Computer Science, Wayne State University, Detroit MI 48202.}

\author{C.~Shen}
\affiliation{Department of Physics and Astronomy, Wayne State University, Detroit MI 48201.}


\author{R.~A.~Soltz}
\affiliation{Department of Physics and Astronomy, Wayne State University, Detroit MI 48201.}
\affiliation{Lawrence Livermore National Laboratory, Livermore, CA 94550.}


\author{G.~Vujanovic}
\affiliation{Department of Physics and Astronomy, Wayne State University, Detroit MI 48201.}
\affiliation{Department of Physics, The Ohio State University, Columbus OH 43210.}

\author{X.-N.~Wang}
\affiliation{Key Laboratory of Quark and Lepton Physics (MOE) and Institute of Particle Physics, Central China Normal University, Wuhan 430079, China.}
\affiliation{Department of Physics, University of California, Berkeley CA 94270.}
\affiliation{Nuclear Science Division, Lawrence Berkeley National Laboratory, Berkeley CA 94270.}

\author{R.~L.~Wolpert}
\affiliation{Department of Statistics, Duke University, Durham NC 27708.}

\author{Y.~Xu}
\affiliation{Department of Physics, Duke University, Durham NC 27708.}

\author{Z.~Yang}
\affiliation{Cyclotron Institute, Texas A\&M University, College Station TX 77843.}
\affiliation{Department of Physics and Astronomy, Texas A\&M University, College Station TX 77843.}

\collaboration{The JETSCAPE Collaboration}

\date{\today}


\begin{abstract}
The JETSCAPE framework is a modular and versatile Monte Carlo software package for the simulation of
high energy nuclear collisions. In this work we present a new tune of JETSCAPE, called PP19, and validate it by comparison to jet-based 
measurements in $p+p$ collisions, including inclusive single jet cross sections, jet shape observables, fragmentation functions,
charged hadron cross sections, and dijet mass cross sections. These observables in $p+p$ collisions provide the 
baseline for their counterparts in nuclear collisions. Quantifying the level of agreement of JETSCAPE results 
with $p+p$ data is thus necessary for meaningful applications of JETSCAPE to A+A collisions. The calculations use the JETSCAPE PP19 tune, 
defined in this paper, based on version 1.0 of the JETSCAPE framework. For the observables discussed in this work calculations using
JETSCAPE PP19 agree with data over a wide range of collision energies at a level comparable to standard Monte Carlo codes. 
These results demonstrate the physics capabilities of the JETSCAPE framework and provide benchmarks for 
JETSCAPE users.
\end{abstract}

\maketitle


\section{Introduction}
\label{sec:introduction}

Monte Carlo (MC) event generators are essential tools in particle and nuclear physics. They are used to create
large numbers of simulated collision events by sampling particles from computed probability distributions using 
Monte Carlo methods. Mature, well callibrated MC event generators are availble for the elementary collision systems 
$e^++e^-$, $e^-+p$ and $p+p$ \cite{Sjostrand:2006za,Sjostrand:2014zea,Bellm:2015jjp,Frixione:2007vw}. 
Of particular importance here is the PYTHIA 8 generator \cite{Sjostrand:2014zea}. It is an integral part of the JETSCAPE 
framework \cite{Putschke:2019yrg}. It is also the Monte Carlo generator that JETSCAPE results are compared to in this paper.

In contrast to the case of elementary collisions, there are no comprehensive MC event generators for high energy nuclear 
collisions which incorporate the soft, hard and electromagnetic sectors consistently. JETSCAPE addresses this issue by providing
a unified framework for current simulation codes \cite{Kauder:2018cdt,Putschke:2019yrg}.
The JETSCAPE framework provides state-of-the-art simulations of the soft sector of nuclear collisions, which refers primarily to 
modes with momentum smaller $\lesssim 2$ GeV/$c$. Those components include modeling of the initial state of the colliding nuclei, 
hydrodynamization and collective dynamics of the quark gluon plasma (QGP) and subsequent hadron gas phase, and freeze-out.
The soft sector includes over $99\%$ of particles in A+A collisions at the Relativistic Heavy Ion Collider (RHIC) and the 
Large Hadron Collider (LHC). 
JETSCAPE also combines several existing jet quenching Monte Carlo codes to model the hard sector
of nuclear collisions. This refers typically to processes with momentum transfer $\gtrsim 2$ GeV/$c$, including QCD jets, high 
transverse momentum partons and hadrons, and heavy quarks or hadrons. 
Hard processes and their final state dynamics evolve together with the soft background, and these processes have to be modeled simultaneously.
Hard probes, through their interaction with quark gluon plasma, can reveal important properties of QGP and have been the main
motivation behind the development of JETSCAPE.
A summary of capabilities of the framework, a detailed description of the framework structure, and instructions for users and 
developers can be found in the manual for JETSCAPE \cite{Putschke:2019yrg}.

In this paper we focus on testing and benchmarking the jet sector for $p+p$ collisions. Jet measurements at RHIC and LHC cover a wide range
of transverse momentum $p_T^{\mathrm{jet}}$ between 10 GeV/$c$ and 1 TeV/$c$ both at central and forward rapidities. Jet radii $R$ used for jet reconstruction in heavy ion physics 
usually vary between 0.2 and 0.5 although values up to 0.7, typical for measurements in $p+p$ collisions, are also used here. 
This paper reports JETSCAPE calculations of the following observables in $p+p$ collisions: inclusive jet cross sections, 
transverse jet shapes, jet fragmentation functions for charged hadrons, hadron cross sections, and dijet mass 
distributions. We carry out calculations at three different center of mass energies, $\sqrt{s}=0.2$, $2.76$ and 7 TeV. 

These calculations calibrate and test a crucial subset of components of JETSCAPE. In brief, the JETSCAPE configuration in $p+p$ mode 
consists of PYTHIA 8 to generate hard processes and to fragment QCD strings, while the final state parton showers are handled by MATTER 
\cite{Majumder:2013re,Cao:2017qpx} and by two string formation procedures developed for JETSCAPE 1.0, Colored and Colorless Hadronization. We
utilize MATTER since it is the default in-medium shower Monte Carlo code used in A+A. Likewise, one of the string formation 
procedures is the default to initialize hadronization in JETSCAPE 1.0 when calculating high momentum observables in A+A collisions. 
For consistency, $p+p$ results that will be used to benchmark A+A results 
will be generated with the same JETSCAPE final state radiation and hadronization modules.
MATTER and the two JETSCAPE string formation procedures will be briefly discussed in the next section. We refer to the
configuration of JETSCAPE used in this paper as the JETSCAPE PP19 tune.
The observables discussed in the previous paragraph probe the transverse and longitudinal structure of jets as well as intra-jet hadronization. 
They provide significant tests of MATTER as a parton shower code and of the JETSCAPE string formation processes.

Leading order (LO) Monte Carlo codes of perturbative QCD processes have limitations. Many of these are shared by the 
MC simulations in the JETSCAPE framework as explained below. LO simulations mimick higher order processes through 
parton showers, but they
are not expected to provide descriptions that fit all aspects of the complex collision 
dynamics equally well. Significant improvements might become available with consistent next-to-leading (NLO) formalisms 
implemented in Monte Carlo simulations for both $p+p$ and A+A collisions. Keeping this in mind it is important to formulate a 
realistic quantitative goal for this paper.
In order to be useful in future studies of A+A collisions, JETSCAPE with MATTER parton showers, in conjunction with JETSCAPE 
hadronization, must provide an overall acceptable description of $p+p$ data sets. To be acceptable results should, broadly 
speaking, be at the same level of agreement with data as comparable leading order Monte Carlo codes, e.g.\ PYTHIA 8. 
For each observable we document the level of agreement between JETSCAPE calculations, 
experimental data and PYTHIA 8, and discuss the possible origins of discrepancies. The results can be used to aid uncertainty 
estimates of future theoretical calculations and experimental analyses in A+A collisions.

The paper is organized as follows: In Sec.~\ref{sec:jetscape} we give a summary of the JETSCAPE 1.0 event generator. 
Subsequently we discuss the JETSCAPE modules used in this work as well as the workflow of JETSCAPE. We define the 
PP19 tune and document its parameter choices. In Sec.~\ref{sec:jetresults} we discuss results from the JETSCAPE PP19 tune. 
We compare JETSCAPE calculations to data and PYTHIA 8 with default parameters. 
We compare to inclusive cross sections, jet shapes, fragmentation functions, charged hadron cross sections and dijet mass
cross sections. We conclude with a discussion and outlook in Sec.\ \ref{sec:discussion}.

\begin{figure}[tb]
	\includegraphics[width=0.9\columnwidth]{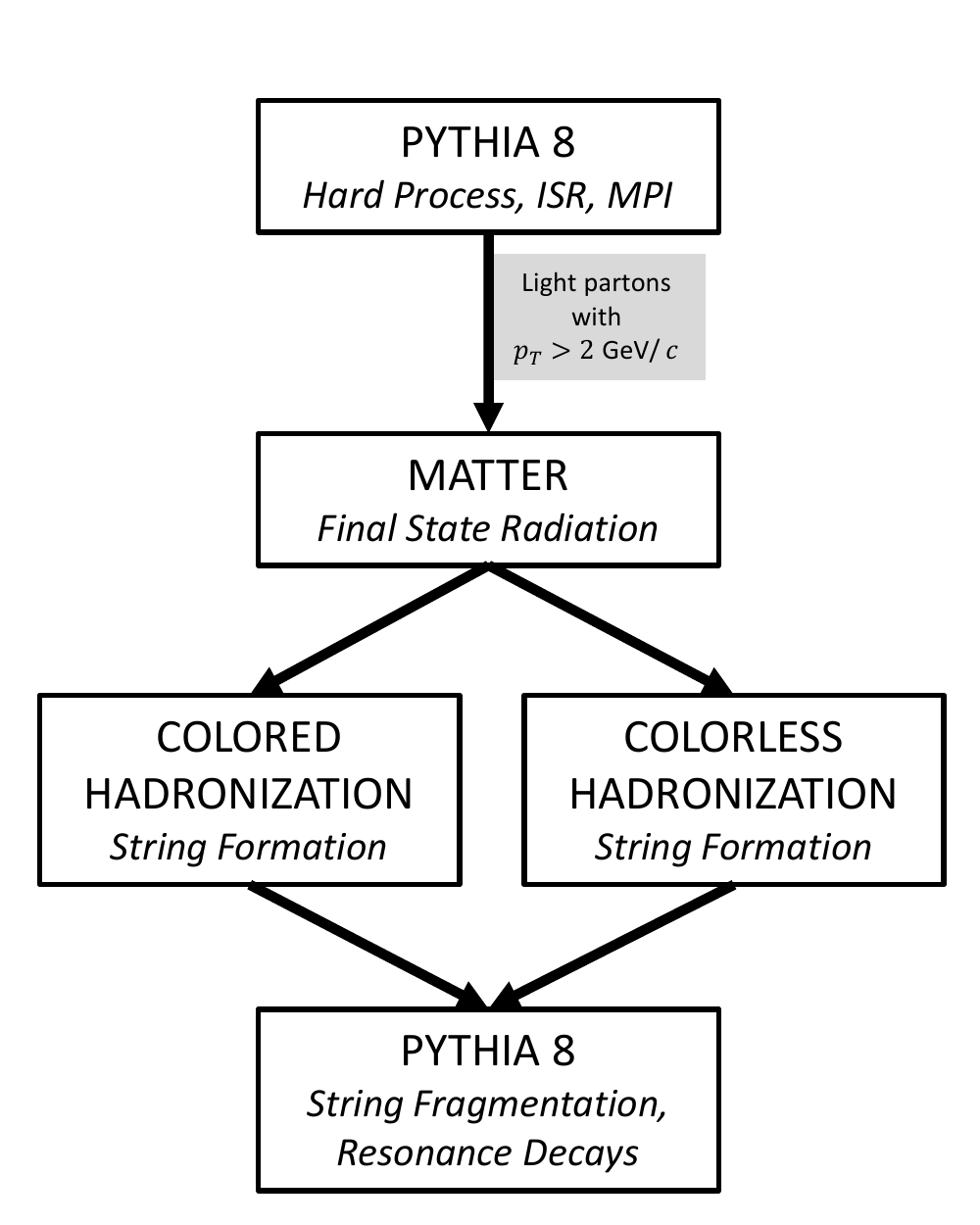}
  \caption{The modules called in JETSCAPE 1.0 for this $p+p$ calculation, using MATTER as the final state parton shower generator. Arrows represent the 
   workflow. If no critera are specified all output from the previous step is used as input for the next module. Either of the two string formation modules can 
   be run in the work flow. We perform calculations using both modules to estimate uncertainties in hadronization.}
  \label{fig:workflow}
\end{figure}

\section{The JETSCAPE Event Generator} 
\label{sec:jetscape}

In this section, we introduce the components of the JETSCAPE 1.0 framework. We then present the JETSCAPE components 
important for the PP19 tune.

\begin{table}[tb]
\begin{tabular}{|l|c|}
\hline
\bf JETSCAPE PP19 \\ \hline \hline
Setting & Value \\ \hline\hline
\multicolumn{2}{|l|}{\bf PYTHIA Hard Processes} \\ \hline
PYTHIA version & default 8.230 \\ \hline
Initial state radiation & ON \\  \hline
Multi parton interactions (MPI) & ON \\  \hline
Final state radiation & OFF \\  \hline
Hard QCD processes & ON \\ \hline
Electroweak processes & OFF \\  \hline
Hadronization & OFF \\  \hline 
Parton distribution function &  \parbox[l][3em][c]{0.3 \columnwidth}{NNPDF2.3 LO $\alpha_s=0.13$}  \\
\hline\hline

\multicolumn{2}{|l|}{\bf PYTHIA to MATTER} \\ \hline
Parton status code & 62 \\ \hline
\parbox[l][3em][c]{0.6 \columnwidth}{\begin{flushleft}Transverse momentum cut for initial partons \end{flushleft}}  & $p_T > 2$ GeV/$c$ \\ \hline
\hline\hline

\multicolumn{2}{|l|}{\bf MATTER}  \\ \hline
Initial shower parton virtuality $Q_\mathrm{ini}$  & $0.5 p_T$ \\ \hline
Medium induced energy loss $\hat q$ & 0 \\ \hline
Final shower parton virtuality $Q_0$ & 1 GeV \\ \hline\hline

\multicolumn{2}{|l|}{\bf Hadronization}  \\ \hline
Hadron decay cutoff $c\tau$  &  1 cm  \\ 
\hline\hline

\multicolumn{2}{|l|}{\bf Others} \\ \hline
QCD scale $\Lambda_\mathrm{QCD}$  & 0.2 GeV \\
\hline
\end{tabular}
\caption{Settings in JETSCAPE PP19. From the top: settings for PYTHIA
as a hard process generator; conditions for partons from PYTHIA to advance to MATTER; settings in MATTER; hadronization and resonance decays; 
general settings. \\
}
\label{tab:settings}
\end{table}

\begin{figure*}[tb]
	\includegraphics[width=0.95\columnwidth]{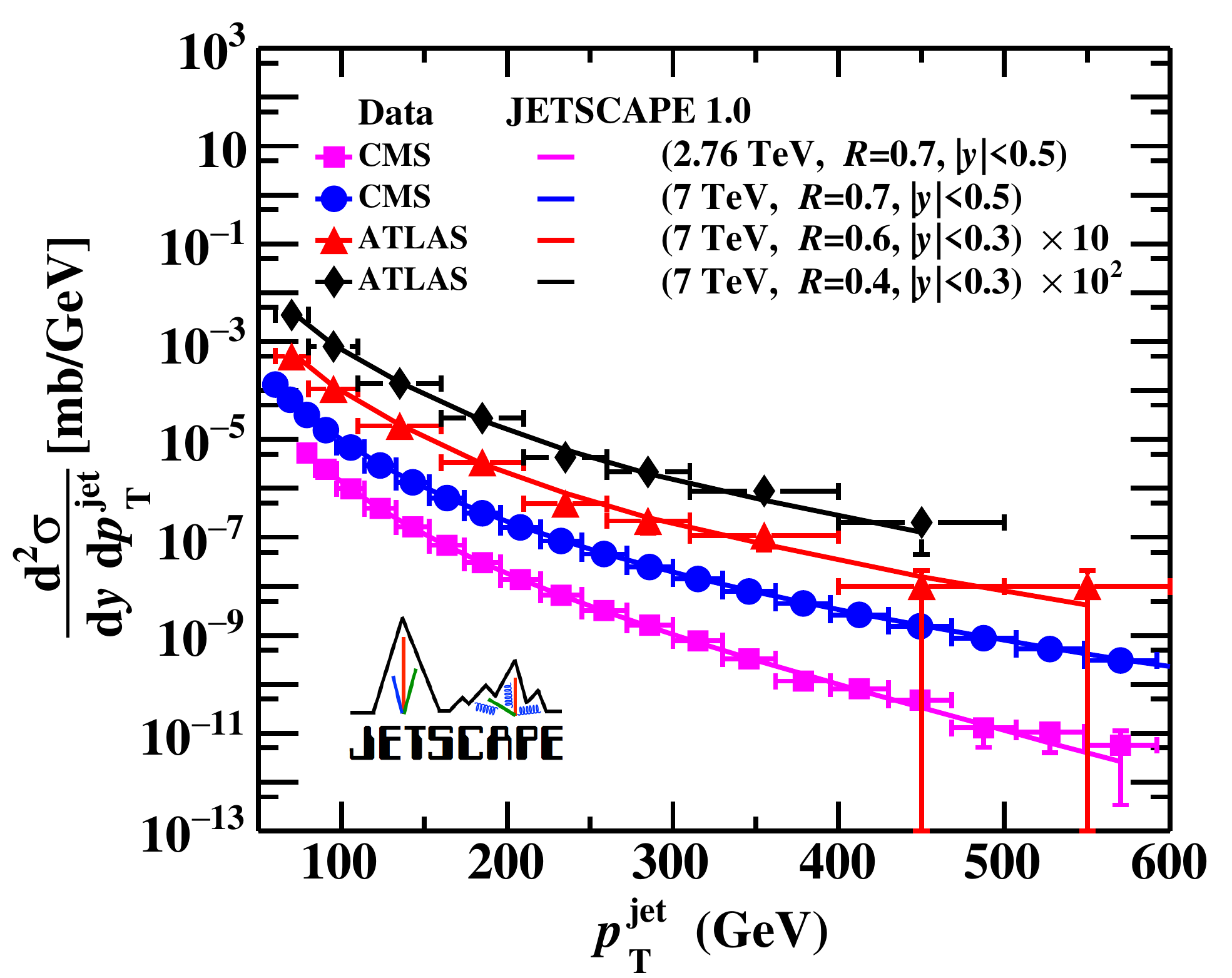}
\hspace{15pt}
	\includegraphics[width=0.95\columnwidth]{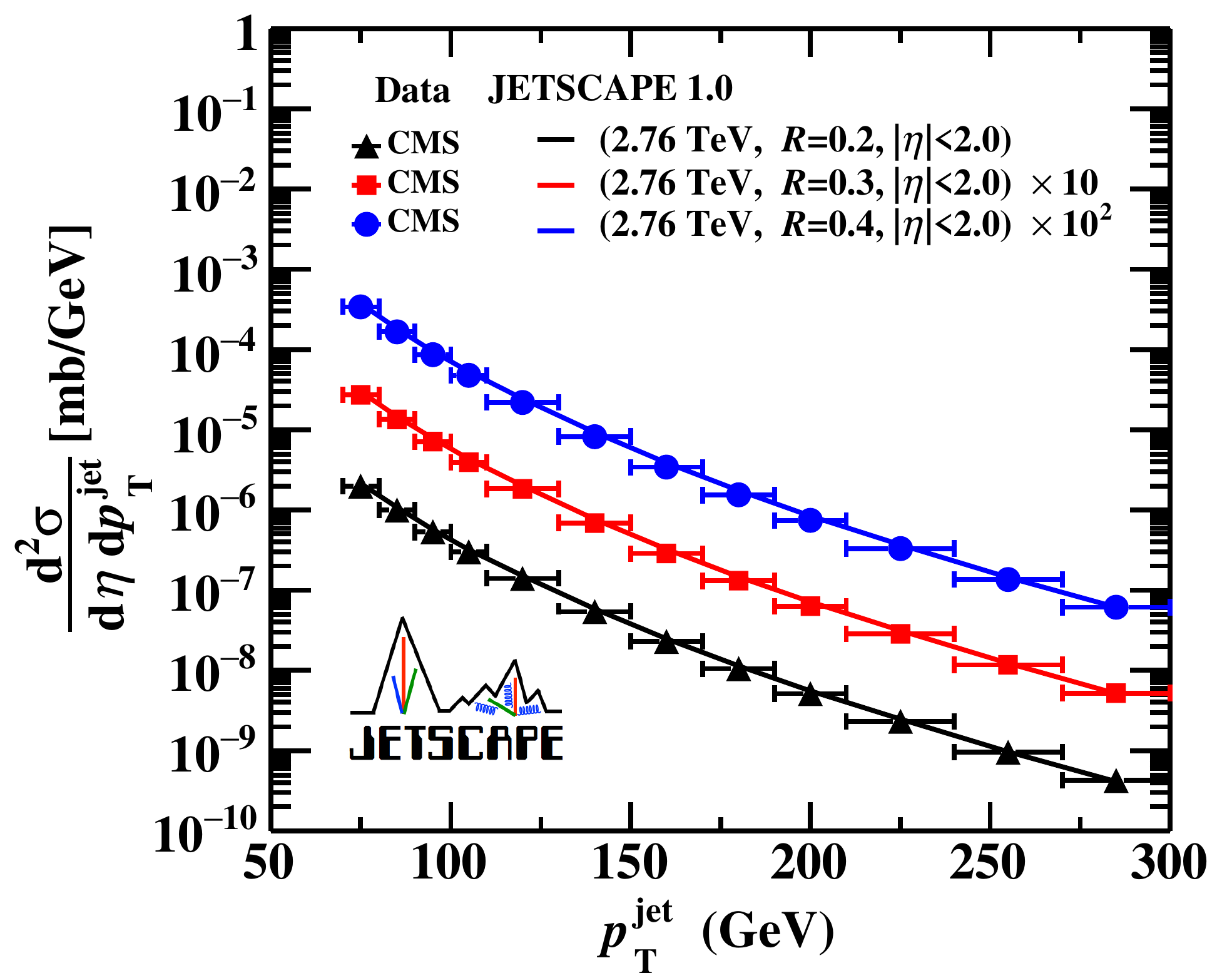}
  \caption{
  Inclusive jet cross sections $d^2\sigma/dp_T^{\rm jet} dy$ 
  vs jet transverse momentum $p_T^{\rm jet}$ for jets at midrapidity calculated with JETSCAPE PP19 
  (solid lines) are compared to LHC measurements (symbols).
Left panel:
Results for jet radius $R=0.7$ and rapidity $|y|<0.5$ at collision energies $\sqrt{s} = 2.76$ TeV (CMS data \cite{Khachatryan:2015luy}: magenta squares) and $\sqrt{s}=7$ TeV (CMS data \cite{Chatrchyan:2012bja}: blue circles). We also show $R=0.6$ (ATLAS data \cite{Aad:2010ad}: red triangles; scaled up by 10) and $R=0.4$ (ATLAS data: black diamonds; scaled up by 100) results at $\sqrt{s}=7$ TeV and $|y|<0.3$. 
Right panel: 
Results for jet radii $R=0.2, 0.3$ and $0.4$ with pseudorapidity $|\eta|<2.0$ 
at collision energy $\sqrt{s} = 2.76$ TeV, compared to CMS data 
\cite{Khachatryan:2016jfl}. 
  \label{fig:jet276+7mid}
  }
\end{figure*}

\subsection{JETSCAPE overview and A+A workflow}
\label{subsec:overview}

JETSCAPE 1.0 is a framework that incorporates several integrated codes for the soft and hard sector in nuclear collisions
working together. In this subsection we give a brief overview of the full event generator before focusing on the $p+p$
mode.
The initial state in nuclear collisions is modelled by the initial state generator TRENTO \cite{Moreland:2014oya}. 
The relativistic fluid dynamic code MUSIC \cite{Schenke:2010nt} is used for subsequent evolution of the soft sector. In the 
hard sector, initial hard scattering is handled by PYTHIA 8 \cite{Sjostrand:2014zea} while final state showers are generated by MATTER \cite{Majumder:2013re,Cao:2017qpx}. The latter code is the default final state radiation simulation in nuclear collisions incorporating 
medium-induced final state radiation. It is therefore also the preferred final state shower Monte Carlo to ensure consistency between simulations of $p+p$ and A+A collisions.
Further propagation of partons through quark-gluon plasma whose evolution is described by MUSIC can be handled by MARTINI \cite{Schenke:2009gb} or LBT \cite{He:2015pra,Cao:2016gvr} which are based on perturbative QCD, or by HyBRID \cite{Casalderrey-Solana:2014bpa,Casalderrey-Solana:2015vaa} which is based on a strong coupling approach. Two hadronization mechanisms, Colored Hadronization and Colorless Hadronization, are used to form string systems from parton showers. In both cases the strings are subsequently handed off to PYTHIA 8 for string fragmentation into hadrons. Decays of resonances are also handled by PYTHIA 8, subject to user settings. 
The soft and hard sectors evolve in space-time and can have mutual interactions in the extended fireball formed in A+A collisions.
In each simulated event  MATTER, MARTINI, LBT and HyBRID are provided the local temperature and collective local flow velocity from fluid dynamics. It is the task of the JETSCAPE framework to call each code at the correct instance, using criteria established by the user. 
At a given time and position, conditions like parton virtuality, parton energy in the local medium rest frame, or local temperature are used to decide the next
step in parton evolution. We refer the reader to the JETSCAPE manual for more information \cite{Putschke:2019yrg}.

\subsection{$\mathbf{p+p}$ work flow}
\label{subsec:ppflow}

In $p+p$ collisions the soft sector of JETSCAPE is inactive. We do not focus on very high multiplicity events in which 
collective effects for soft particles might occur \cite{Khachatryan:2010gv,Aad:2015gqa}. However, soft processes do occur in $p+p$ and create
an underlying event (UE) of soft partons.
For the PP19 tune we use PYTHIA 8.230 in JETSCAPE to generate the primary hard processes, together with the underlying event. 
The latter is modelled in PYTHIA 8 by including multi-parton interactions (MPIs) and initial state radiation (ISR). 
After the hard process and the underlying event are generated, but before the generation of any final state radiation (FSR), all objects in the 
PYTHIA event record are extracted. Gluons and light quarks (up, down, strange) with transverse momentum $p_T > 2$ GeV/$c$
are retained while all other objects are discarded. The remaining partons will be referred to as hard partons
from here on. 

These cuts are implemented for two reasons. The momentum cut omits partons for which MATTER does not create final state radiation, and
the JETSCAPE 1.0 version of MATTER is set up only for light partons. 
The momentum cut discards part of the underlying event. For PP19 we have compensated for this by adjusting the available parameters in MATTER to describe the single inclusive jet cross section at midrapidity directly, without any underlying event subtraction (see next paragraph). 
Other observables might differ in their behavior, as discussed at the end of this section.

Since MATTER is the default final state radiation module in JETSCAPE we describe the physics of MATTER and its 
implementation in JETSCAPE in more detail in the next subsection. Only the partons selected in the previous step are used to generate final 
state radiation and passed from PYTHIA 8 to MATTER by the framework. These partons will in general have a virtuality. However, PYTHIA 8, which is based on a dipole formalism, does not provide a calculation of their initial virtuality. The virtuality is therefore generated in the final state generator MATTER. The virtuality leads to the emission of final state radiation and the build up of a parton shower. MATTER, a virtuality ordered generator, starts by ascribing an initial virtuality to each 
parton. A parameter that has to be specified by the user at this stage is the maximum virtuality $Q_\text{ini}$ allowed for each hard 
parton as input. In PP19 $Q_\text{ini}$ for a hard parton is chosen to be $Q_\text{ini} = p_T/2$
for a hard parton with transverse momentum $p_T$. MATTER then generates parton showers through repeated QCD splitting processes 
until all partons have residual virtualities smaller than a scale $Q_0$, whose value is $1$ GeV in the tune PP19.

After the creation of final state parton showers for all hard partons the framework forms QCD strings through one of two string formation
processes. These modules, called Colored Hadronization and Colorless Hadronization, have been developed specifically for the JETSCAPE framework and are discussed in detail
below. 
After all partons are assigned to color singlet string systems, PYTHIA 8 is called a final time to hadronize the string systems 
and to handle hadron decays. Fig.\ \ref{fig:workflow} gives an overview of the JETSCAPE modules used in this tune, and the work flow utilizing them.

\begin{figure*}[tb]
  \includegraphics[width=0.64\columnwidth]
{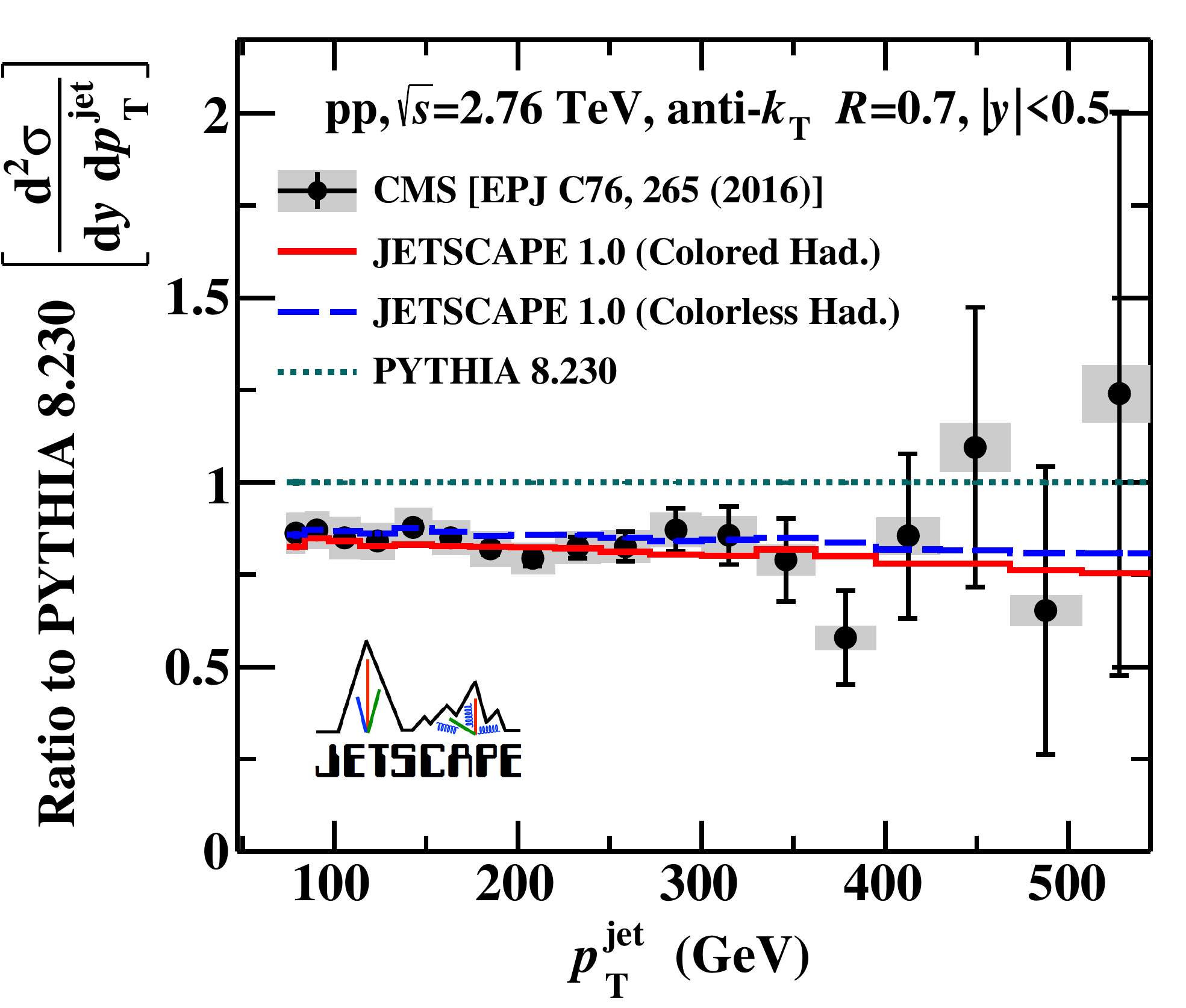}
\hspace{15pt}
 \includegraphics[width=0.64\columnwidth]
{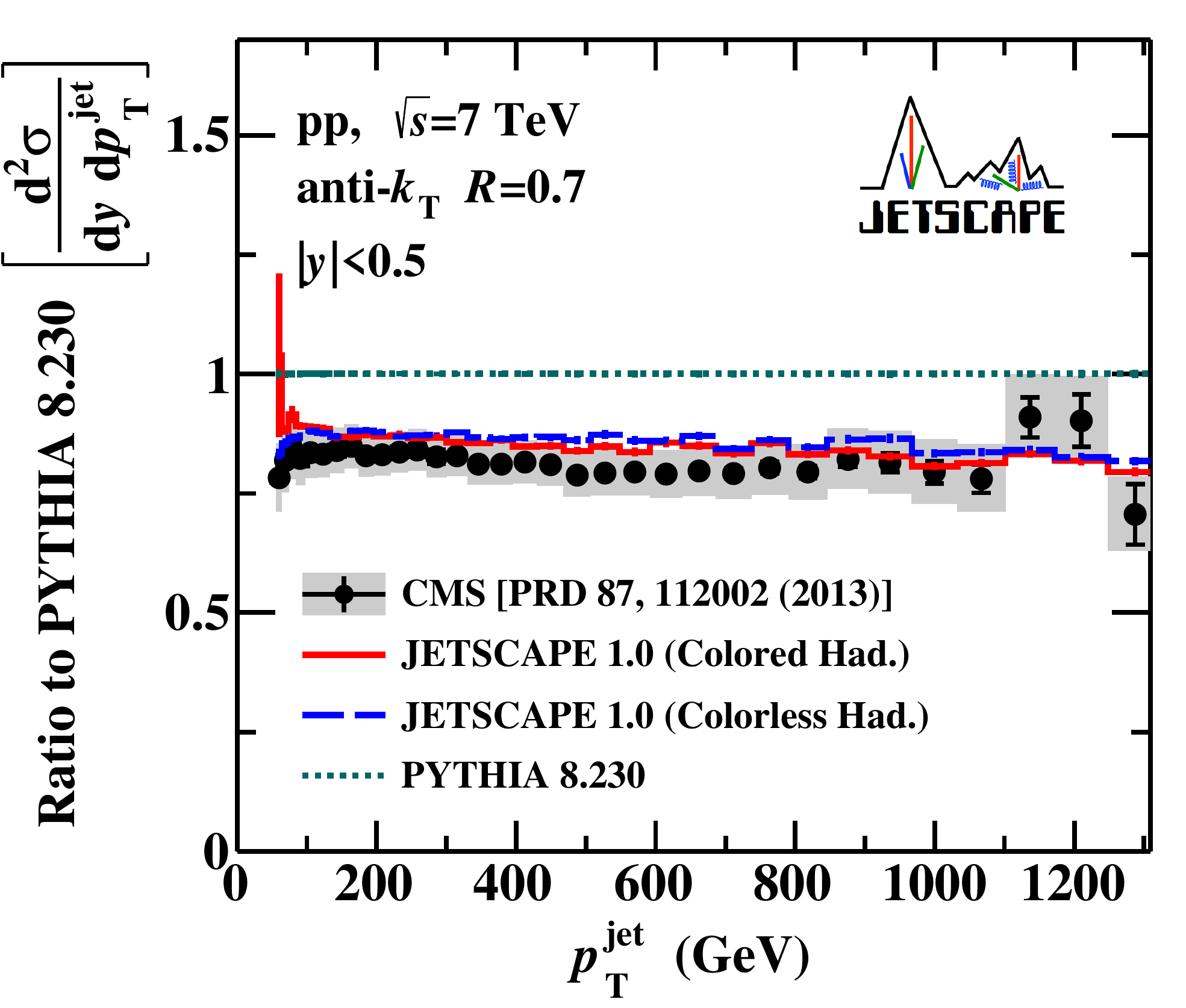}
\includegraphics[width=0.64\columnwidth]{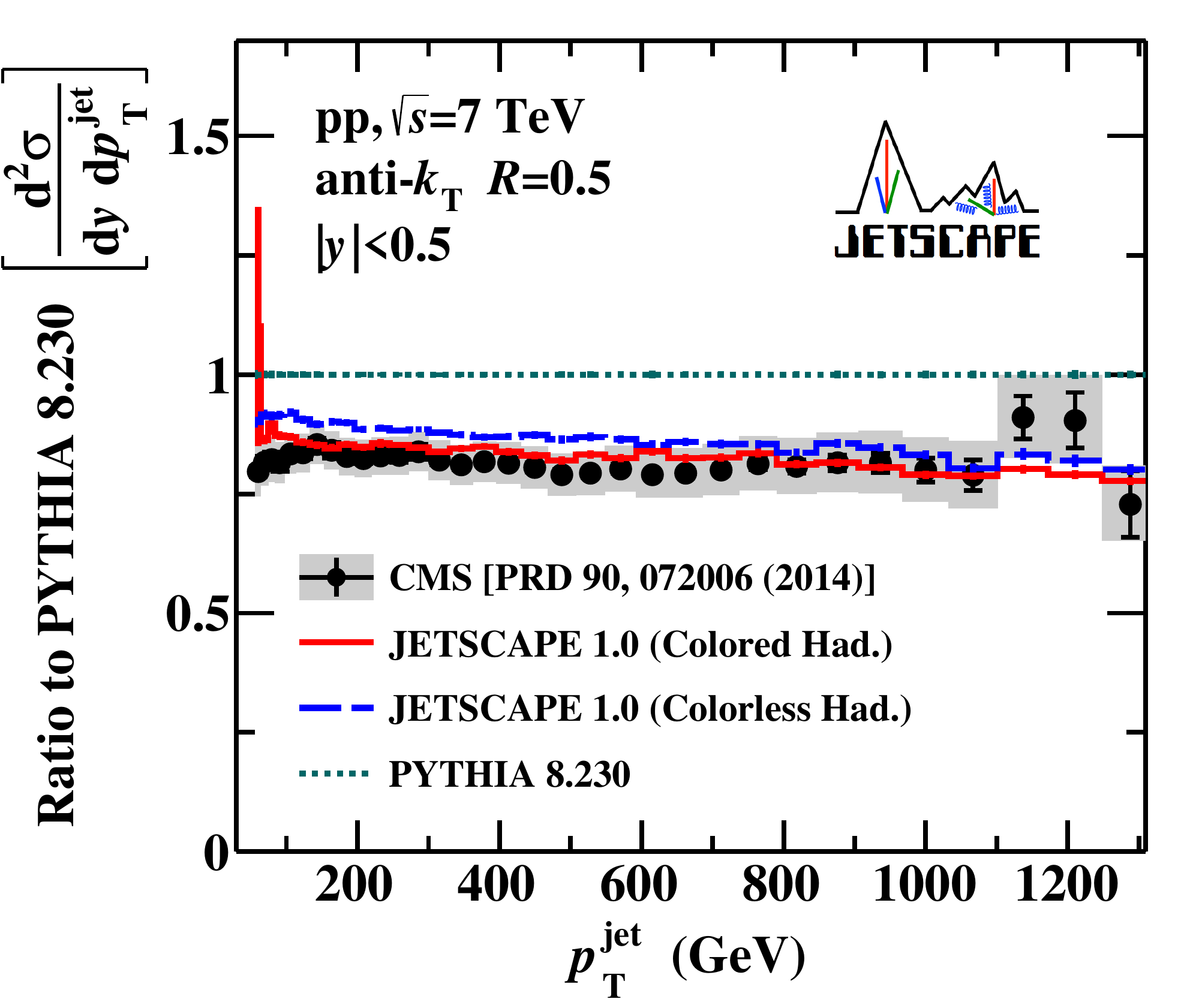}
  \caption{
  Ratios of full jet cross sections for jet radius $R=0.7$ and rapidity $|y|<0.5$ 
  (shown in the left panel of Fig.\ \ref{fig:jet276+7mid}), and for radius $R=0.5$,
  to PYTHIA 8 Monte Carlo results. Three different Monte Carlo calculations are presented:
  JETSCAPE Colored Hadronization (solid red line), 
  JETSCAPE Colorless Hadronization (dashed blue line), and PYTHIA 8 (dotted green line). Statistical errors (black error bars) and systematic uncertainties
  (grey bands) are plotted with the experimental data. 
  Left panel: Calculations at $\sqrt{s} = 2.76$ TeV, compared to CMS data \cite{Khachatryan:2015luy}.
  Center and right panels: Calculations at $\sqrt{s} = 7$ TeV, compared to CMS data \cite{Chatrchyan:2012bja}. 
  \label{fig:jet276+7midratio1}
  }
\end{figure*}

\subsection{MATTER Parton Showers}
\label{subsec:MATTER}

In this subsection we describe the MATTER parton shower generator and its integration into the JETSCAPE framework. 
The MATTER shower generator calculates parton showers in both vacuum and medium. In the following we 
will only describe the generator in vacuum. We will focus here on light flavors. 
The MATTER generator in its native setup is described in detail in Refs.~\cite{Majumder:2013re,Majumder:2014gda,Cao:2017qpx}. 
MATTER is exclusively a virtuality ordered shower generator. On the other hand JETSCAPE is a time ordered framework. As a result the life-time (or split-time) of every emission in MATTER has to be determined and the emission executed at the appropriate time as determined by the framework. 

Given a parton with a near on-shell four-momentum $(E,\mathbf{p})$, where $E = \sqrt{ p^2 + m_2 }\,\,  (p = |\mathbf{p}|)$, the MATTER generator calculates 
the two light-cone momenta $p^+ = \left[ E + p \right]/\sqrt{2}$ and $p^- = \left[ E - p \right]/\sqrt{2}$. Given the maximum and minimum allowed values of the virtuality $Q_0<t<t_{\text{max}}$ for a parton, its virtuality $t$ is estimated by sampling the Sudakov form factor, 
\bea
\Delta (t_{\text{max}}, t) = \exp\left[  -\int_{t}^{t_{\text{max}}} \frac{dt'}{t'} \alpha_S(t') \int_{y_{\text{min}}}^{y_{\text{max}}} dyP(y)        \right]. 
\eea
In this equation $P(y)$ is the splitting function for the parent parton with light-cone momentum $p^+$ to split into two partons, with light cone momenta $yp^+$ (for daughter particle 1) and $(1-y)p^+$ (for daughter particle 2). The limits of the splitting function integral are  
$y_{\text{min}} = Q_{0}/t'$ and $y_{\text{max}} = 1 - y_{\text{min}}$. For the shower-initiating parton the maximum virtuality is
$t_\text{max}=Q_\text{ini}$.

Once the virtuality $t$ is determined, the $p^-$ momentum of the parent parton is rescaled to $p^- = [ t + p_\perp^2 ]/2p^+$. 
At this stage the $(+)$-light-cone momenta of the daughter partons are determined by a sampling of the splitting function $P(y)$. 
In the next step, the virtualities $t_1,t_2$ of the daughter partons are determined using $\Delta(y^2 t, t_1)$ and $\Delta((1-y)^2 t, t_2)$. Subsequently, 
the two outgoing partons are assigned transverse momenta $\pm \mathbf{k}_\perp$ relative to the parent, using
\bea
k_\perp^2 = y(1-y) t - y t_2 - (1-y) t_1.
\eea

The above process is repeated iteratively until all partons reach virtuality $t=Q_0$, at which point the shower terminates. Unlike PYTHIA 8, MATTER 
also tracks the location of each of the partons. While this information is critical for the case of jets in a medium, it currently plays no role for jets in 
vacuum. 
MATTER also maintains the color information of the shower within the large $N_c$ (arbitrary number 
of colors) approximation, as is the case in PYTHIA 8. At the end of the shower, the color of the entire jet is equivalent to the color of the shower-initiating 
parton. The momentum and color information of all the final state partons with $t=Q_0$ are passed to a hadronization routine.

\begin{figure*}[tb]

  \includegraphics[width=0.9\columnwidth]{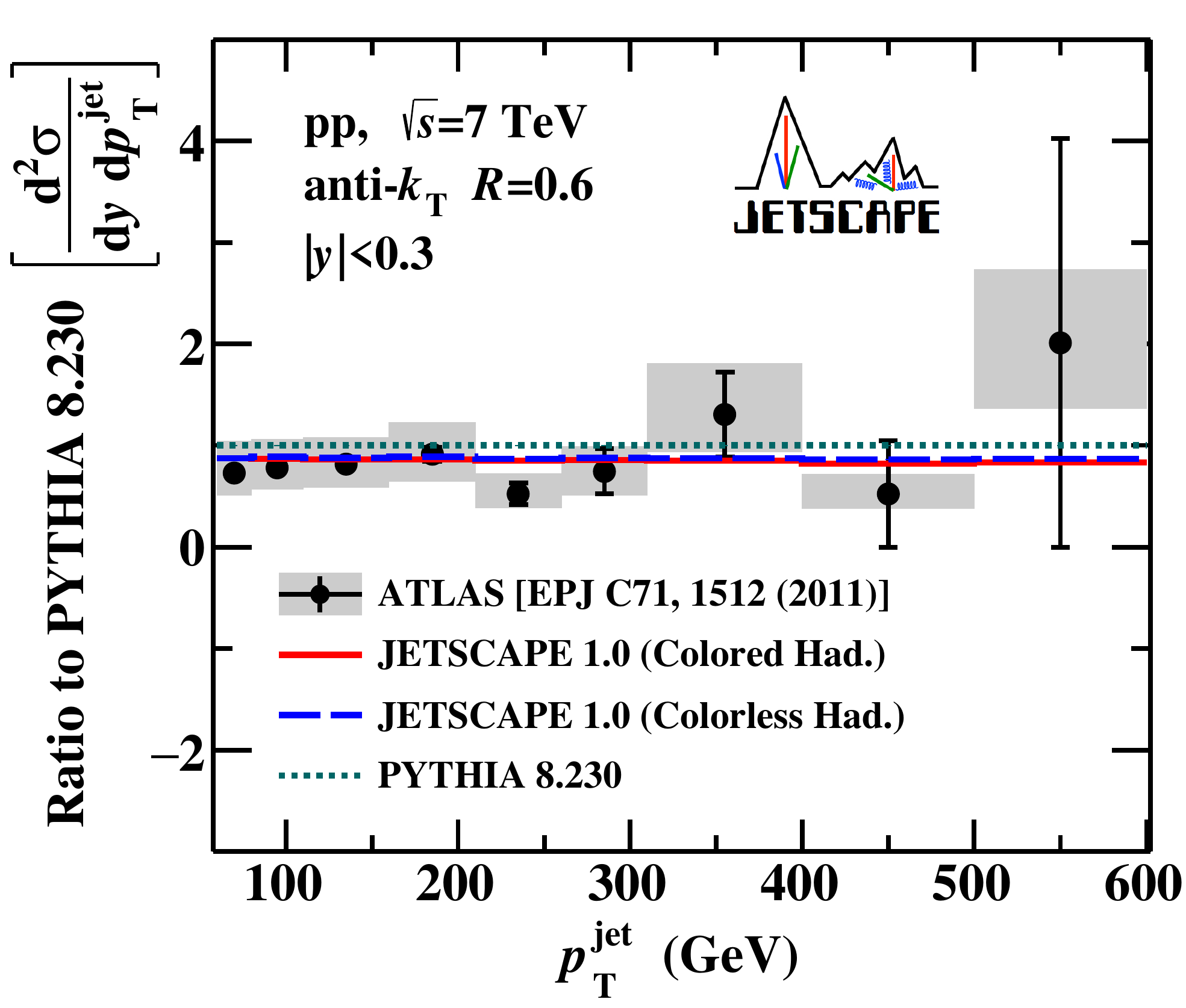}
  \hspace{15pt}
  \includegraphics[width=0.9\columnwidth]{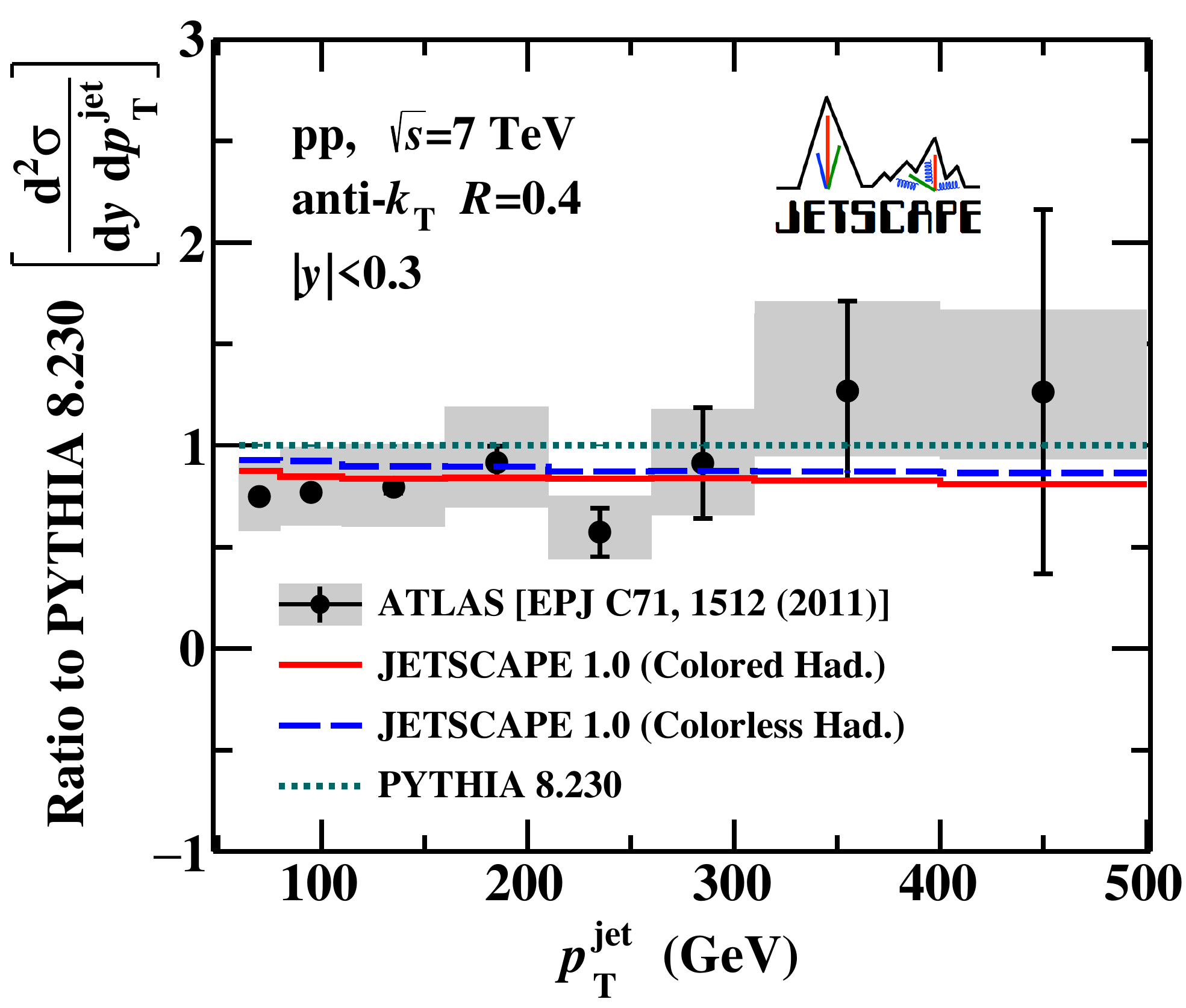}
  \caption{
  Same as Fig.\ \ref{fig:jet276+7midratio1} 
  for full jets with rapidity $|y|<0.3$ at collision energy $\sqrt{s} = 7$ TeV. 
   (shown in the left panel of Fig.\ \ref{fig:jet276+7mid}), compared to ATLAS data \cite{Aad:2010ad}. 
   Left panel: $R=0.6$. 
   Right panel: $R=0.4$. 
  \label{fig:jet276+7midratio2}
  }
\end{figure*}

\subsection{String hadronization}
\label{subsec:chm}

JETSCAPE 1.0 uses default string hadronization provided by PYTHIA 8. 
in-medium parton shower modules, is used just before hadronization, strings need to be defined through
a protocol and then handed over to PYTHIA 8 string fragmentation. In this subsection we describe two alternative
algorithms for string formation, Colored Hadronization and Colorless Hadronization.

The labels ``colored" and ``colorless" describe whether or not color flow information is utilized for the string formation process.
These alternatives are provided in JETSCAPE because hadronization in-medium can occur through 
channels that are not active in vacuum. In particular, color can be exchanged with the medium and color coherence can be lost
in the parton shower \cite{CasalderreySolana:2011rz}. Most in-medium shower Monte Carlo codes thus do not explicitly track color flow
in parton showers. Therefore, Colorless Hadronization should be the default choice for jets in a medium. However, in vacuum 
the MATTER shower generator assigns color to each radiated parton, utilizing the large-$N_c$ approximation.
Thus, in $p+p$ collisions the Colored Hadronization module is a more physical alternative. Nevertheless,
Colorless Hadronization should be studied for $p+p$ calculations as well for consistency.
In this paper we use both string formation models to estimate uncertainties in the treatment of hadronization.

The Colored Hadronization module requires color tags to be assigned to all the partons in a shower. In that
case the addition of a single ``external" parton can make the shower a color singlet.
Given a system with $n$ hard, shower-initiating partons, each shower is assigned such an external parton with the correct color tags needed to make the selected
shower a color singlet. External partons approximate the effects of beam remnants that are present in $p+p$ events. They are given longitudinal momentum 
of magnitude $\sqrt{s}/6$ and a transverse momentum of order $\sim 1$ GeV. The signs of the longitudinal momenta of the $n$ showers in the event are chosen 
to alternate between positive and negative. Each of these color-singlet systems then represents a string that can be handled by PYTHIA 8. They are handed over to PYTHIA 8
string fragmentation as input. Due to the chosen color and string configuration each of these showers hadronizes independently from the others.

It should be noted that although Colored Hadronization is realistic in its attempt to keep as much color information as possible, it treats showers as independent. 
In reality showers are color correlated. 
In other words, there could be a single string that connects several final state showers with a single high rapidity parton in each beam direction. 
A different assumption is made in Colorless Hadronization, where only one fake parton is introduced for a system of multiple showers. As a result, in most observables the Colored Hadronization model will produce a larger yield at lower $p_T$ than the Colorless Hadronization module.

The Colorless Hadronization module disregards any color flow information present, and constructs strings based on a minimization 
criterion. Specifically, the module minimizes the distance 
\begin{equation}
  \label{eq:dr}
  \Delta R=\sqrt{{(\Delta \eta)}^2 + {(\Delta \phi)}^2}
\end{equation}
using pseudorapidity $\eta$ and azimuthal angle $\phi$ of partons.
Strings are not established shower by shower. Instead the the full recorded parton event output from final state shower Monte Carlos is used, and strings can include partons from different showers in the same event. 
The following algorithm is applied:
\begin{enumerate}
\item Find the number of strings by counting quarks and antiquarks together. If an odd number of quarks is found, an external quark 
with momentum along the beam direction is added similar to the Colored Hadronization case. If the number is even
two external quarks in opposite directions along the beamline are added.
\item Find quark pairs whose $\Delta R$ is minimal. This procedure establishes pairs of string endpoints.
\item Go through the list of gluons and find the string which minimizes the quantity ${[{(\Delta R)}_1+{(\Delta R)}_2]/2}$, where ${(\Delta R)}_{1,2}$ 
are the distances $\Delta R$ between the gluon and the first and second endpoint, respectively, of a string. Assign the gluon to that string.
\item Decide the order of gluons inside each string. Starting from one of the endpoints, the gluon in this string with the smallest $\Delta R$ with respect 
  to that endpoint is placed next to it. Then of the remaining gluons in this string the one with the smallest $\Delta R$ with respect to the first gluon is 
  placed next to it, and so on. Continue until all gluons for that string are placed. Repeat for each string.
\item With the order of partons in a string established, assign proper color tags. Feed the string system into PYTHIA 8
for string fragmentation.
\end{enumerate}

Note that neither hadronization module currently handles junctions or more complicated string objects.

\begin{figure*}[tb]
  \includegraphics[width=0.64\columnwidth]{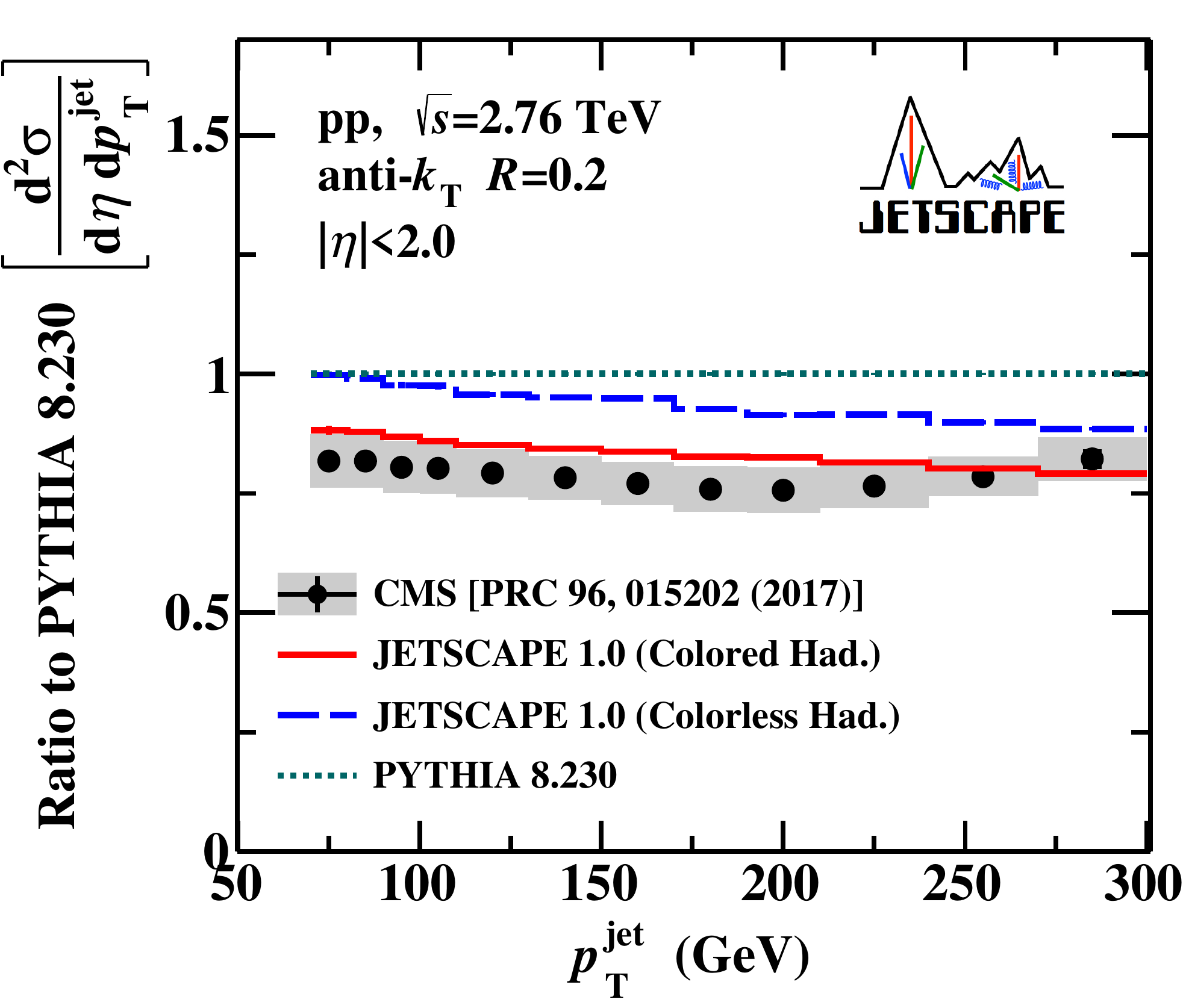}
  \hspace{10pt}
  \includegraphics[width=0.64\columnwidth]{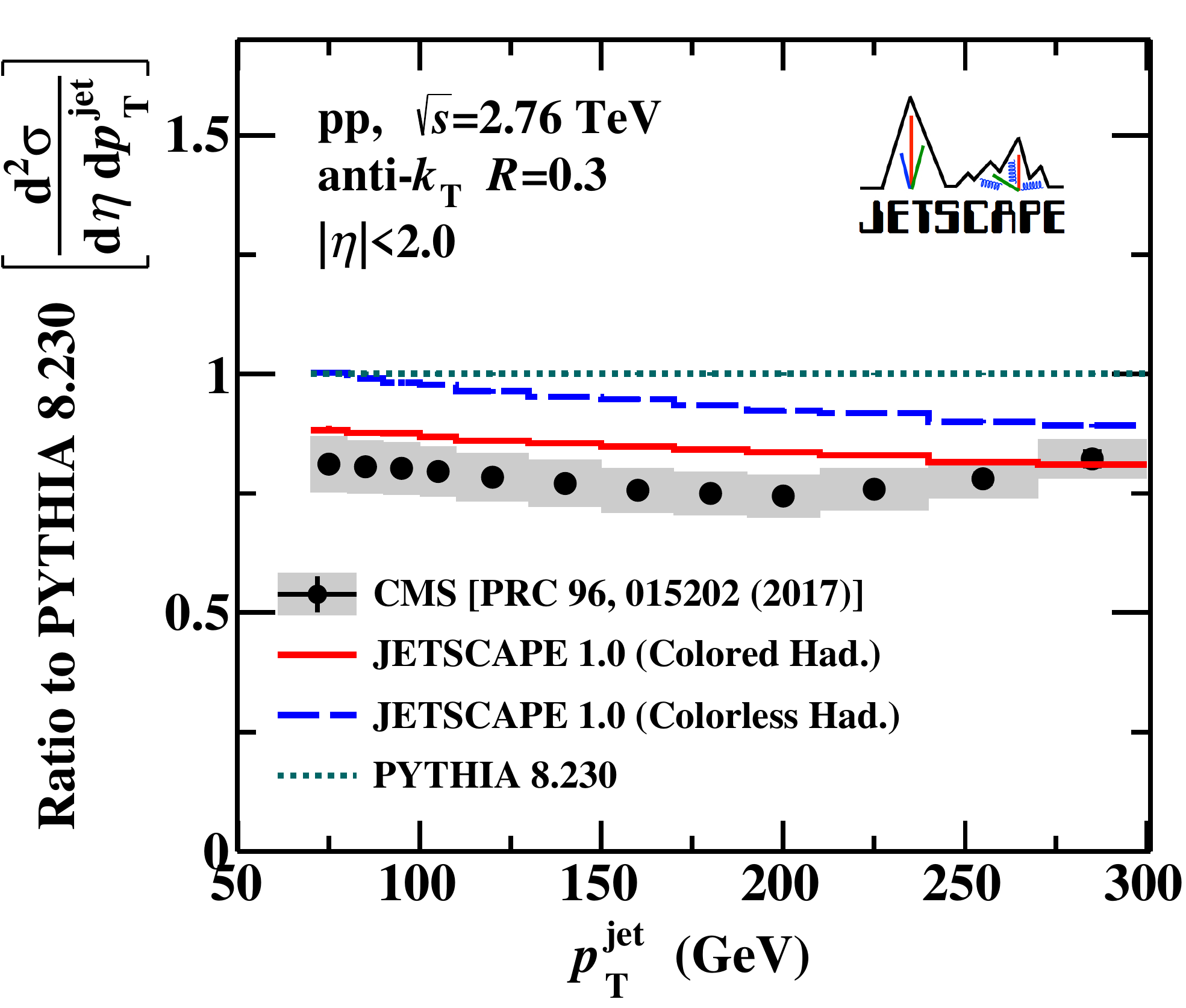}
  \hspace{10pt}
  \includegraphics[width=0.64\columnwidth]
{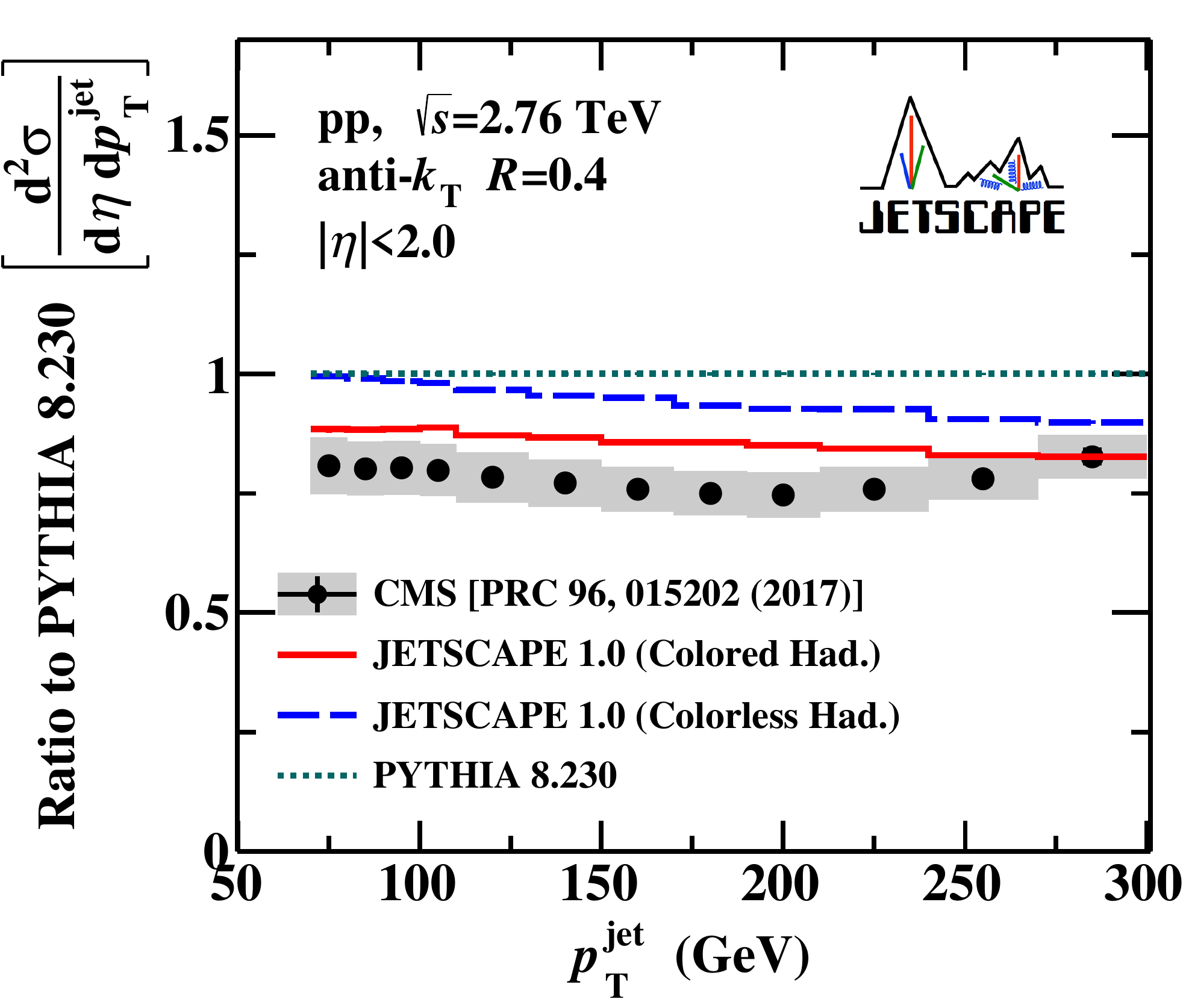}
  \caption{
    Same as Fig.\ \ref{fig:jet276+7midratio1} 
   for full jet with pseudorapidity $|\eta|<2.0$ at collision energy $\sqrt{s} = 2.76$ TeV. 
   (shown in the right panel of Fig.\ \ref{fig:jet276+7mid}), compared to CMS data \cite{Khachatryan:2016jfl}. 
   Left panel: $R=0.2$. 
   Center panel: $R=0.3$. 
   Right Panel: $R=0.4$. 
   \label{fig:jet276midratio}
}
\end{figure*}

\begin{figure*}[tb]
\hspace{-9pt}
\includegraphics[width=0.95\columnwidth]
{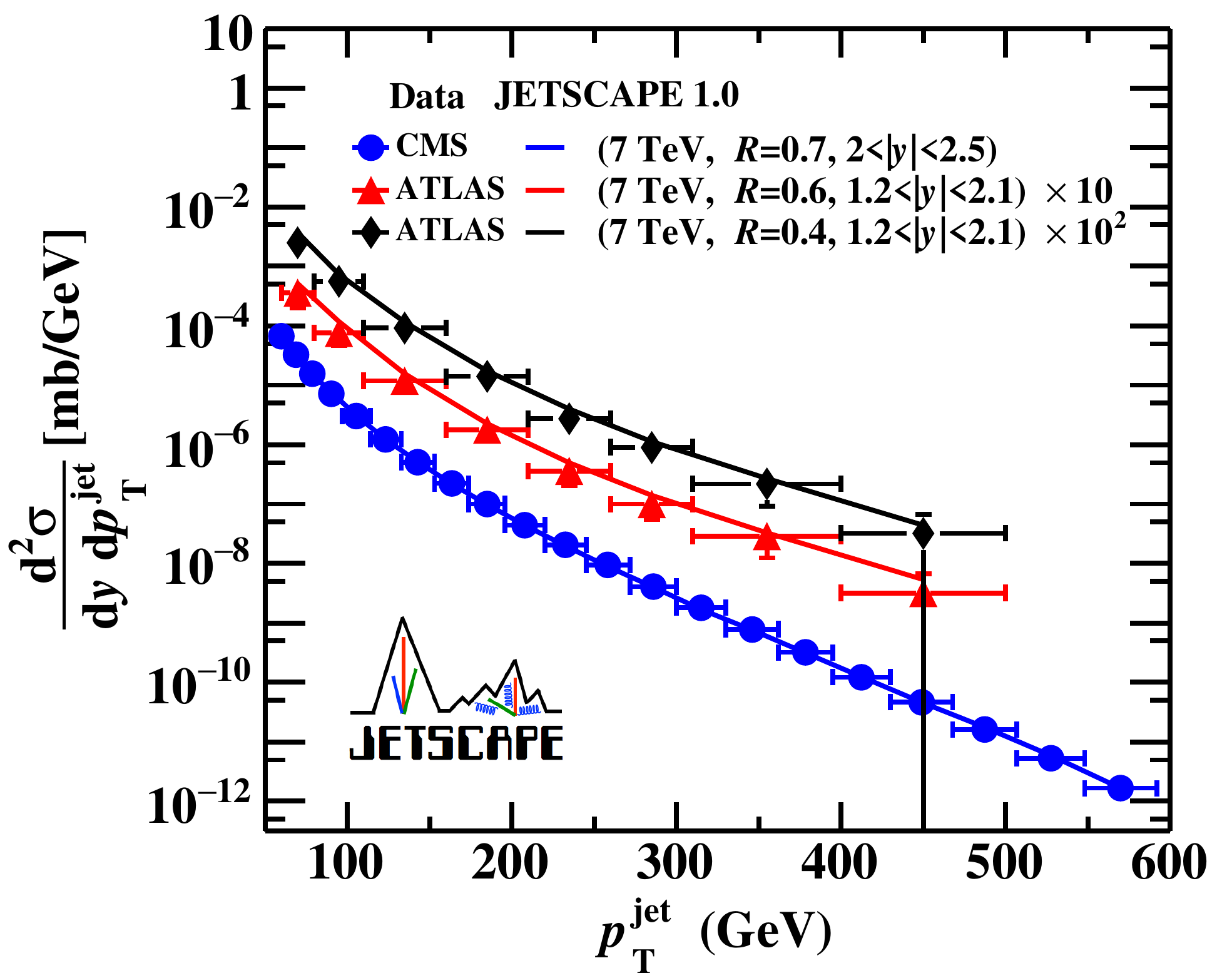}
\hspace{10pt}
\includegraphics[width=0.9\columnwidth]
{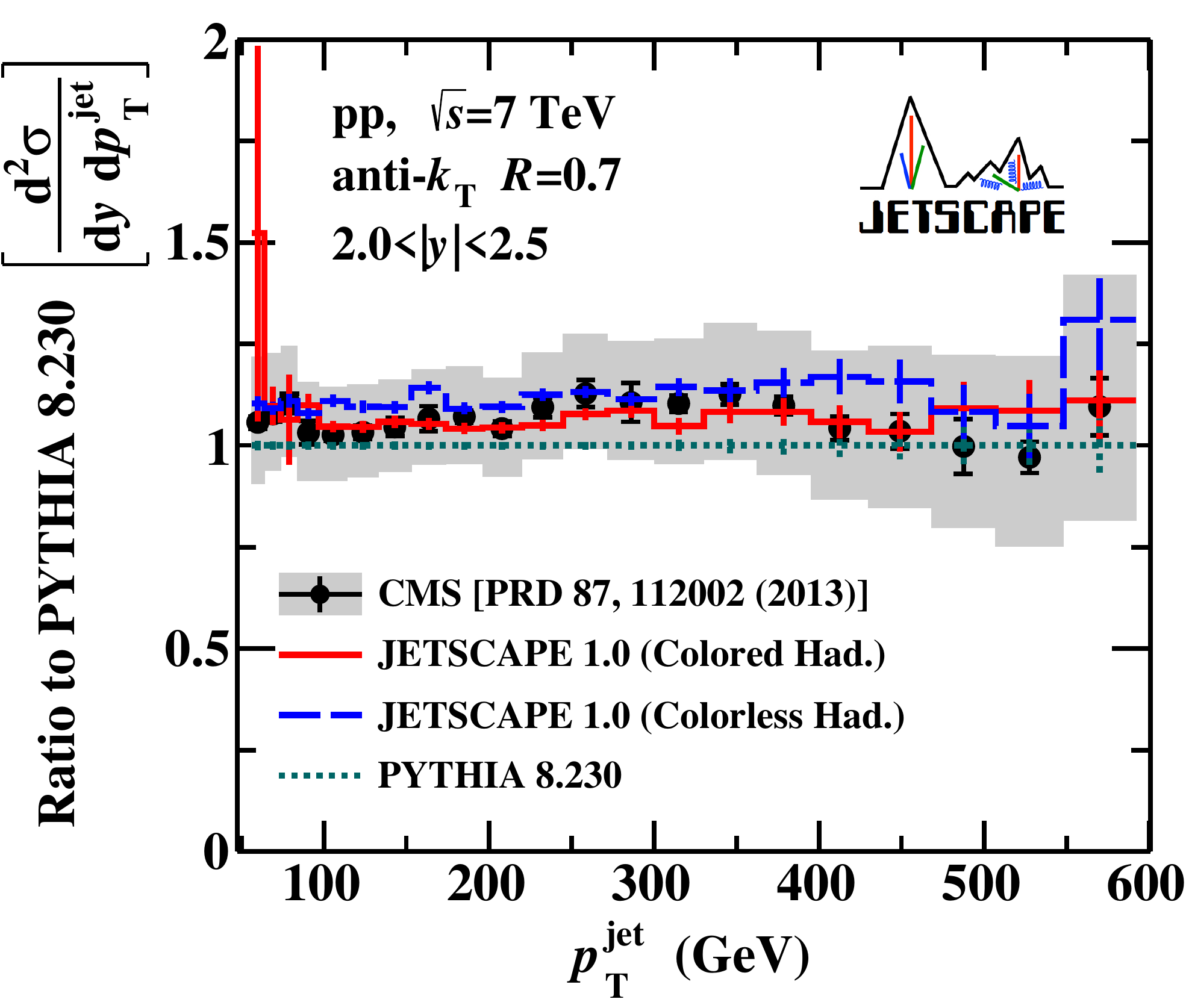}
\\\vspace{15pt}
\includegraphics[width=0.9\columnwidth]
{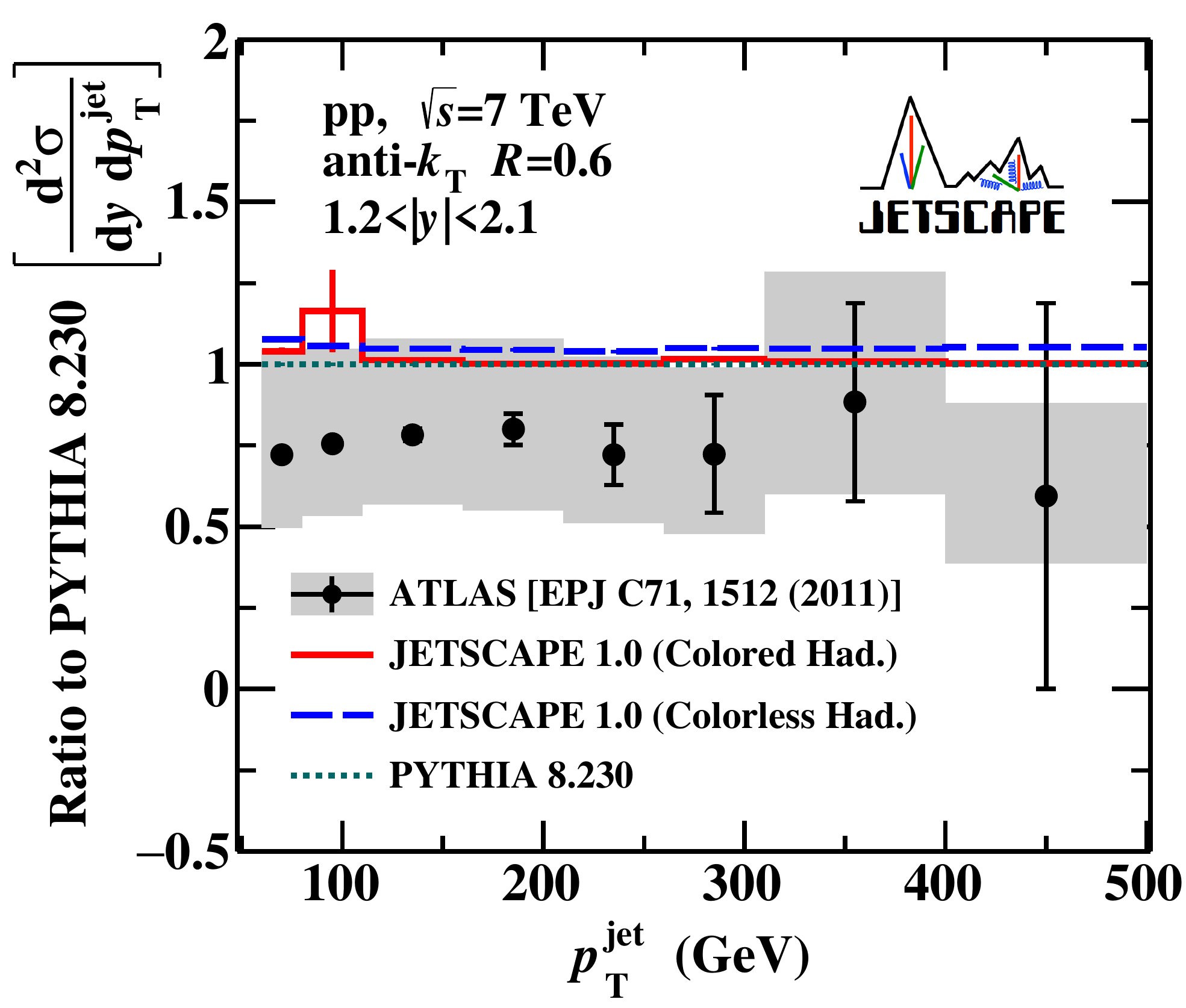}
\hspace{15pt}
\includegraphics[width=0.9\columnwidth]
{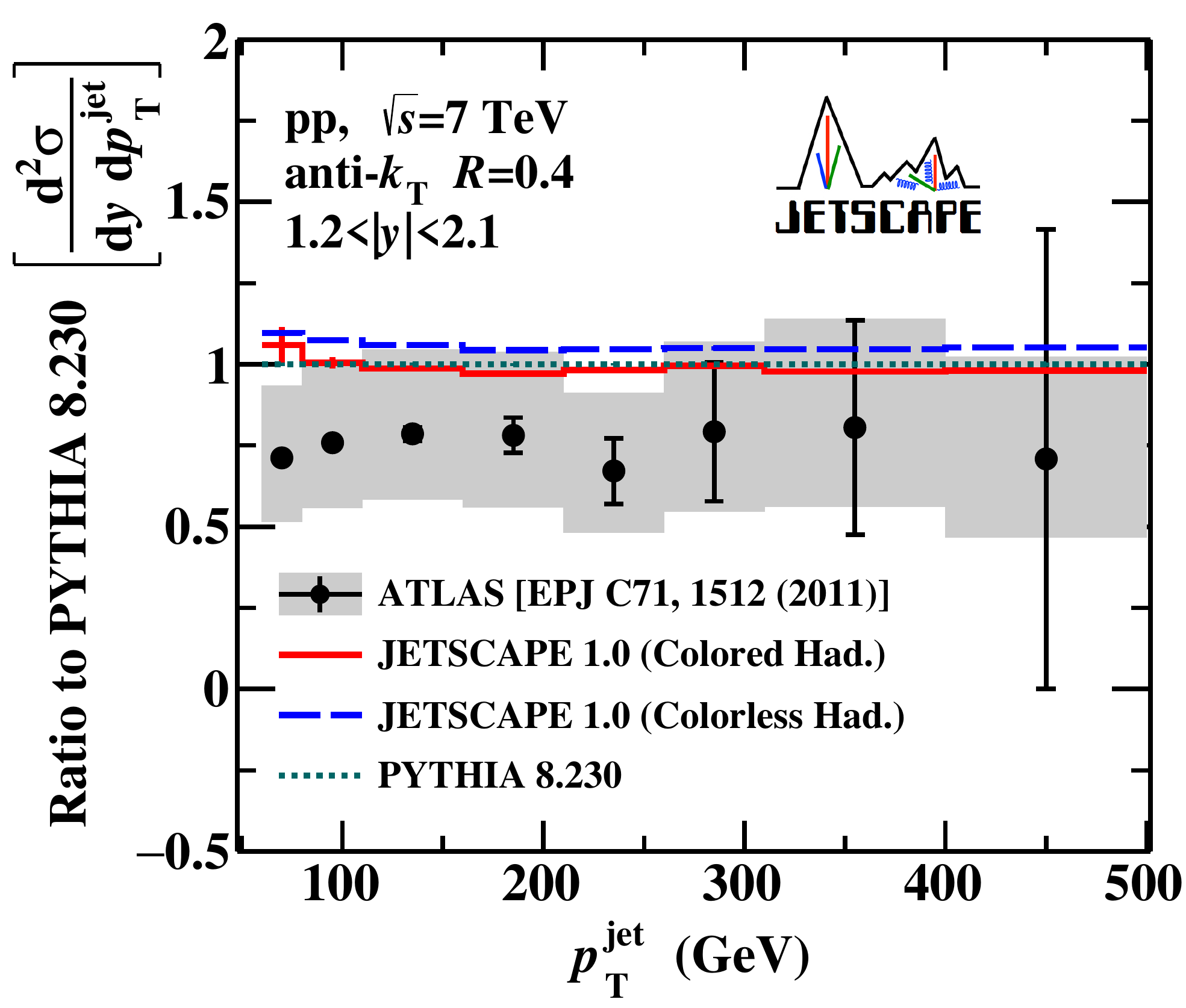}
\caption{
  Inclusive jet cross sections for full jets at forward rapidity for p+p collisions at 
   $\sqrt{s}=7$ TeV and their ratios to PYTHIA 8 Monte Carlo. 
  Upper left panel: JETSCAPE 1.0 jet cross sections (Colored Hadronization only) and data. Other panels: Ratios of these JETSCAPE calculations and 
  data to PYTHIA 8.
\label{fig:jet7offmid}
}
\end{figure*}

\begin{figure*}[tb]
\hspace{-9pt}
\includegraphics[width=0.64\columnwidth]
{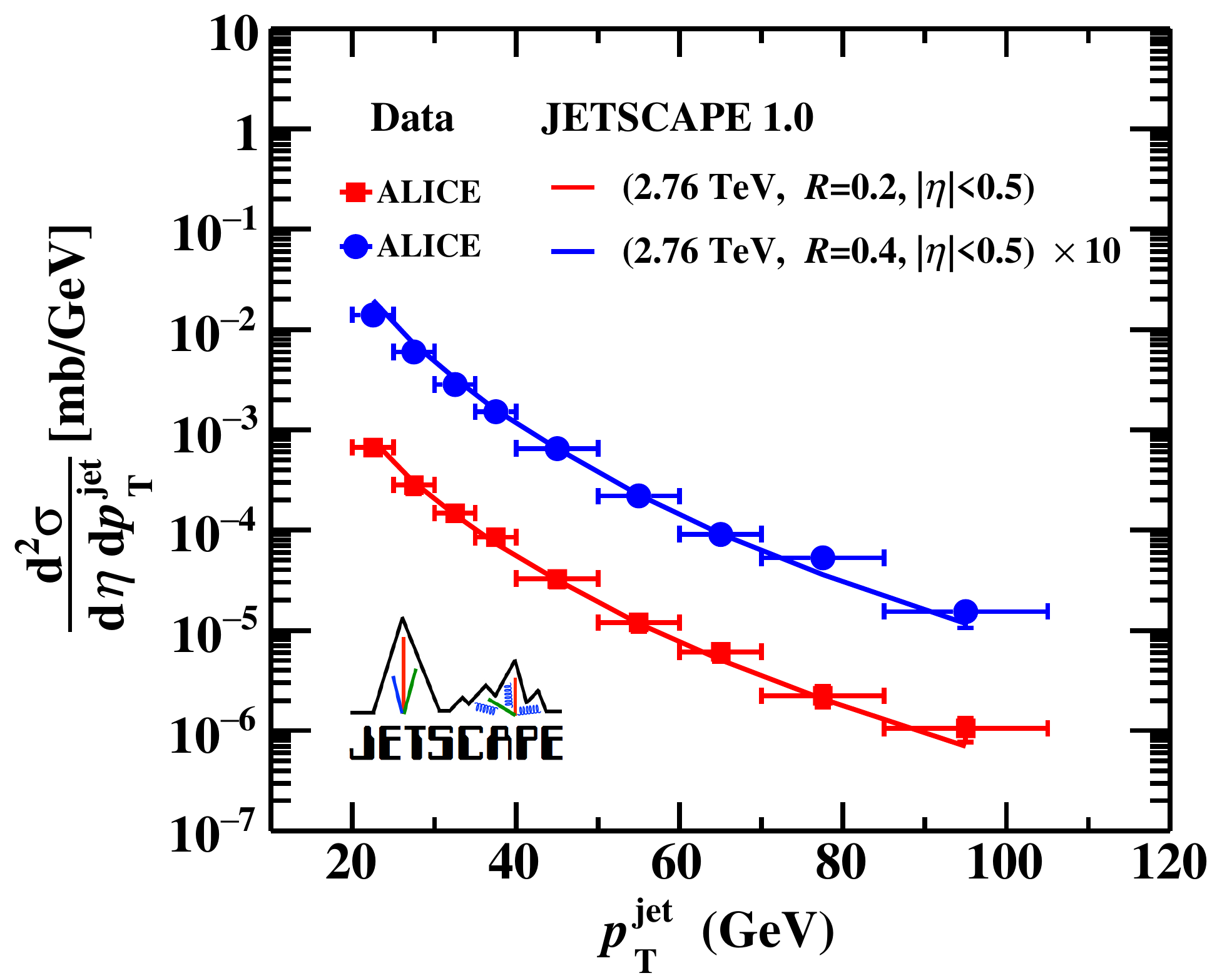}
\includegraphics[width=0.64\columnwidth]
{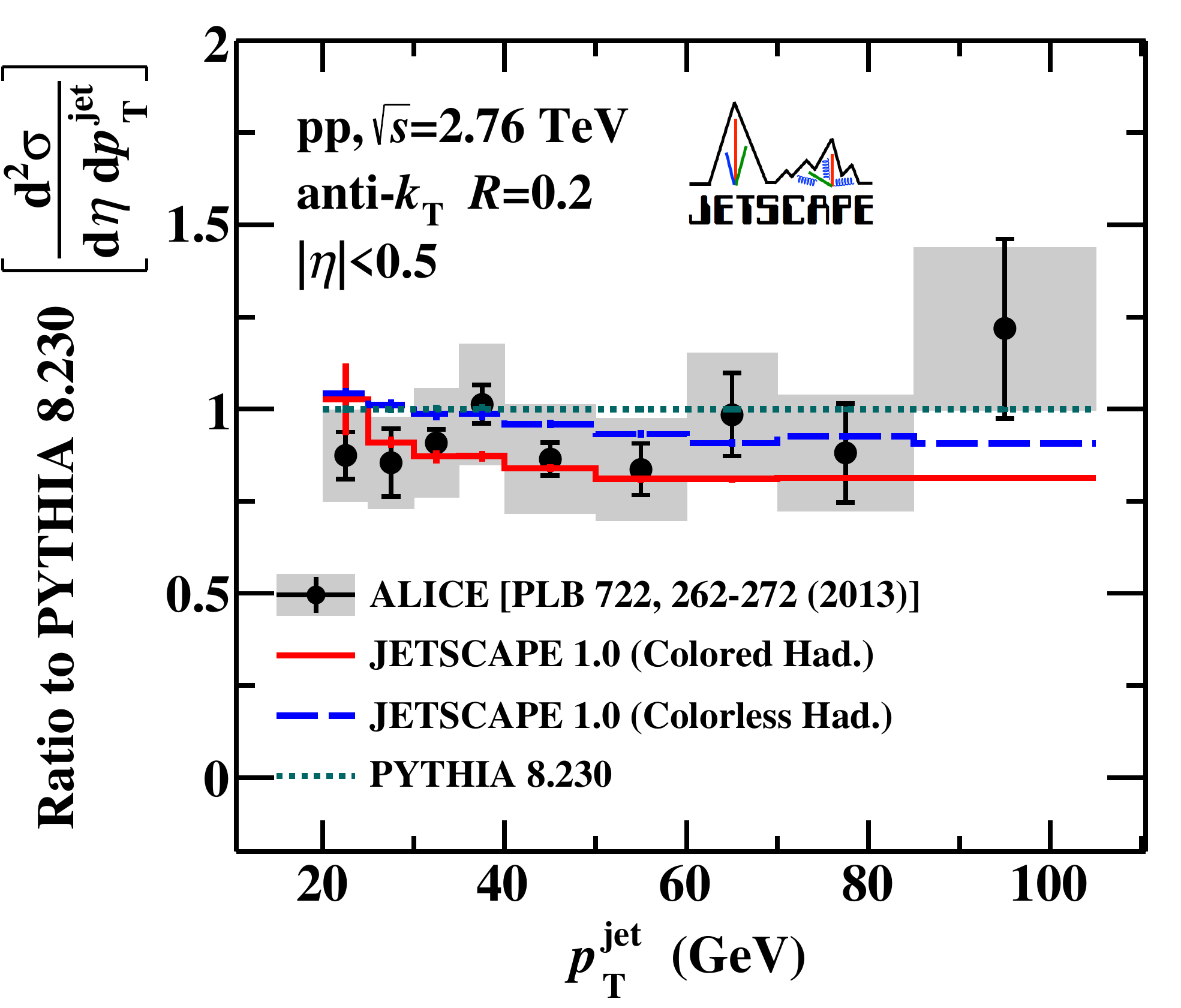}
\includegraphics[width=0.64\columnwidth]
{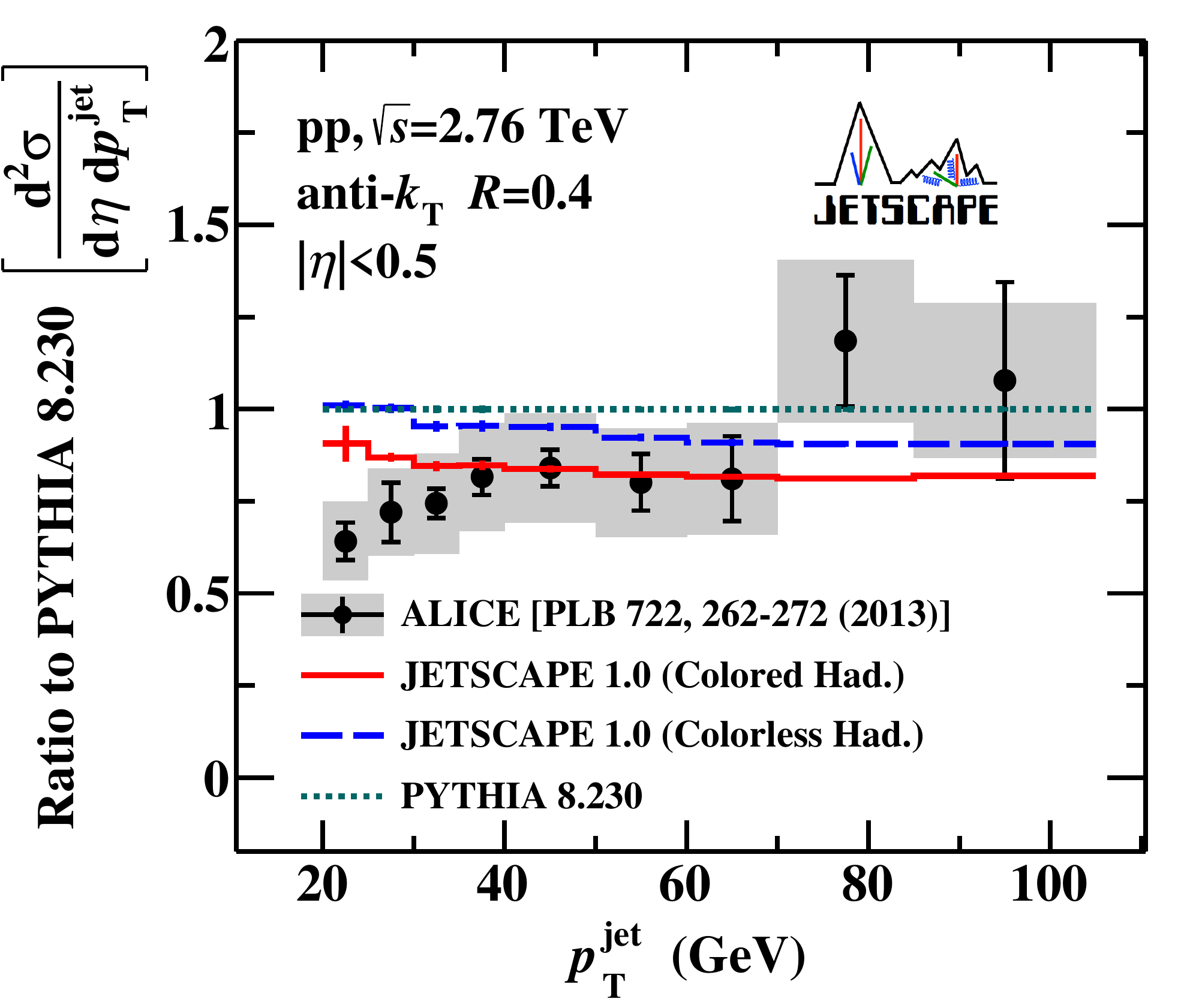}
\caption{
  Inclusive jet cross sections for full jets around midrapidity for p+p collisions at 
   $\sqrt{s}=2.76$ TeV for jet radii $R=0.2$ and $R=0.4$. Data is taken from the ALICE experiment \cite{Abelev:2013fn}
  Left panel: JETSCAPE cross sections (Colored Hadronization only) and ALICE data. Center and right panels: Ratios of various calculations and 
  data to PYTHIA 8 for $R=0.2$ and $R=0.4$, respectively.
\label{fig:jet276lowpt}
}
\end{figure*}

\subsection{The PP19 tune}
\label{subsec:tune}

The JETSCAPE PP19 tune is defined as the workflow in Fig.\ \ref{fig:workflow}, with two choices for string hadronization, and the list of
parameters settings given in Tab.\ \ref{tab:settings}, which have been optimized for $p+p$ calculations. 
PYTHIA 8.230 parameters that are not mentioned explicitly are kept at the default values.

In JETSCAPE PP19 hard QCD processes are initialized by PYTHIA 8.230. Multi-parton interactions (MPI) and initial state radiation (ISR) are 
switched on. Electroweak processes are switched off at present. Final state radiation in PYTHIA 8 is switched off to allow MATTER to take over that task. The complete event record at this stage is extracted. 
Gluons and light quarks ($u$, $d$, $s$) with transverse momentum $p_T >2$ GeV/$c$ are retained.

Each parton is assigned an initial maximum virtuality $Q_\mathrm{ini} = 0.5 p_T$ when handed over to MATTER. Parton showers in MATTER are evolved to a virtuality cutoff set to $Q_0 = 1$ GeV. All partons from MATTER output are handed over to one of the two string formation modules. The resulting string systems are fed back into PYTHIA 8 for string fragmentation. 
The cutoff for the decay length $c\tau$ is set to 1 cm, appropriate for comparison to measurements of jets and unidentified charged hadrons.
Identified hadrons might require different decay settings which should be chosen to reflect conditions in experimental data taking and analysis.

There are two parameters in MATTER that are explicitly optimized for PP19, the proportionality constant between the initial maximum 
virtuality $Q_\mathrm{ini}$ and parton $p_T$, and the value for the QCD scale parameter $\Lambda_\mathrm{QCD}$. 
Values of $Q_\mathrm{ini}/p_T = 0.5$ and $\Lambda_\mathrm{QCD} = 0.2$ GeV provide the best description
of single inclusive jet cross sections and other results to be discussed in the next section.
We have opted not to retune PYTHIA 8 for use in JETSCAPE 1.0.
It is possible that simultaneous tuning of PYTHIA 8, MATTER, and JETSCAPE hadronization could give results in better agreement with 
data than that achieved by tune PP19.

We can classify the uncertainties of JETSCAPE calculations in the following way: (i) Uncertainties shared with PYTHIA 8, e.g.\ from the leading order treatment of hard processes, 
uncertainties in parton distribution functions (PDFs), etc. (ii) Uncertainties from the MATTER shower Monte Carlo. (iii) Uncertainties from 
hadronization. (iv) Uncertainties from the treatment of the underlying event. We estimate these uncertainties by comparing 
JETSCAPE calculations with different hadronization options, and by comparing with PYTHIA 8.230 and data. Specifically, the 
comparison of Colored and Colorless Hadronization, which make different assumptions about string formation, give an estimate of the
uncertainty due to our incomplete knowledge of the hadronization process (iii). Comparison of JETSCAPE PP19 results with 
PYTHIA 8 results show in addition the differences in final state shower Monte Carlos and UE treatment. Hence they can provide an 
estimate of combined uncertainties of type (ii) and (iv). Lastly, the comparison of both JETSCAPE and PYTHIA 8 to data allows us to 
assess the combined uncertainties (i)-(iv). In some cases we also add observables calculated with the final parton output before hadronization to 
show the absolute size of hadronization effects.
As an example, if the two JETSCAPE calculations agree within experimental errors but deviate significantly from PYTHIA 8 and data, we may infer that 
for this particular observable uncertainties in modeling hadronization are small, but variations in details of the shower Monte Carlo and underlying event treatment have an effect that is larger than experimental uncertainties.

NLO calculations and beyond have been carried out for many observables either analytically, with hadronization effects estimated or
added by Monte Carlo \cite{Khachatryan:2015luy}, or by using NLO Monte Carlo event generators, e.g.\ POWHEG \cite{Frixione:2007vw}. 
Uncertainties in the case of analytic calculations are usually determined by scale variations and propagation of PDF uncertainties. 
They can be comparable to or exceed experimental uncertainties. We show analytic calculations for some important observables
to indicate the size of their uncertainties. Leading order MC event generators mimic NLO effects to an extent that make them
successful for some observables. 
Nevertheless, there can be differences between LO and NLO calculations that can not be directly assessed 
experimentally. One prominent example is the ratio of quark to gluon jets. This ratio is relevant in A+A collisions because of the different quenching for quark and gluons. The ratio of quark to gluon jets depends foremost on the hard matrix element and parton
distribution functions which are calculated here using PYTHIA 8. In addition there can be a dependence on final state radiation and hadronization 
which leads to a dependence of the ratio on jet radius. This issue has been studied using PYTHIA 8 in Ref.\ 
\cite{He:2018xjv} and analytically, e.g., in \cite{Dasgupta:2014yra,Qiu:2019sfj}. 
A more careful study of this issue in $p+p$ and A+A using JETSCAPE is useful but lies beyond the scope of this work.

\section{Results}
\label{sec:jetresults}

In this section we discuss results obtained with JETSCAPE PP19 for several observables of jets and high momentum hadrons.
We focus on three collision energies: $\sqrt{s} = 2.76$ TeV and 7 TeV for which data are available from LHC experiments ATLAS, CMS and ALICE, and $\sqrt{s} = 200$ GeV for which the STAR and PHENIX experiments have taken data at RHIC. 
We use both Colored and Colorless Hadronization in order to estimate uncertainties from hadronization. We also perform
the same calculation with PYTHIA 8.230 as defined in the previous section for comparison.

\subsection{Inclusive jet cross sections}
\label{subsec:jetxsec}

\begin{figure*}[tb]
\hspace{-9pt}
\includegraphics[width=0.95\columnwidth]{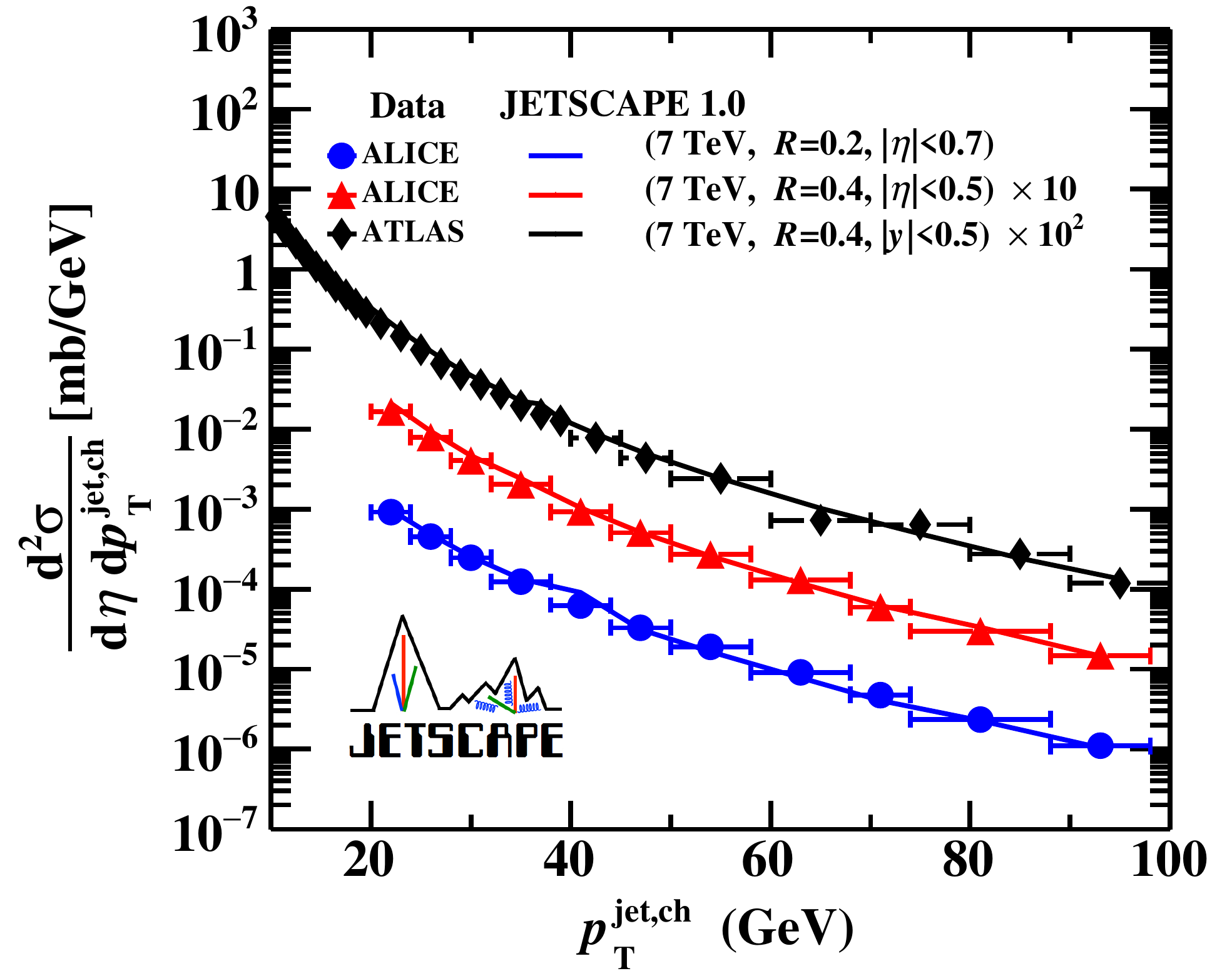}
\hspace{10pt}
\includegraphics[width=0.9\columnwidth]
{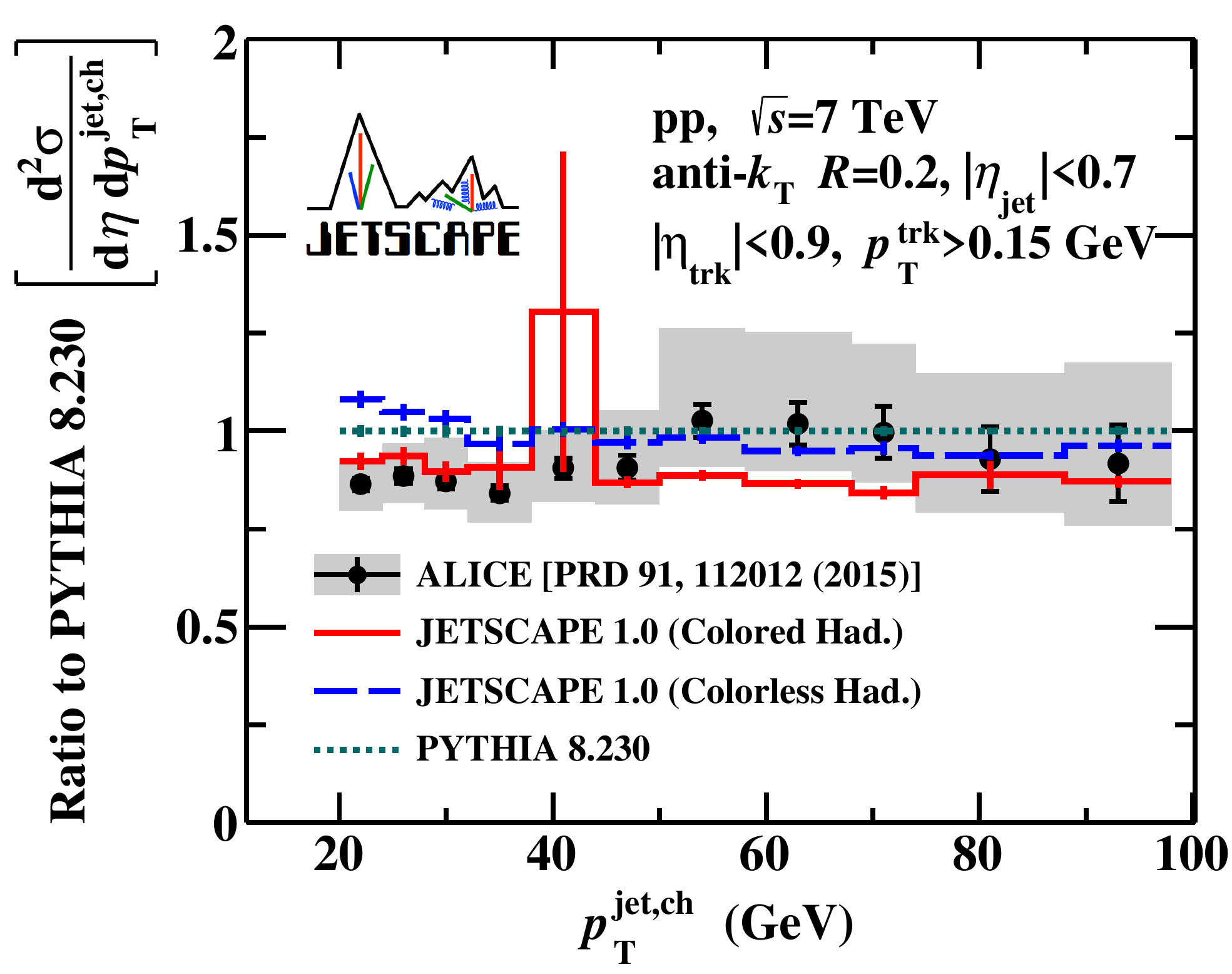}
\\\vspace{15pt}
\includegraphics[width=0.9\columnwidth]
{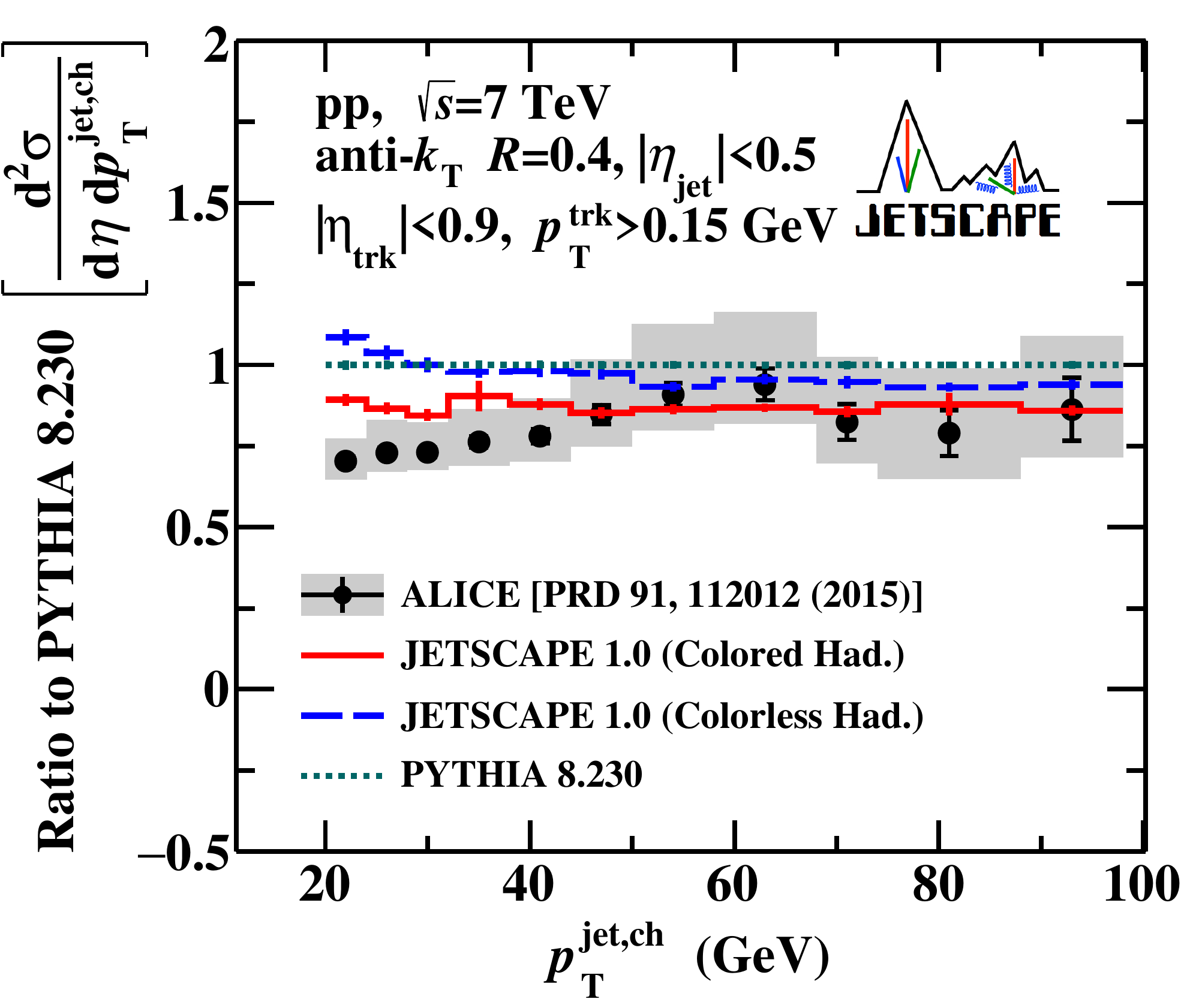}
\hspace{15pt}
\includegraphics[width=0.9\columnwidth]
{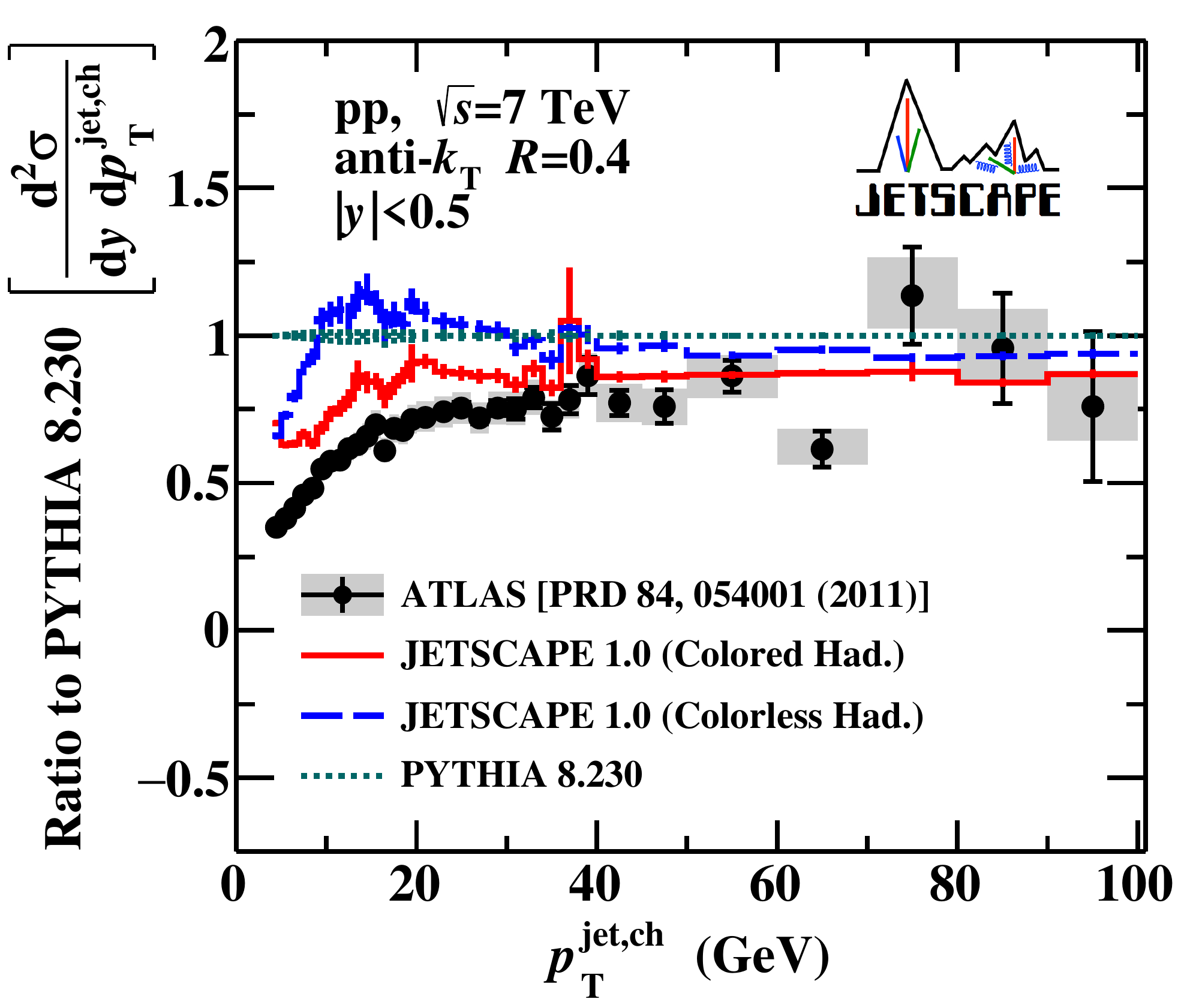}
  \caption{
  Inclusive jet cross sections $d^2\sigma/dp_Td\eta$ vs jet transverse momentum $p_T$ for charged jets at collision energy $\sqrt{s} = 7$ TeV and 
  their ratios to the PYTHIA 8 Monte Carlo results. Results of jets for $R=0.2$ and $|\eta|<0.7$, $R=0.4$ with $|\eta|<0.5$ are compared with the ALICE 
  data \cite{ALICE:2014dla}. We also calculate charged jets with $R=0.4$ with $|y|<0.5$ compared with ATLAS data \cite{Aad:2011gn}. 
Upper left panel: Differential cross section. Upper right panel: Ratios to default PYTHIA 8 in the $R=0.2$ and $|\eta|<0.7$ case and ALICE data.
Lower left panel: The ratios for $R=0.4$ with $|\eta|<0.5$ and ALICE data. Lower right panel: Ratios using $R=0.4$ with $|y|<0.5$ and ATLAS data.
 \label{fig:chjet7}}
\end{figure*}

\begin{figure*}[tb]
\includegraphics[width=0.9\columnwidth]{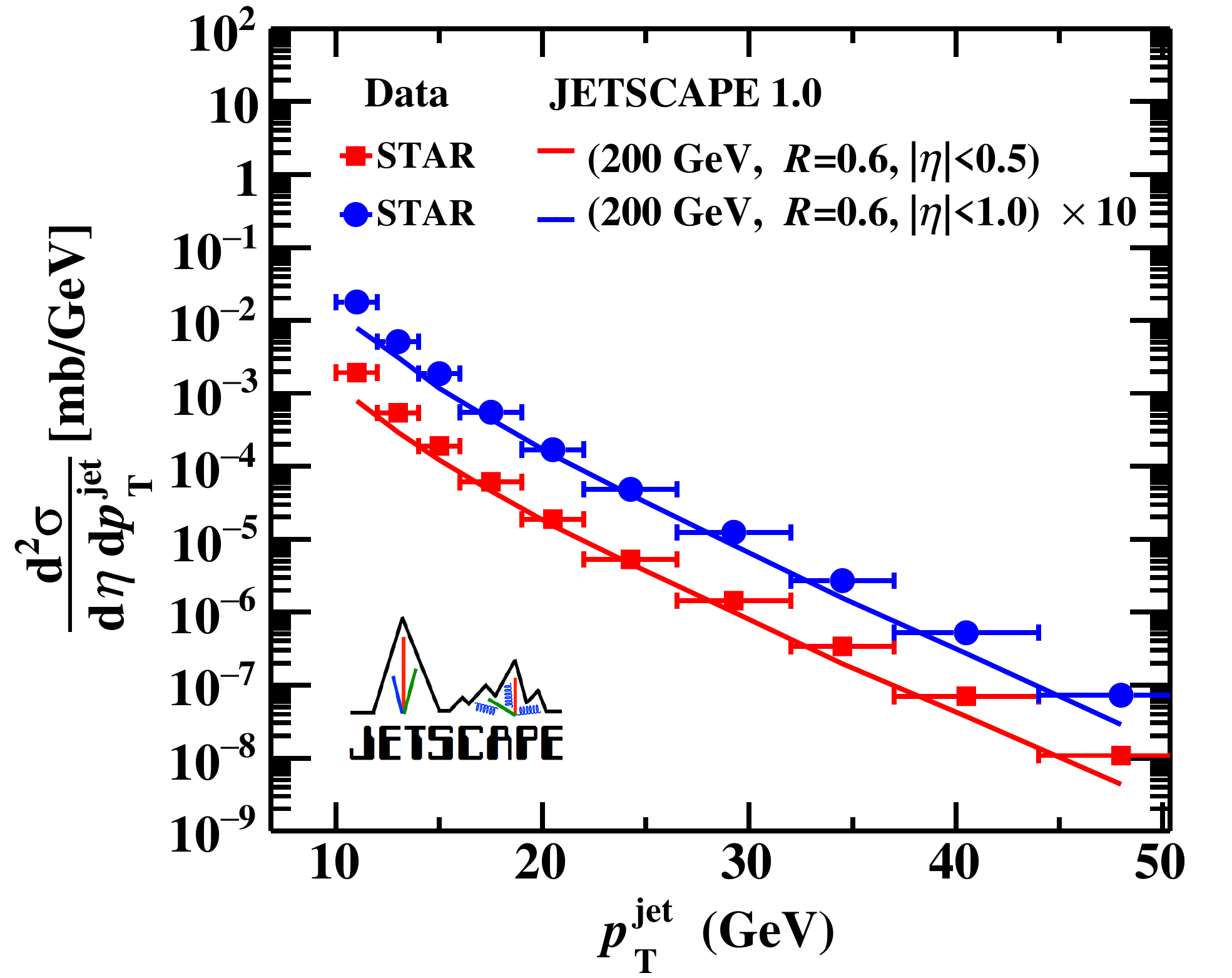}
\\\vspace{15pt}
\includegraphics[width=0.9\columnwidth]
{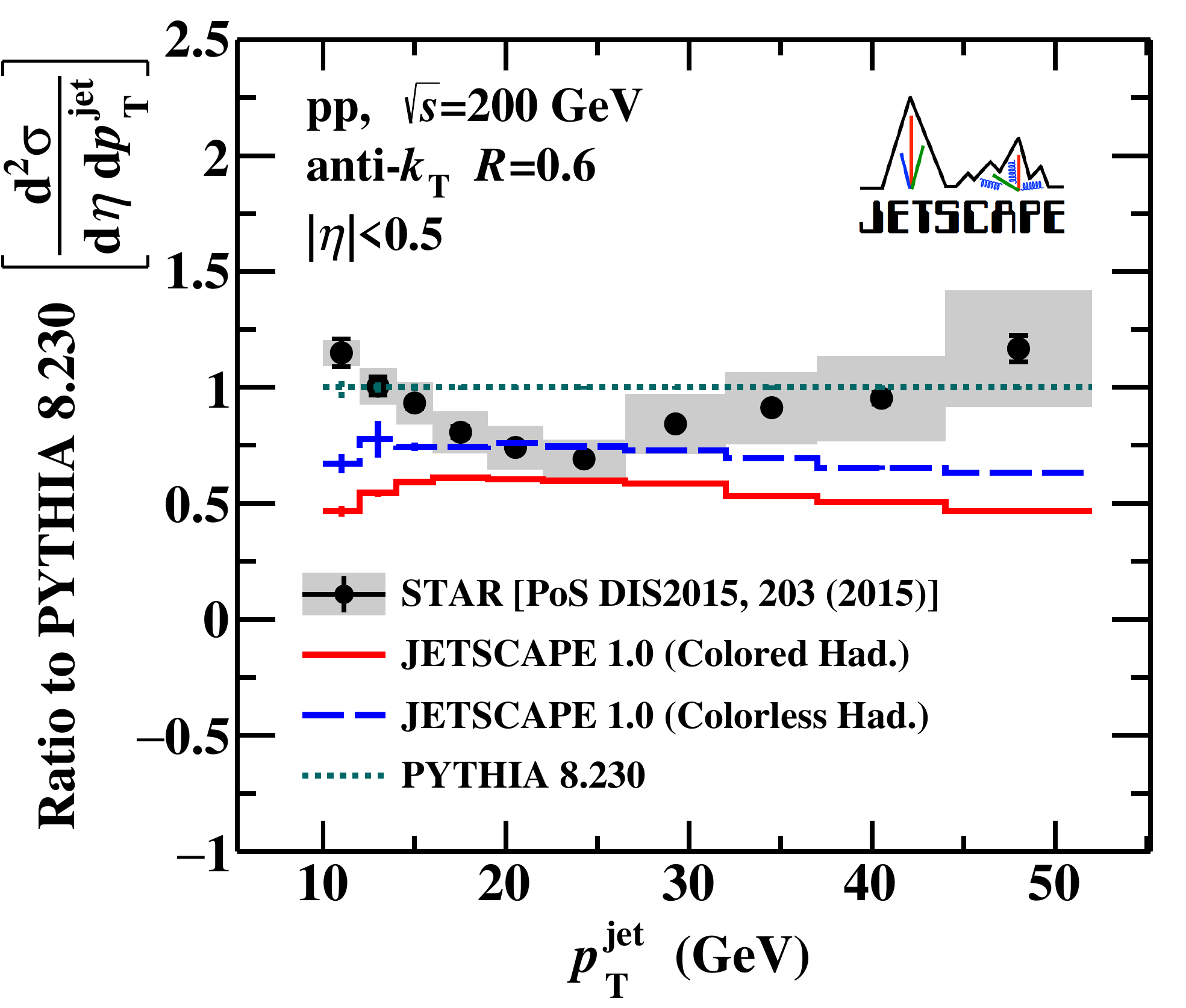}
\hspace{15pt}
\includegraphics[width=0.9\columnwidth]
{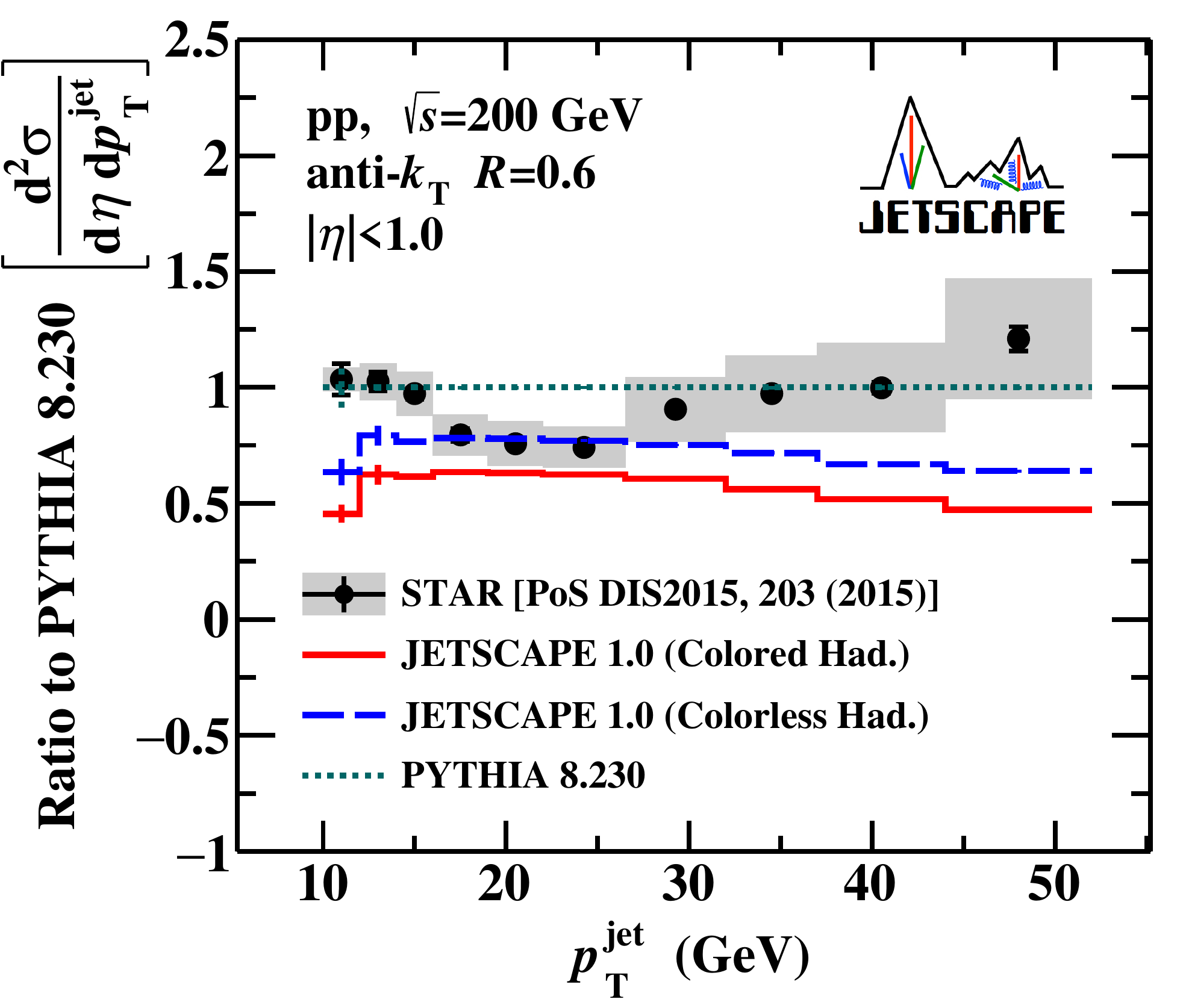}
    \caption{Inclusive jet cross sections $d^2\sigma/dp_Td\eta$ vs jet transverse momentum $p_T$ for full jets at $\sqrt{s}=200$ GeV and 
    their ratios to the PYTHIA 8 Monte Carlo results. The results are shown for jets radii $R=0.6$ with two rapidity ranges: 
    $|\eta|<0.5$ and $|\eta|<1.0$, and compared with the preliminary data from STAR \cite{Li:2015gna}.
   Upper panel: Differential cross section. Lower panels: Ratios to PYTHIA 8 with $|\eta|<0.5$ (left) and $|\eta|<1.0$ (right).
  \label{fig:jet200}}
\end{figure*}

Single inclusive cross sections of jets 
have been measured at various energies at the LHC, and by the 
STAR experiment at RHIC. 
We use the anti-$k_T$ algorithm \cite{Cacciari:2008gp} implemented in the FASTJET package
\cite{Cacciari:2011ma,Cacciari:2005hq} to define jets based on the hadronic finals state, consistent with experiments. 
First, we check the performance of JETSCAPE PP19 for jets measured at LHC energies for jet transverse momentum 
up to several hundreds of GeV/$c$. We compare to CMS data at $\sqrt{s}=2.76$ TeV \cite{Khachatryan:2015luy} and 
7 TeV \cite{Chatrchyan:2012bja} around midrapidity, 
and to ATLAS data at $\sqrt{s}=7$ TeV at both midrapidity and forward rapidity \cite{Aad:2010ad}.  
We then focus on comparisons to data sets which emphasize jet momenta below 100 GeV/$c$. Those are available for fully-reconstructed jets 
from ALICE \cite{Abelev:2013fn}, and for charged jets from ALICE \cite{ALICE:2014dla} and ATLAS \cite{Aad:2011gn}.
Lastly we present calculations for RHIC energies compared to data from STAR \cite{Li:2015gna} .

Calculations using JETSCAPE PP19 for jets around midrapidity at the LHC are shown in 
Figs.\ \ref{fig:jet276+7mid} through \ref{fig:jet276midratio} together with data, compared to the reference calculation using PYTHIA 8.
In the cross sections plots in Fig.\ \ref{fig:jet276+7mid} we only show results for Colored Hadronization; Colorless Hadronization 
and PYTHIA 8 reference results are indistinguishable. Data from CMS and ATLAS \cite{Khachatryan:2015luy,Chatrchyan:2012bja,Khachatryan:2016jfl,Aad:2010ad} 
are overlaid for comparison. 
Figs.\ \ref{fig:jet276+7midratio1} through \ref{fig:jet276midratio} show ratios
of JETSCAPE results for both hadronization models, and data with the PYTHIA 8 reference calculation.
Figure \ref{fig:jet7offmid} 
shows the differential cross sections for jets at forward rapidity two ($2<|y|<2.5$ for $R=0.7$, and $1.2<|y|<2.1$ for
$R=0.6$, $0.4$), and the ratios of JETSCAPE calculations and data to PYTHIA 8. Data from CMS and ATLAS 
\cite{Chatrchyan:2012bja,Aad:2010ad} are used for comparison.

For single inclusive jet cross sections the two JETSCAPE string formation models give compatible results, typically
with less than 10\% deviation. Deviations are smaller at midrapidity and for larger jet radii.
The discrepancy is typically within the uncertainties of available data except for very small jet radii where deviations between the two hadronization models reach 20\%.
This is consistent with expectations that hadronization effects are largest for small jet radii.

Results from JETSCAPE PP19 are compatible within uncertainties with data from CMS with $R=0.7$ for all energies and rapidities considered. For ATLAS data this 
is the case at midrapidity, but JETSCAPE calculated cross sections are displaced from the central values for ATLAS data at forward rapidities, though still within uncertainties.
Results from PYTHIA 8 tend to be similar to JETSCAPE PP19 results but deviations are generally larger than the difference between JETSCAPE hadronization models, suggesting 
the importance of differences in final state shower Monte Carlos and underlying event treatment.
PYTHIA 8 and JETSCAPE are consistent within the uncertainties of the ATLAS data, but the smaller uncertainties of the CMS data seem to prefer JETSCAPE results.

\begin{figure*}[tb]
	\includegraphics[width=0.9\columnwidth]{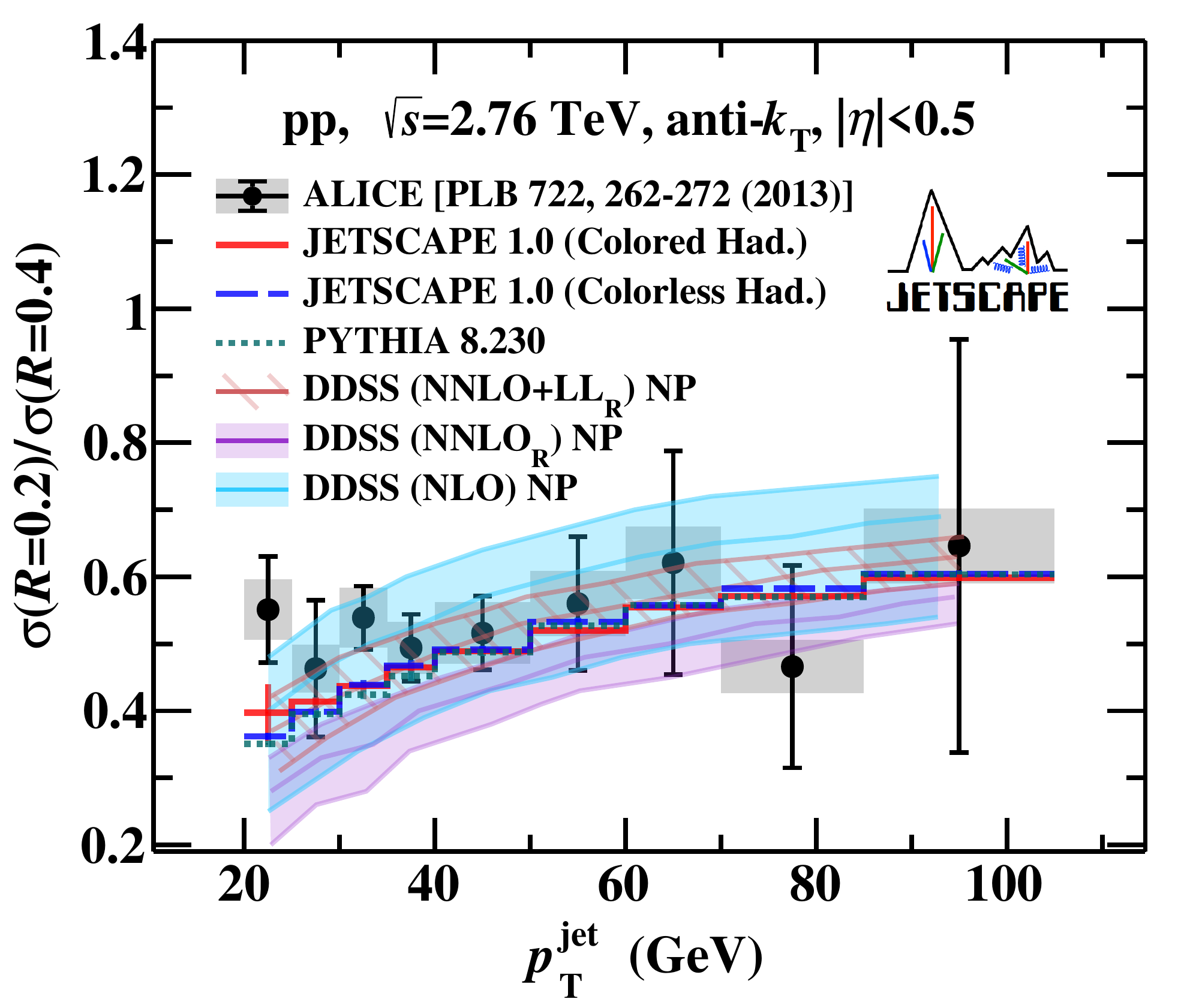}
  \includegraphics[width=0.9\columnwidth]{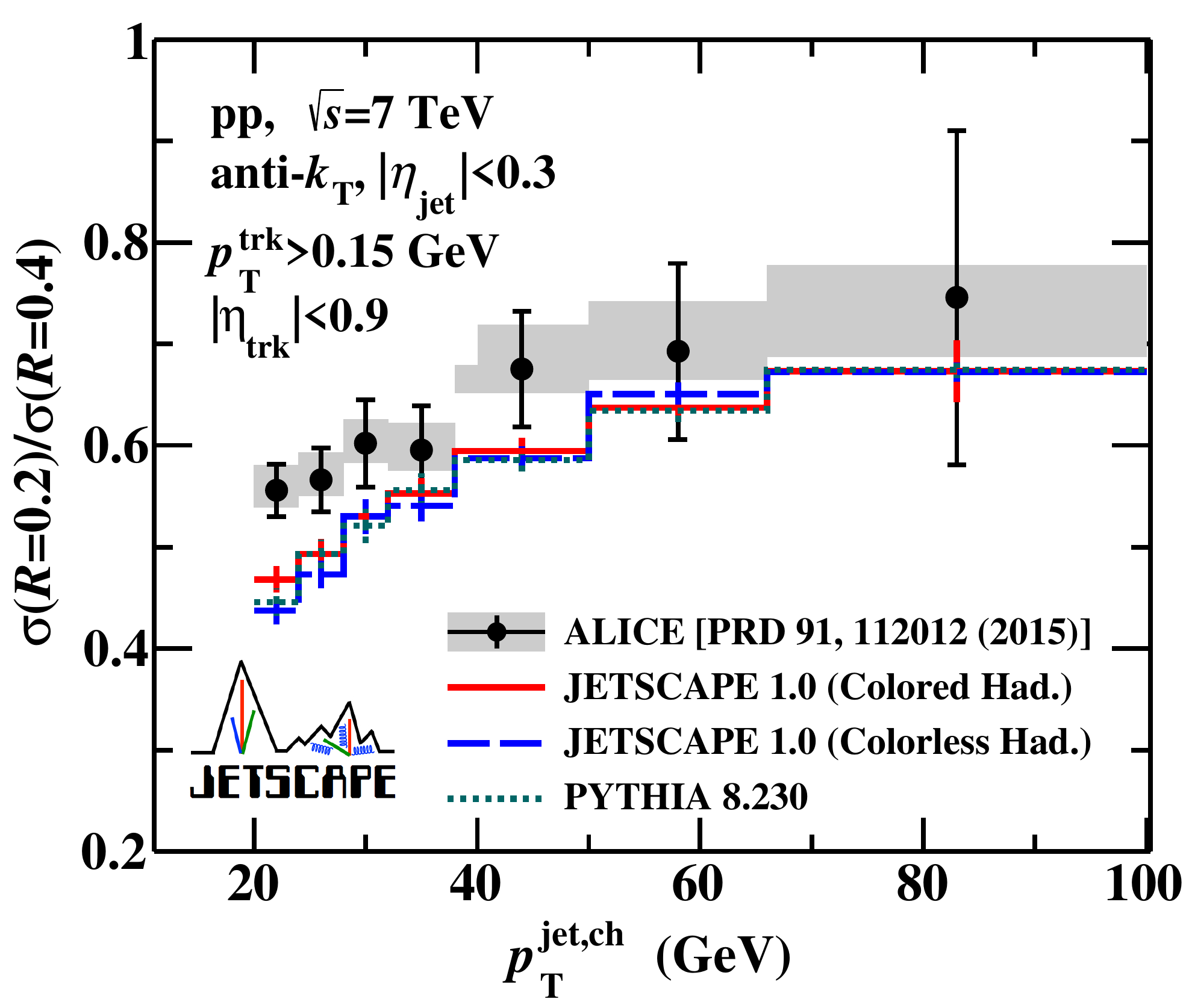}
  \caption{The ratio of jet cross sections for radii $R=0.2$ and $R=0.4$ at jet momenta smaller than 100 GeV/$c$.
   Left panel: Full jets at collision energy $\sqrt{s}=2.76$ TeV and rapidity $|\eta|<0.5$ compared to ALICE data \cite{Abelev:2013fn}. Analytic calculations 
   (DDSS) \cite{Dasgupta:2016bnd} are included for comparison and shown as bands. Right panel: Charged jets at $\sqrt{s}=7$ TeV and rapidity $|\eta|<0.3$ 
   compared to ALICE data \cite{ALICE:2014dla}  
}
  \label{fig:lowptjetratio7R02R04}
\end{figure*}

\begin{figure*}[b]
  \includegraphics[width=0.9\columnwidth] {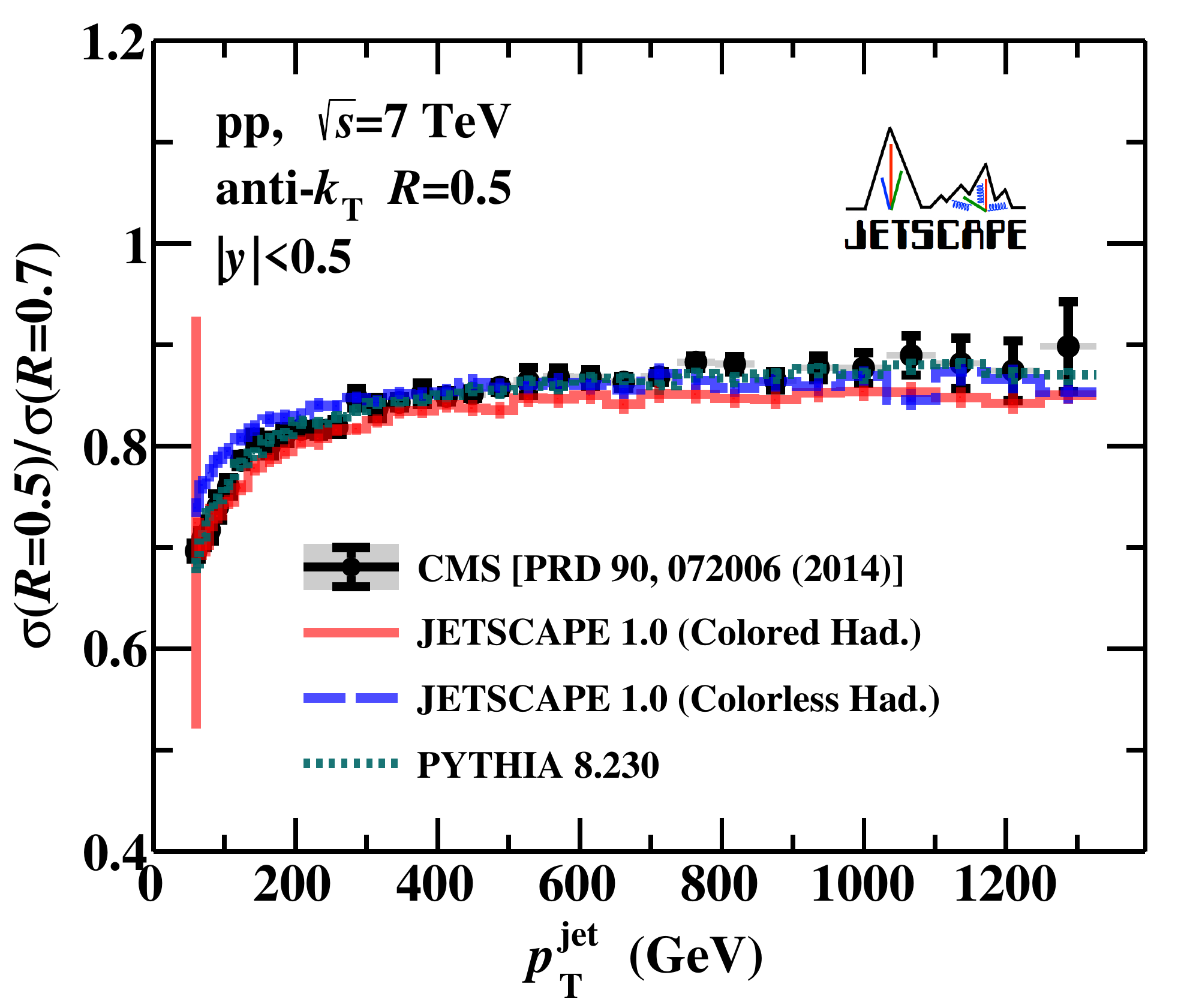}
  \hspace{15pt}
    \includegraphics[width=0.9\columnwidth] 
{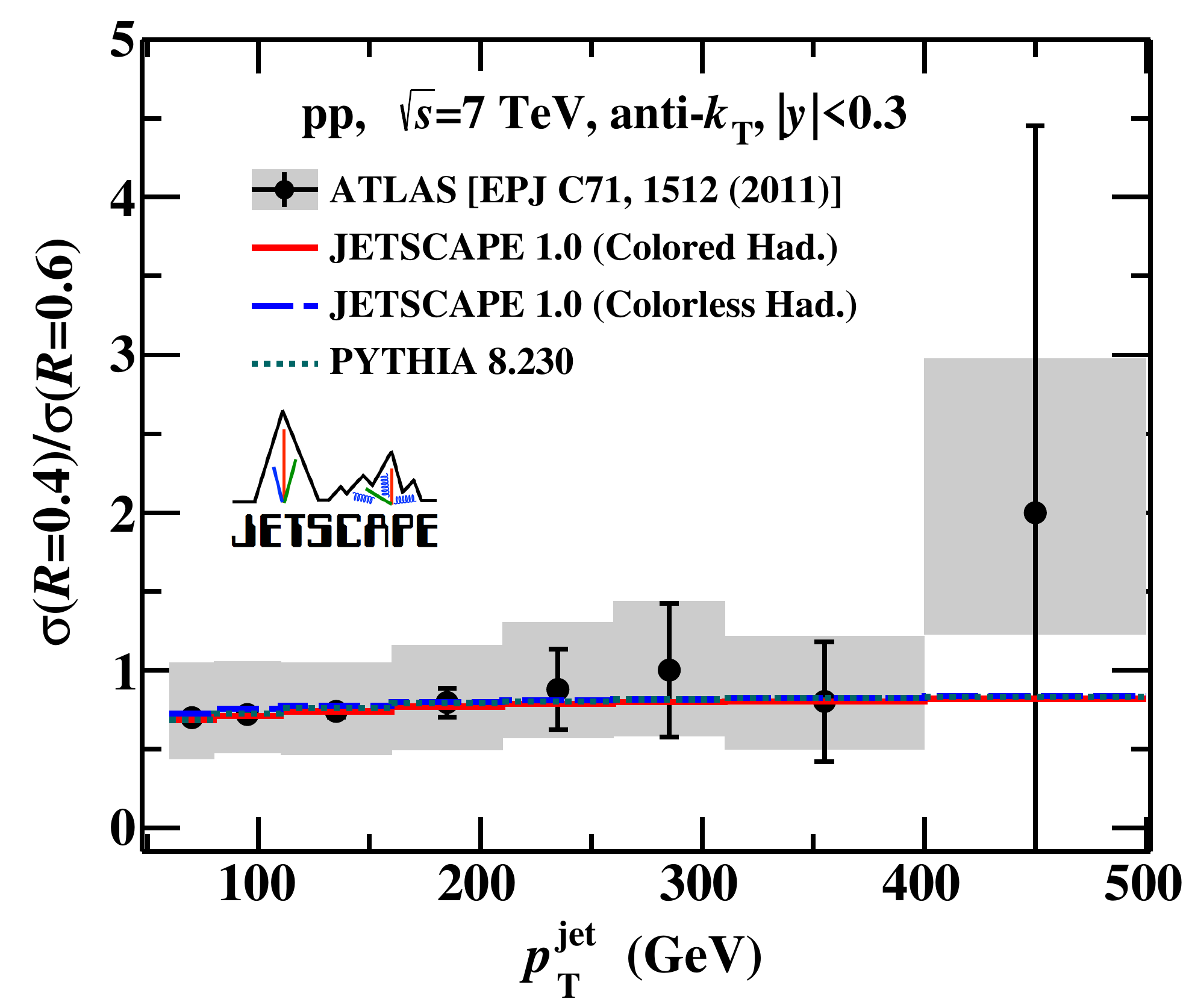}
    \caption{The same as Fig.\ \ref{fig:lowptjetratio7R02R04} for full jet cross sections at 7 TeV. 
Left panel: Ratio $R=0.5$ over $R=0.7$ and rapidity $|y|<0.5$ measured by CMS \cite{Chatrchyan:2014gia}. 
Right panel: Ratio $R=0.4$ over $R=0.6$ for rapidity $|\eta|<0.3$ measured by ATLAS \cite{Aad:2010ad}.
  \label{fig:jetratio276+7R02R04R06}}
\end{figure*}

Inclusive jet cross section measurements for jet transverse momenta down to 20 GeV/$c$ are available from the ALICE collaboration 
at $\sqrt{s}= 2.76$ TeV \cite{Abelev:2013fn}. 
A comparison of JETSCAPE calculations with these data sets for $R=0.2$ and $R=0.4$, spectra and ratios to PYTHIA 8, are presented in Fig.\ \ref{fig:jet276lowpt}. 
Similar low momentum data are available for charged jets at LHC energies. 
Fig.\ \ref{fig:chjet7} shows the single inclusive differential cross section for charged jets 
at collision energies $\sqrt{s}= 7$ TeV around midrapidity. The cross sections for radii $R=0.2$ and $R=0.4$ are calculated with JETSCAPE PP19
and compared to data from ALICE \cite{ALICE:2014dla} and ATLAS \cite{Aad:2011gn}. 
Fig.\ \ref{fig:chjet7} also shows ratios of results obtained with both hadronization models and of data with PYTHIA 8.

The uncertainty attributable to hadronization is similar to that at large momenta. Deviations between hadronization models 
increase toward smaller jet momenta. At the smallest $p_T^\mathrm{jet}$ and for small jet radii differences between JETSCAPE hadronizaton models become significant, up to 20-30\% around $p_T = 20$ GeV/$c$. 
This sensitivity to hadronization effects is expected: For small jet radii $R$
the average shift in transverse momentum is estimated to be $\langle p_T^{\rm jet}\rangle \sim - C_i/R$ where $C_i=4/3$, $3$ is
the appropriate color charge for quark or gluon jets \cite{Dasgupta:2007wa}. This momentum shift can reach several GeV/$c$ for $R=0.2$.

PYTHIA 8 calculations are similar to JETSCAPE results, with differences compatible with the uncertainties from hadronization.
All three Monte Carlo calculations are consistent with ALICE and ATLAS data within experimental uncertainties for $\sqrt{s}=7$ TeV for
$p_T^\mathrm{jet}\gtrsim 40$ GeV/$c$. Below 40 GeV/$c$ all three Monte Carlo calculations overestimate the measured jet cross sections, 
with JETSCAPE Colored Hadronization typically being closest to data.

We finally turn to $p+p$ collisions at $\sqrt{s}=0.2$ GeV. Fig.\ \ref{fig:jet200} shows the single inclusive differential cross sections for jets with jet radius $R=0.6$ for a narrow ($|\eta|<0.5$) and a wide ($|\eta|<1.0$) rapidity interval. Preliminary data from STAR is taken from Ref.\ \cite{Li:2015gna}. 
In the same figure we plot the ratios of differential cross sections from both JETSCAPE hadronization models and data to PYTHIA 8.
We find that differences between JETSCAPE hadronization models and between JETSCAPE and PYTHIA 8 are typically on the level of 20-30\%, 
similar to previous results at low jet transverse momentum at LHC. The spread between Monte Carlo results is larger than 
the size of the STAR uncertainties. STAR data fall between PYTHIA 8 and JETSCAPE Colorless Hadronization results with JETSACPE Colored Hadronization 
being disfavored by STAR preliminary data.

\subsection{Transverse jet structure}
\label{subsec:jetshape}

The distributions of energy or particles transverse to the jet axis gives insights into the structure of QCD parton showers. 
They are also sensitive to hadronization effects. For jets in a medium they can be used to explore the interplay of jets with quark 
gluon plasma. In this section we discuss the baseline that we obtain with MATTER showers and string hadronization in $p+p$ collisions. 

Two distinct approaches can be found in the literature. The first one compares differential jet cross sections for different jet radii by 
taking ratios of those cross sections. The second approach defines the jet transverse profile $\rho(r)$ as the $p_T$ of all particles at 
a certain distance $r$ from the jet axis, divided by the total $p_T$ in the jet. This is practically achieved in bins of size $\delta r$ 
in radius,
\begin{equation}
  \rho(r) =  \frac{1}{\delta r} \frac{\sum_{i \in (r\pm \delta r/2)}p^i_{T}}{\sum_{i \in (0,R)}p^{i}_{T}}  \, ,
\end{equation}
which is then averaged over jets with cone size $R$. We will discuss examples of both approaches in this subsection.

\subsubsection{Jet cross section ratios}

Figs.\ \ref{fig:lowptjetratio7R02R04} and \ref{fig:jetratio276+7R02R04R06} show ratios of jet cross sections with different jet radii. In Fig.\ 
\ref{fig:lowptjetratio7R02R04} the ratio $R=0.2$ over $R=0.4$ is taken for full jets at $\sqrt{s}=2.76$ TeV and charged jets at $7$ TeV, and 
compared to ALICE data \cite{Abelev:2013fn,ALICE:2014dla}.
Results from both JETSCAPE hadronization models and PYTHIA 8 are consistent with each other, with small deviations below
$p_T^\mathrm{jet} \lesssim 30$ GeV/$c$. Deviations of all 3 Monte Carlo results from ALICE data are more pronounced, 
with Monte Carlo calculations consistent with data only above 40 GeV/$c$. At smaller $p_T^\mathrm{jet}$ 
Monte Carlos predict that cross sections decrease faster with jet cone radius $R$ than observed in data.
In the left panel of Fig.\ \ref{fig:lowptjetratio7R02R04} we include results of analytic calculations by Dasgupta et al.\ \cite{Dasgupta:2016bnd}
‘(denoted by DDSS in the figure legend) at NLO, next-to-next-to-leading order (NNLO), and NNLO with resummation of leading logarithm in small jet radii ($\mathrm{LL}_\mathrm{R}$), supplemented 
by estimates of non-perturbative (NP) effect, together with their estimated uncertainty bands. All three Monte Carlo results are compatible 
with NNLO+$\mathrm{LL}_\mathrm{R}$ calculations. Further improvements in theory are necessary to distinguish between Monte Carlo calculations
using analytic calculations.

In contrast to the previous figure most of the data and calculations in Fig.\ \ref{fig:jetratio276+7R02R04R06} cover jet momenta above 100 GeV/$c$.
The left panel discusses the ratio of $R=0.5$ over $R=0.7$ for full jets at $\sqrt{s}=7$ TeV and rapidity $|y|<0.5$ compared to CMS data
\cite{Chatrchyan:2014gia}. The right panel uses jets ($R=0.4$ and $R=0.6$) at $\sqrt{s}=7$ TeV 
and rapidity $|y|<0.3$ with ATLAS data \cite{Aad:2010ad}. In the latter case all three Monte Carlo calculations are consistent with data and with each other. This is also true for the comparison with CMS data above $p_T^\mathrm{jet} \approx 300$ GeV/$c$. However at smaller momenta the deviations between Colored and Colorless Hadronization exceeds the size of the CMS error bars. Overall, there are indications that very precise experimental data on ratios of inclusive jet cross sections can be a good discriminator between different theoretical calculations.

\begin{figure*}[tb]
\includegraphics[width=0.9\columnwidth]{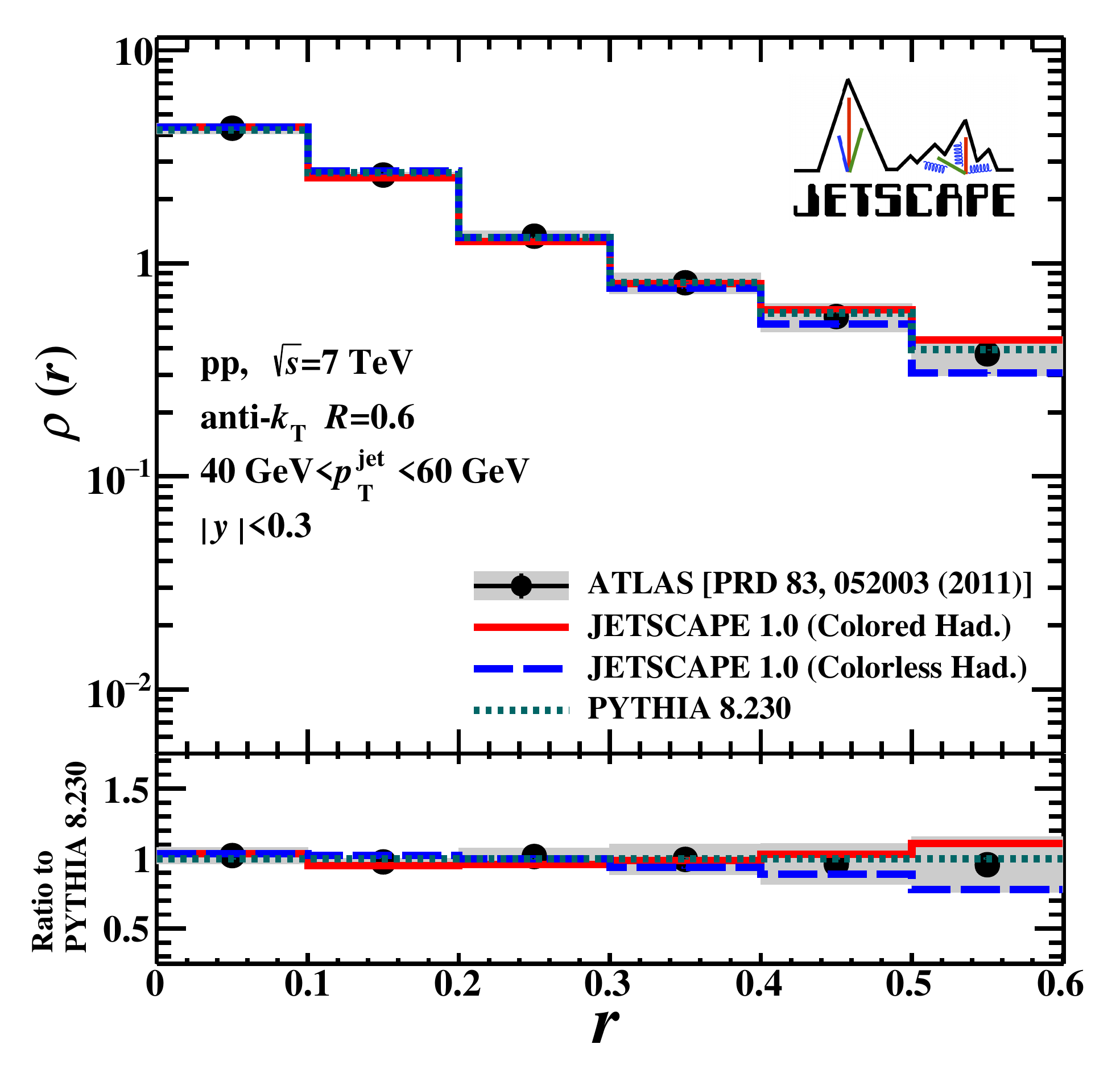}
\hspace{15pt}
\includegraphics[width=0.9\columnwidth]
{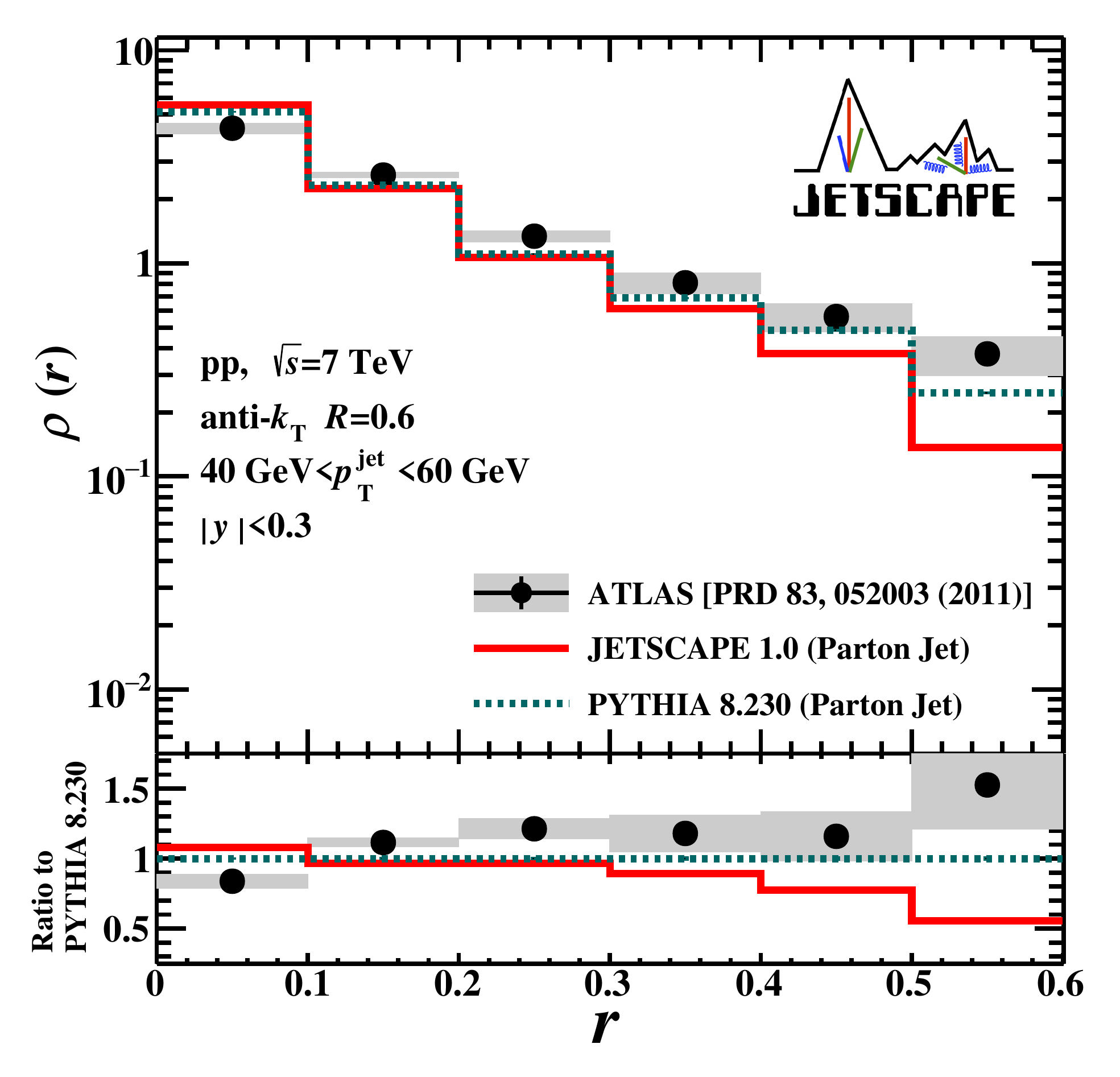}
      \caption{Jet transverse profile $\rho(r)$ for full jets of radius $R=0.6$ in $p+p$ collisions at 7 TeV 
      and its ratio to the PYTHIA 8 Monte Carlo results. 
  Jets are required to have transverse momenta $p_T$ between 40 and 60 GeV/$c$ and rapidities $|y|<0.3$.
  Data points in both panels are from the ATLAS collaboration \cite{Aad:2011kq} (black circles). Monte Carlo
  results are from JETSCAPE Colored Hadronization 
  (solid red line), JETSCAPE Colorless Hadronization (dashed blue line), and PYTHIA 8 (dotted green line). 
  Statistical errors (black error bars) 
  and systematic errors (gray bands) are plotted with the data. The statistical errors of the Monte Carlo calculations are negligible. 
  Left panel: Jets in Monte Carlo reconstructed from hadrons. Right Panel: Jets in Monte Carlo reconstructed from partons (JETSCAPE partons = solid red line).
  \label{fig:jetshape7R06pt4060}}
\end{figure*}
\begin{figure*}[tb]
\includegraphics[width=0.9\columnwidth]
{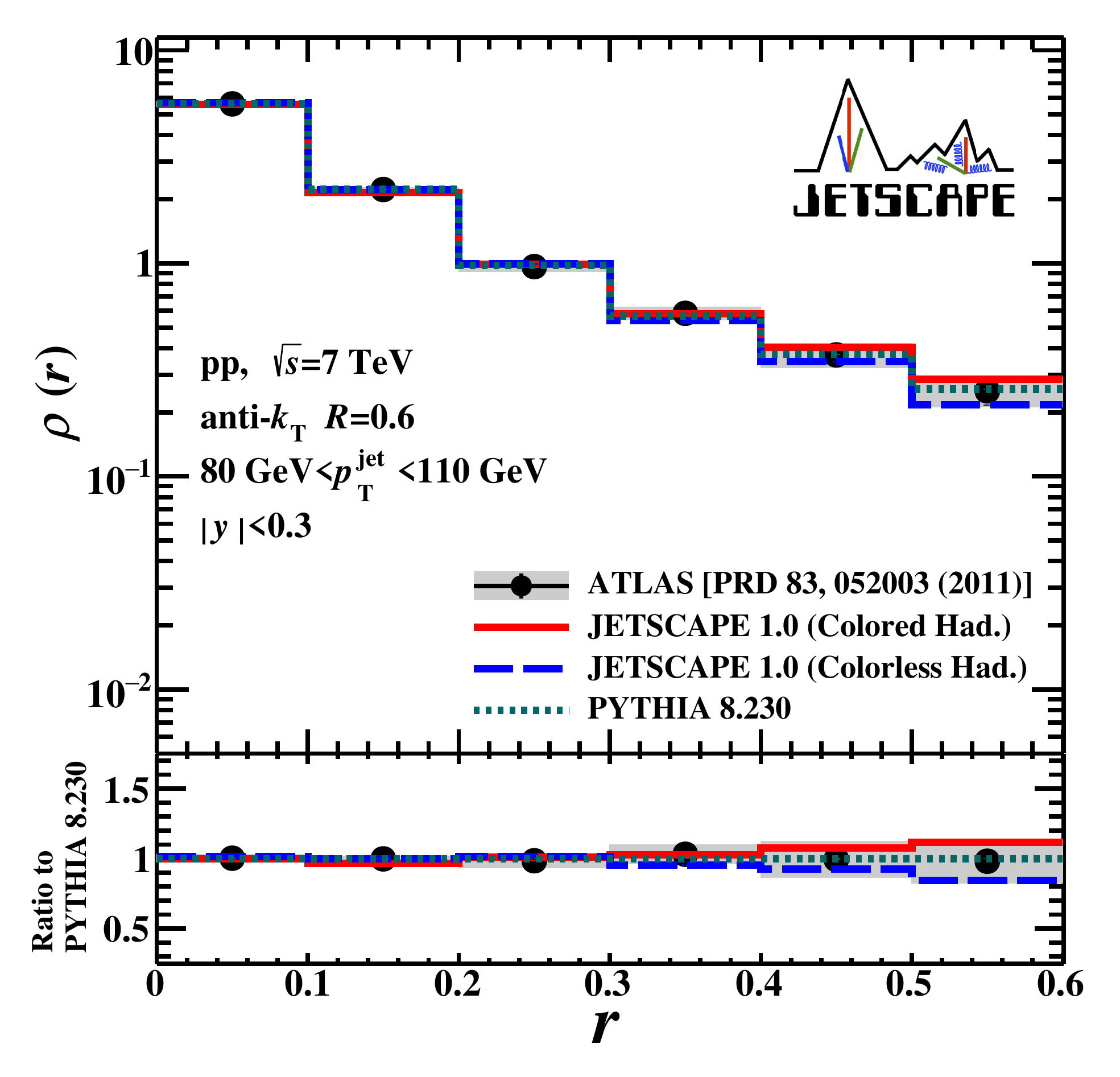}
\hspace{15pt}
\includegraphics[width=0.9\columnwidth]
{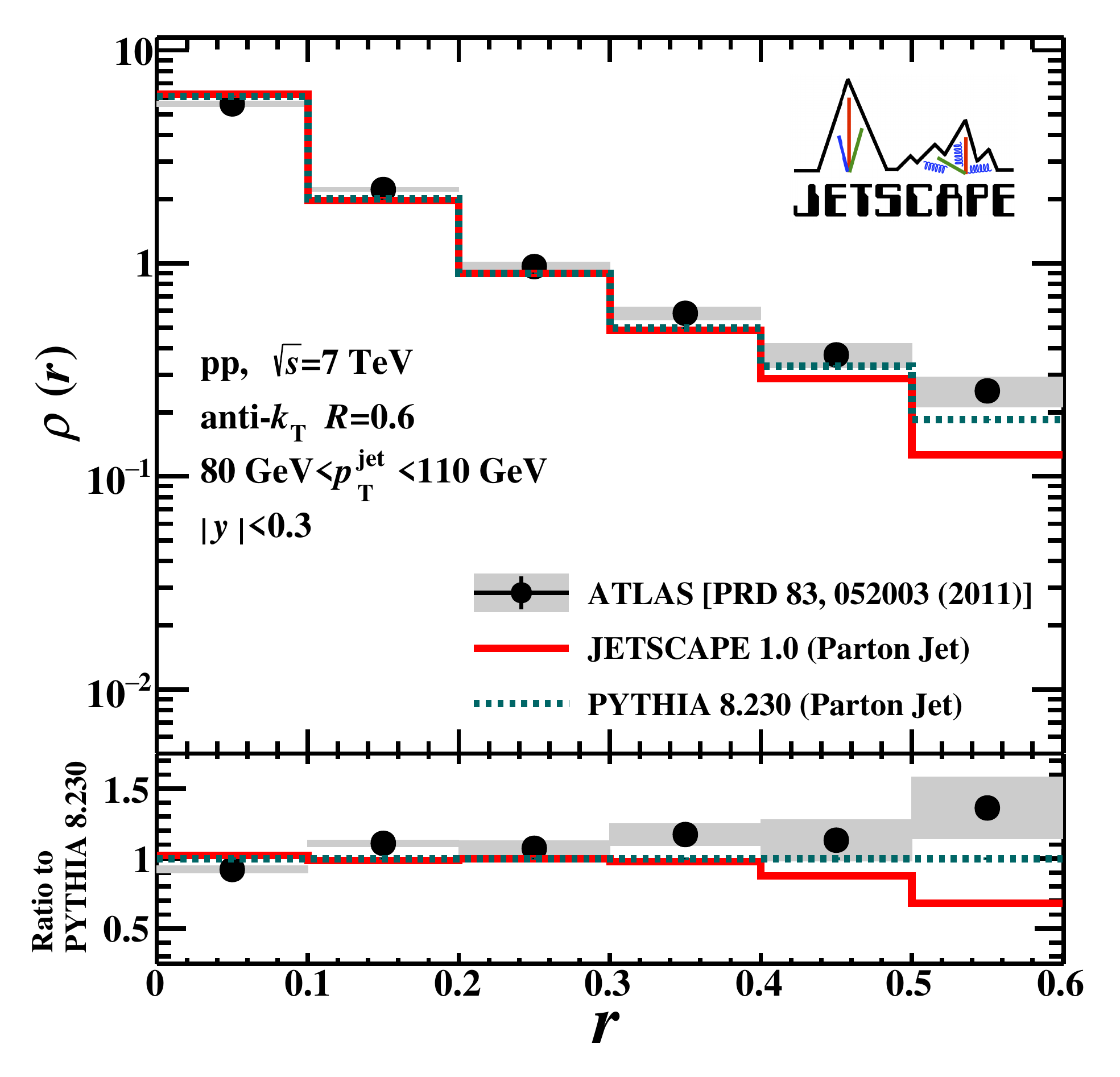}
 \caption{Same as Fig.\ \ref{fig:jetshape7R06pt4060} for jets with $p_T$ between 80 and 110 GeV/$c$  and rapidities $|y|<0.3$.
  \label{fig:jetshape7R06pt80110}}
\end{figure*}
\begin{figure*}[tb]
\includegraphics[width=0.9\columnwidth]
{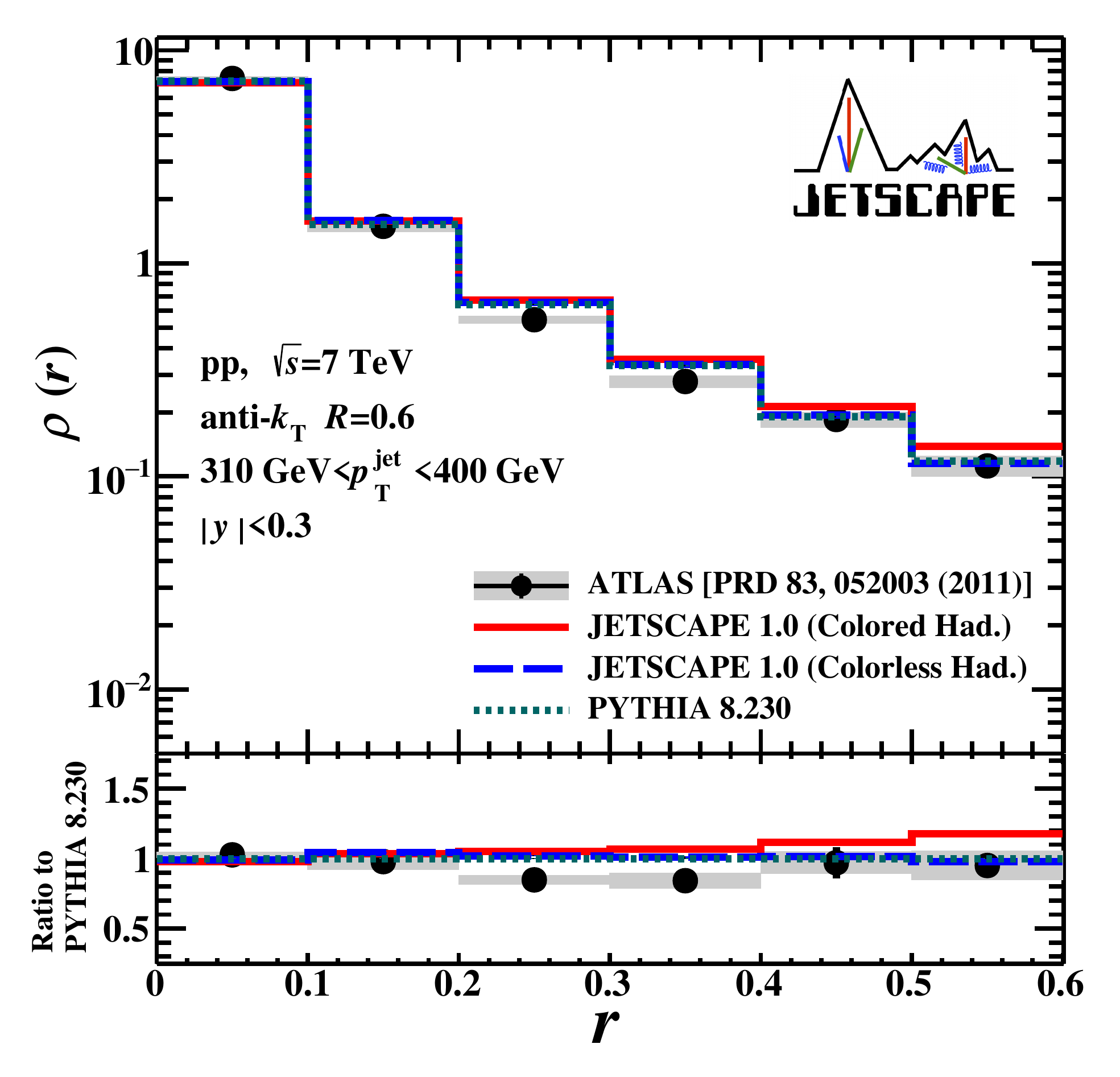}
\hspace{15pt}
\includegraphics[width=0.9\columnwidth]
{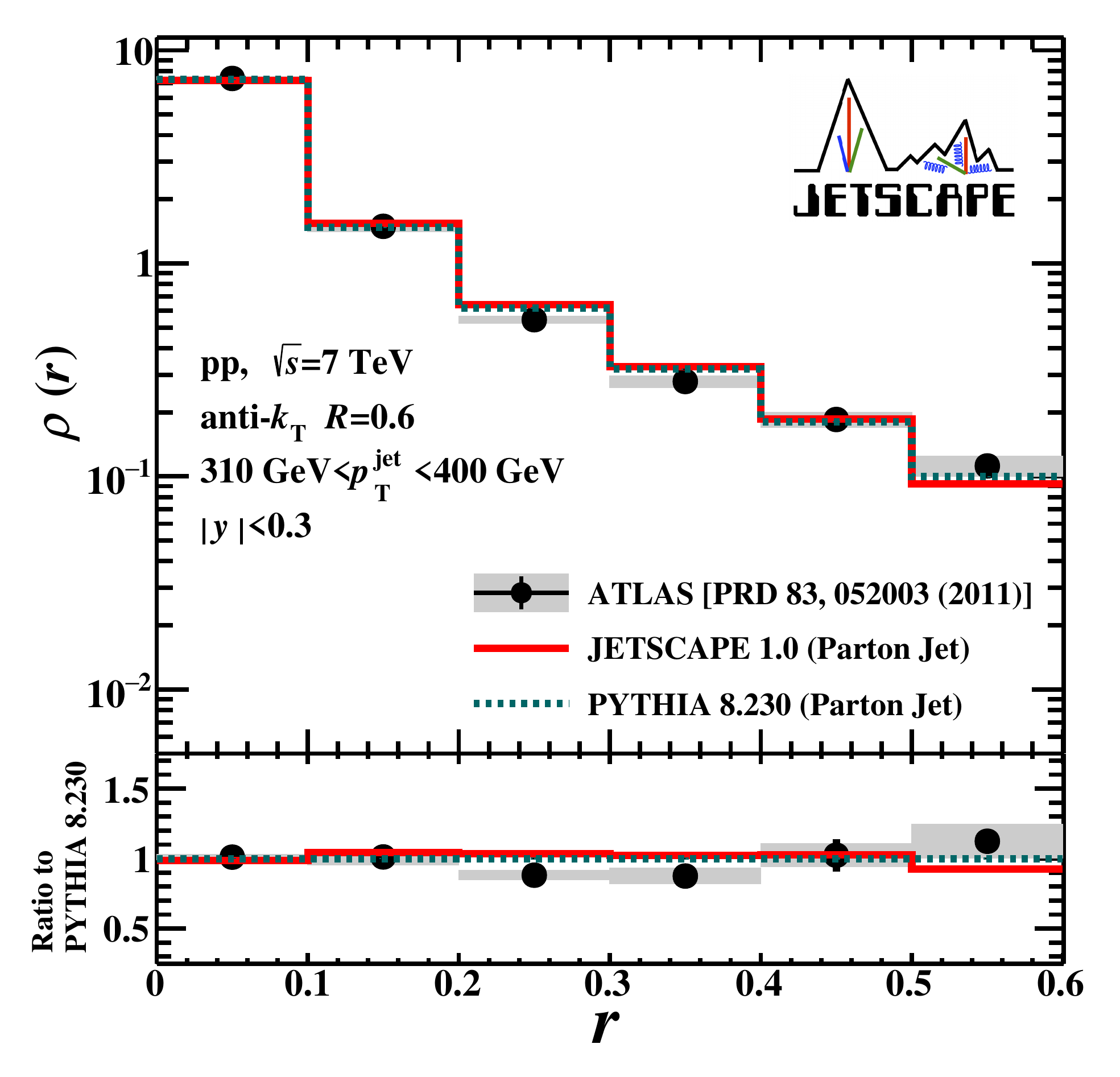}
 \caption{Same as Fig.\ \ref{fig:jetshape7R06pt4060} for jets with $p_T$ between 310 and 400 GeV/$c$ and rapidities $|y|<0.3$.
  \label{fig:jetshape7R06pt310400}}
\end{figure*}
\begin{figure*}[tb]
\includegraphics[width=0.9\columnwidth]
{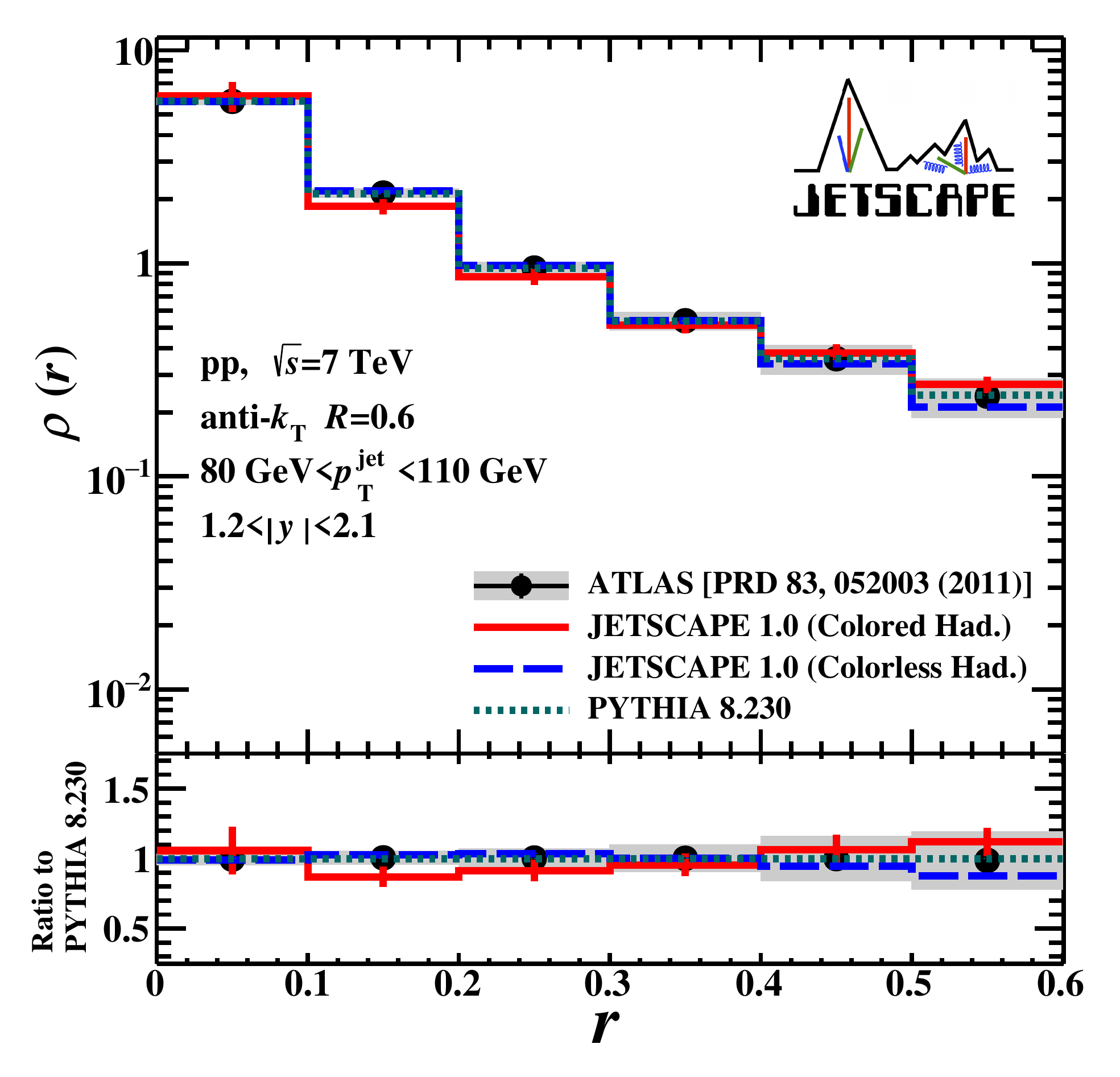}
\hspace{15pt}
\includegraphics[width=0.9\columnwidth]
{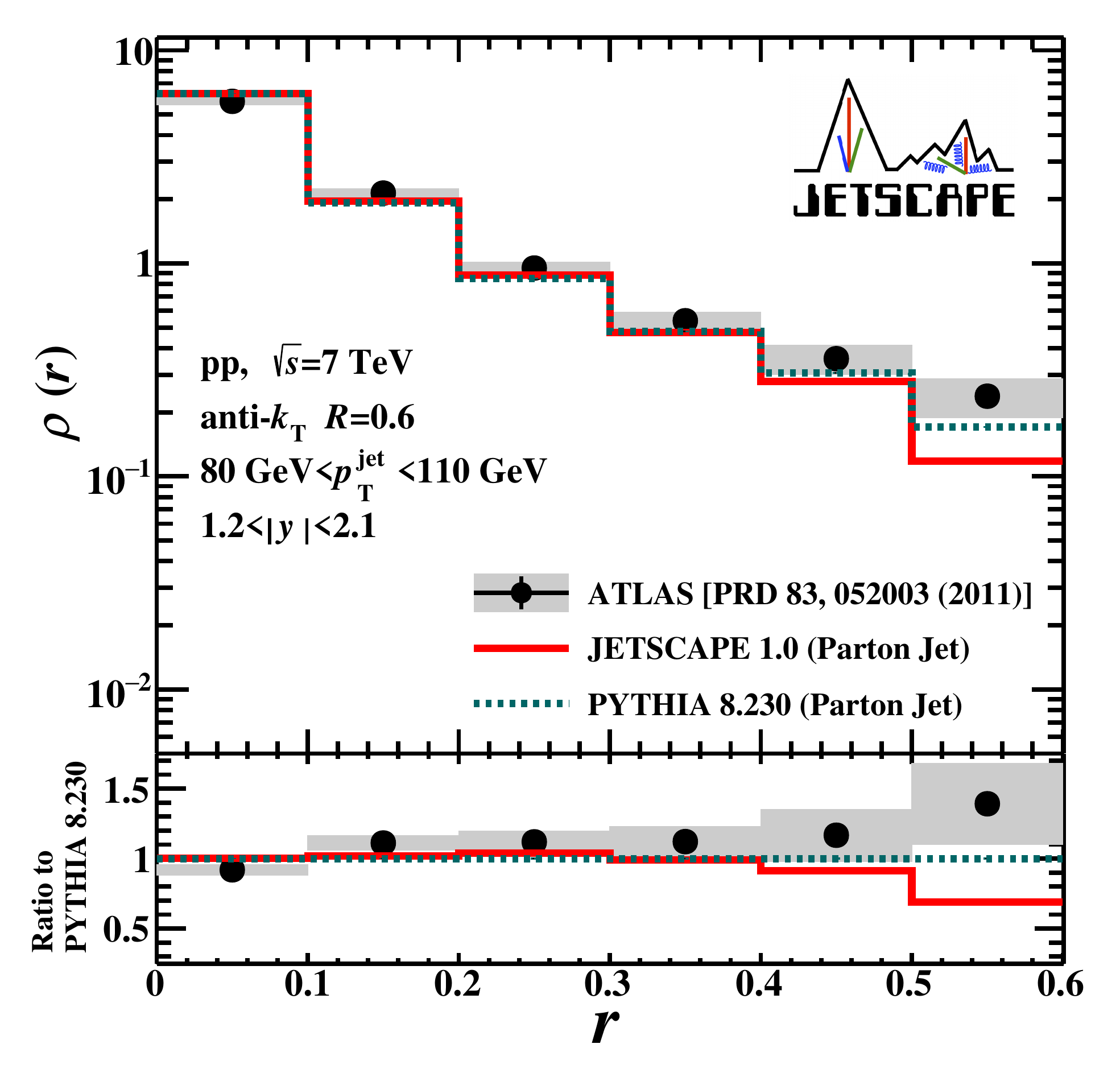}
 \caption{Same as Fig.\ \ref{fig:jetshape7R06pt4060} for jets with $p_T$ between 80 and 110 GeV/$c$ and rapidities $1.2<|y|<2.1$.
   \label{fig:jetshape7R06pt80110y1221}}
\end{figure*}

\subsubsection{Jet transverse profile}

Turning to transverse jet profile, we have calculated $\rho(r)$ for 7 TeV collisions for $R=0.6$ jets and for a large number of jet transverse momentum 
and rapidity bins for which the ATLAS experiment has provided data \cite{Aad:2011kq}. We show a small selection of these results in this publication.
Monte Carlo results in Figs.\ \ref{fig:jetshape7R06pt4060} through 
\ref{fig:jetshape7R06pt310400} are calculated for jets around midrapidity ($|y|<0.3$)
and three different bins for jet transverse momentum, 40-60 GeV/$c$, 80-110 GeV/$c$ and 310-400 GeV/$c$. In the left panels we show the result obtained with
the two JETSCAPE hadronization models and with PYTHIA 8 together with ATLAS data. The right panel shows parton jets from JETSCAPE and
PYTHIA 8 for comparison. For the latter calculations the string formation and hadronization steps are omitted and partons are directly clustered with FASTJET.
Fig.\ \ref{fig:jetshape7R06pt80110y1221} adds results at larger rapidity $1.2<|y|<2.1$ for transverse jet momenta 80-110 GeV/$c$.

We observe that the three Monte Carlo calculations with hadron jets generally agree very well with each other and with data, within 
experimental error bars. Differences between hadronization models in JETSCAPE start to play a role for $r$ close to the jet cone radius in
accordance with expectations. Hadronization transfers particles close to the jet cone boundary in or out of the jet as defined at the partonic level. This observation 
is confirmed when parton jets are compared to hadron jets.
The results for parton jets from both JETSCAPE and PYTHIA 8 are consistent with data for $r \ll  R$ but underestimate the transverse jet profile for 
large $r$. Generally, differences between data and Monte Carlo are larger for jets with smaller tranverse momentum.
In addition we note a systematic trend between Colorless and Colored Hadronization which bracket the PYTHIA 8 result in most cases.
However, JETSCAPE results are typically within experimental error bars.

To summarize, JETSCAPE PP19 does well with transverse jet shape observables when compared to data above $p_T^\mathrm{jet} \gtrsim 40$ GeV/$c$.
For smaller momenta deviations from data occur but are in line with analytic calculations of jet cross section ratios.
Differences between hadronization models become visible for smaller momenta and close to the jet periphery for jet shape variables.

\subsection{Jet fragmentation functions}
\label{subsec:jetff}

Fragmentation functions $D_\mathrm{jet}(z)$ describe the longitudinal structure of the jet by counting particles in the jet according to their momentum 
fraction $z$ with respect to the full jet momentum. $z$ for a particle with momentum $\boldsymbol{p}_{\rm particle}$ with respect to a jet with 
momentum $\boldsymbol{p}_{\rm jet}$ is defined as
\begin{equation}
  z = \frac{\boldsymbol{p}_{\rm jet}\cdot\boldsymbol{p}_{\rm particle}}{|\boldsymbol{p}_{\rm jet}|^{2}}  \, .
\end{equation}
Fragmentation functions for large momentum jets in vacuum are well understood. In a medium the distribution of hadrons is expected to be modified due to quenching effects.. Fragmentation functions $D_\mathrm{jet}(p_T)$ measured as a function of absolute particle transverse momentum $p_T$ are an alternative way to plot fragmentation functions. This is particularly interesting to find violations of scaling with $z$ through the presence of momentum scales given by the medium.

\begin{figure*}[tb]
\includegraphics[width=0.85\columnwidth]{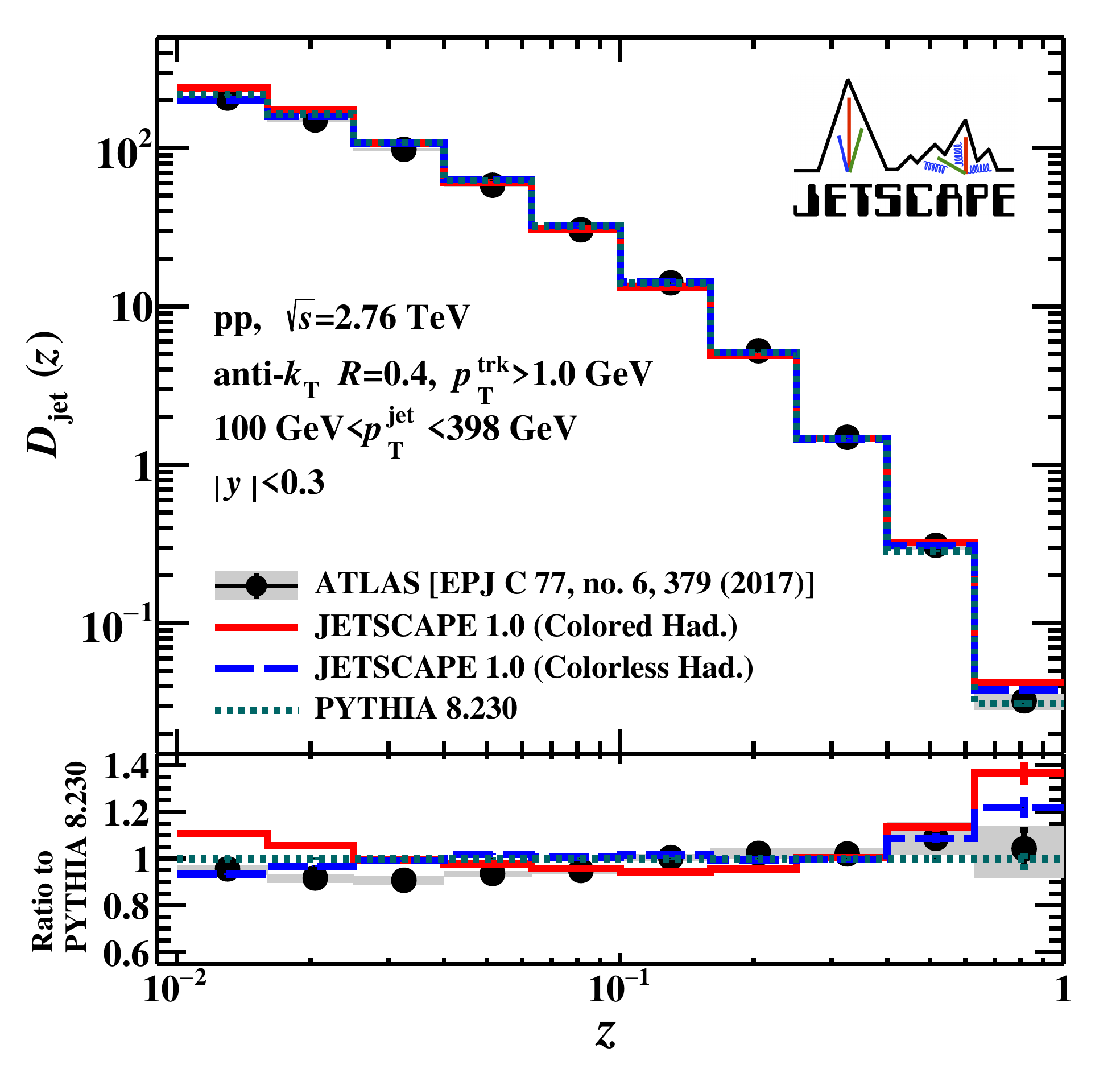}
\hspace{15pt}
\includegraphics[width=0.85\columnwidth]{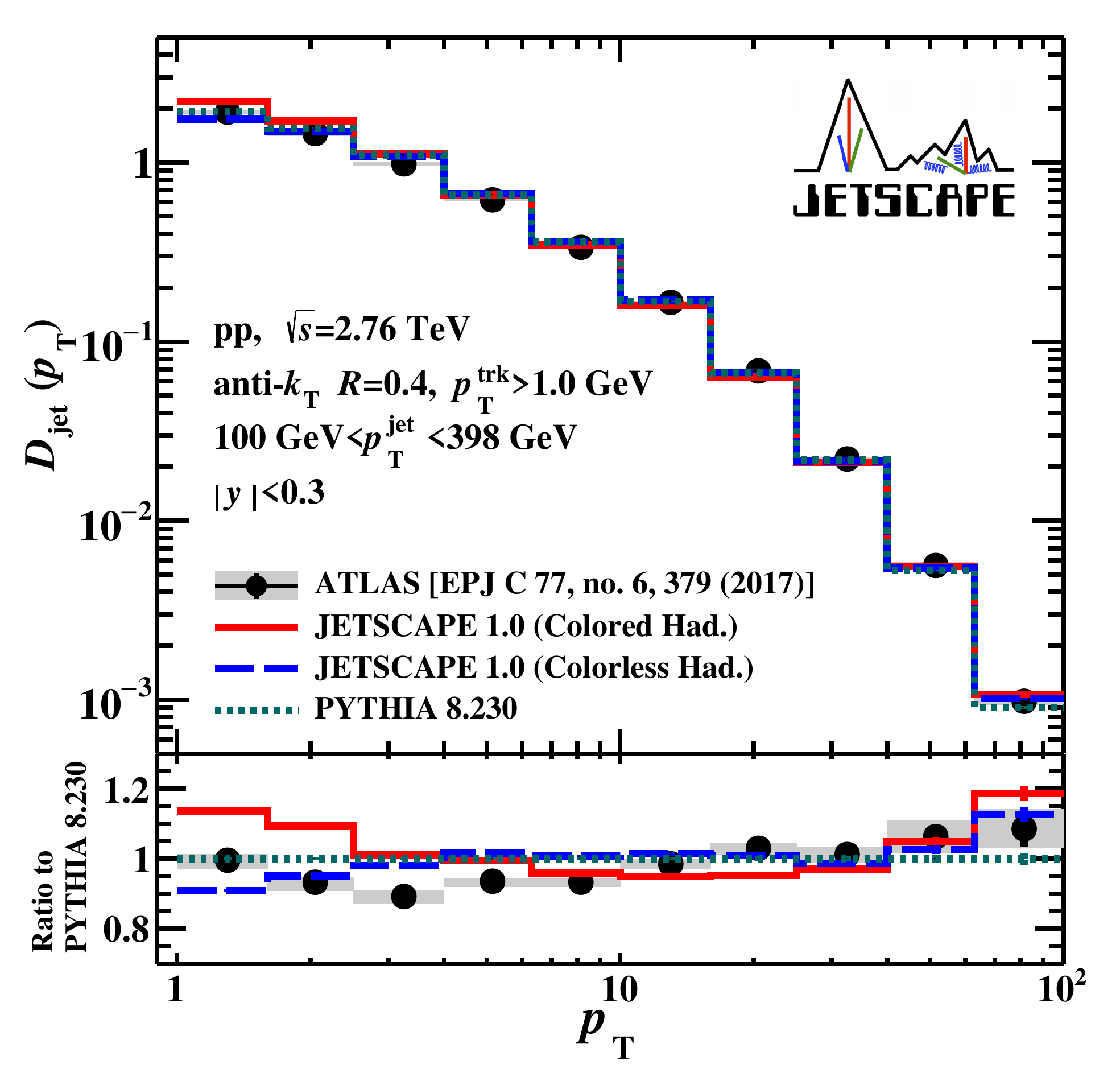}
  \caption{Jet fragmentation function $D_\mathrm{jet}$ for charged hadrons in jets of radius $R=0.4$ in $p+p$ collisions at 2.76 TeV, and its ratio to the PYTHIA 8 Monte Carlo results.
  Jets are required to have transverse momenta $p_T$ between 100 and 398 GeV/$c$ and rapidities $|y|<0.3$.
  Data points in both panels are from the ATLAS collaboration \cite{Aaboud:2017bzv} (black circles). Monte Carlo
  results are from JETSCAPE Colored Hadronization 
  (solid red line), JETSCAPE Colorless Hadronization (dashed blue line), and PYTHIA 8 (dotted green line). 
  Statistical errors (black error bars) 
  and systematic errors (gray bands) are plotted with the data. 
  Left panel: $D(z)$ as a function of momentum fraction $z$. Right Panel: $D(p_T)$ as a function of hadron transverse momentum.
  \label{fig:frag276R04had}}
\end{figure*}

\begin{figure*}[tb]
	\includegraphics[width=0.85\columnwidth]{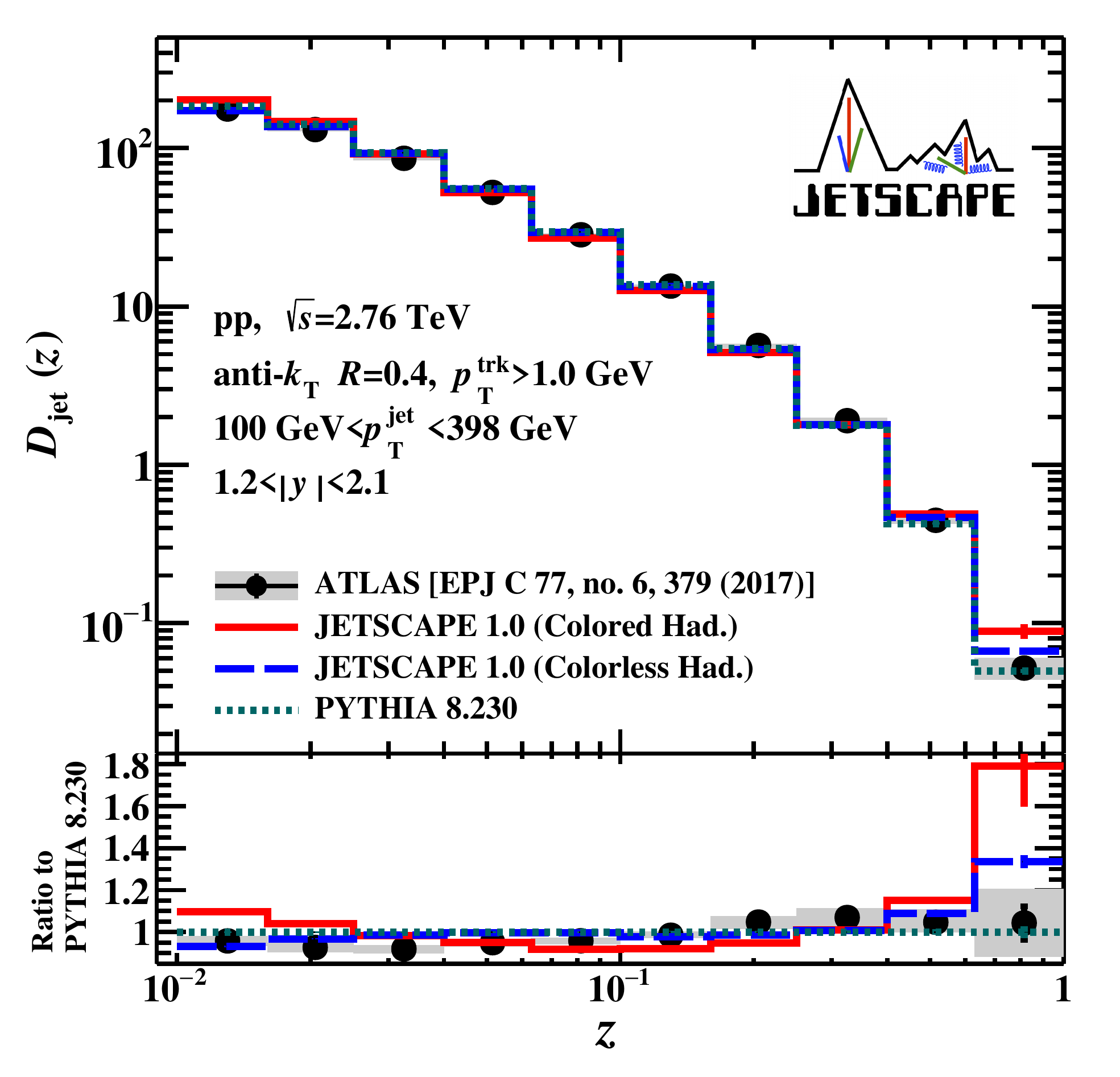}
	\hspace{15pt}
  \includegraphics[width=0.85\columnwidth]{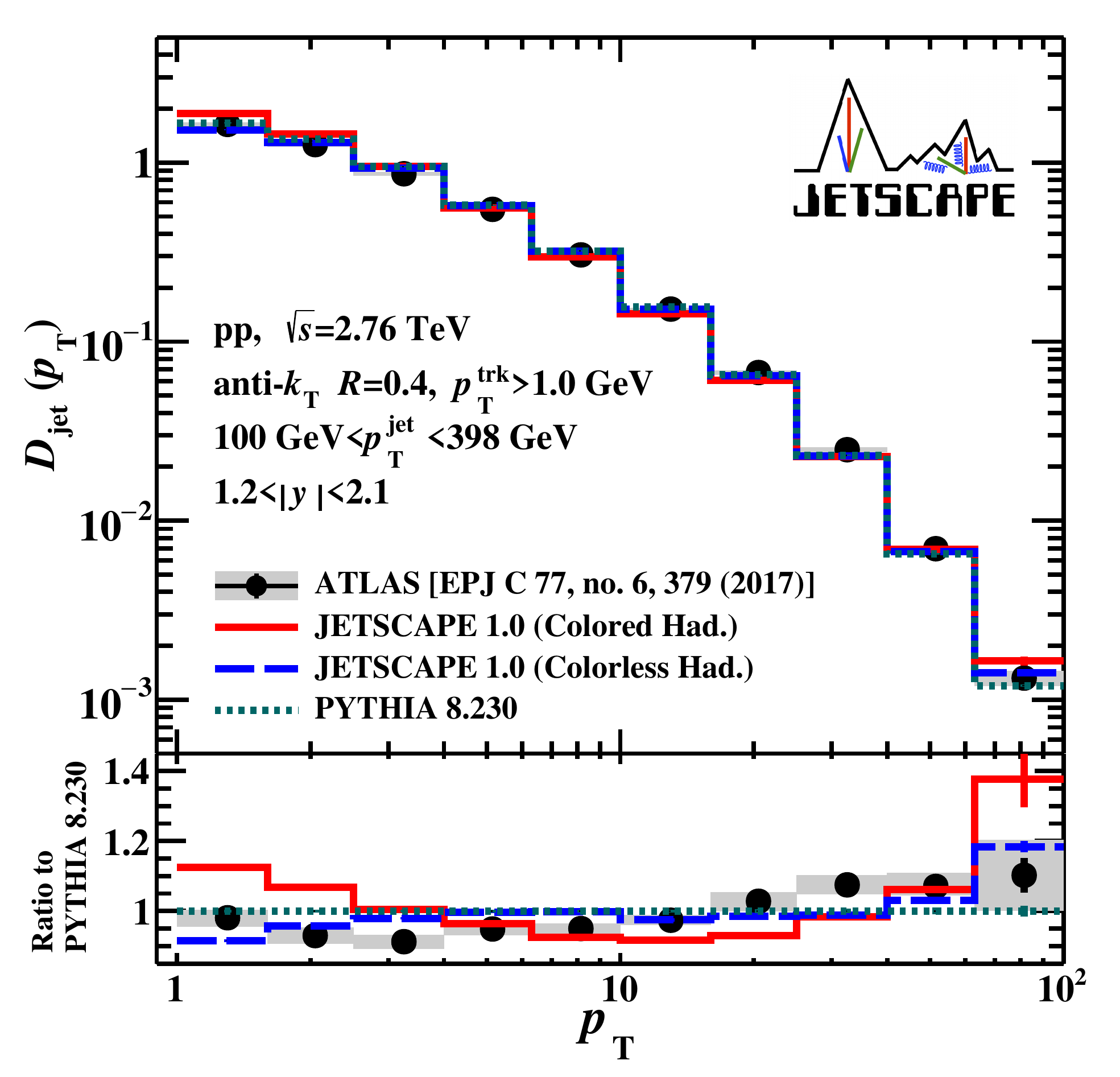}
  \caption{Same as Fig.\ \ref{fig:frag276R04had} for jets with rapdity $1.2<|y|<2.1$.
  \label{fig:frag276R04hady12-21}}
\end{figure*}

\begin{figure*}[tb]
\includegraphics[width=0.85\columnwidth]{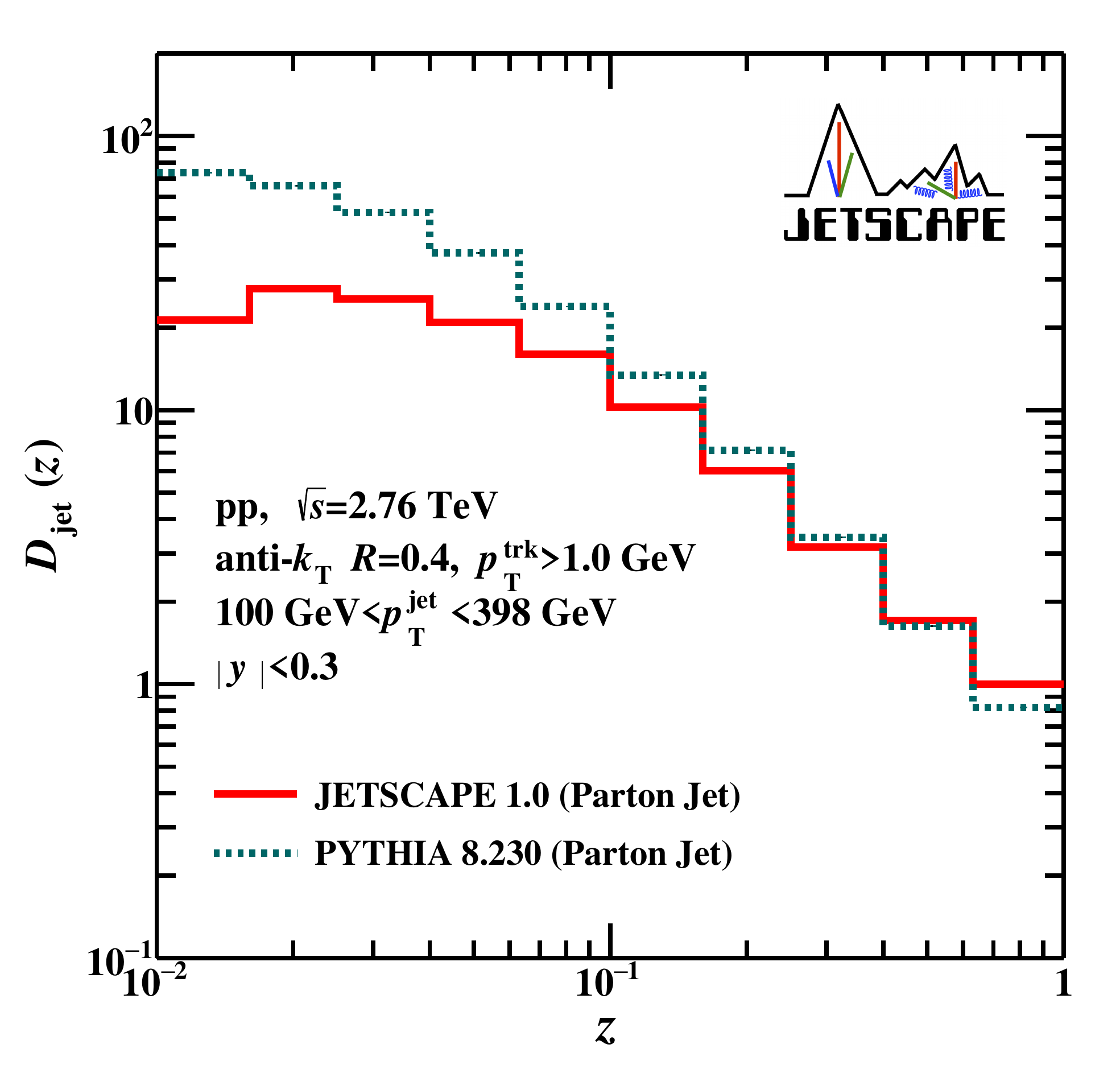}
\hspace{15pt}
\includegraphics[width=0.85\columnwidth]{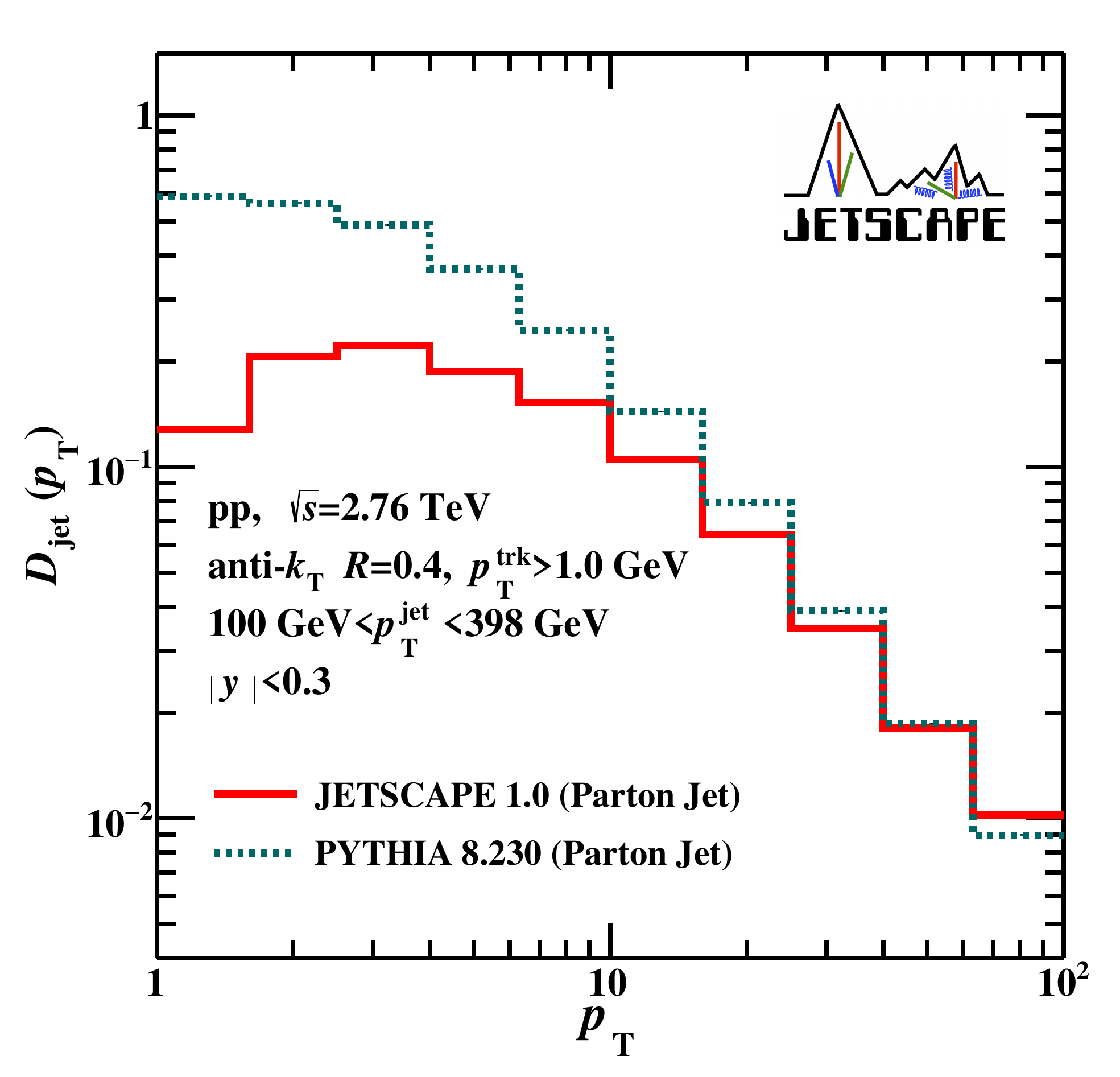}
  \caption{Same as Fig.\ \ref{fig:frag276R04had} for the momentum fraction and transverse momentum of partons (quarks, anti-quarks and gluons) within parton jets 
  (JETSCAPE = solid red line, default PYTHIA 8 = dotted green line). 
  \label{fig:frag276R04part}}
\end{figure*}

\begin{figure*}[tb]
	\includegraphics[width=0.85\columnwidth]{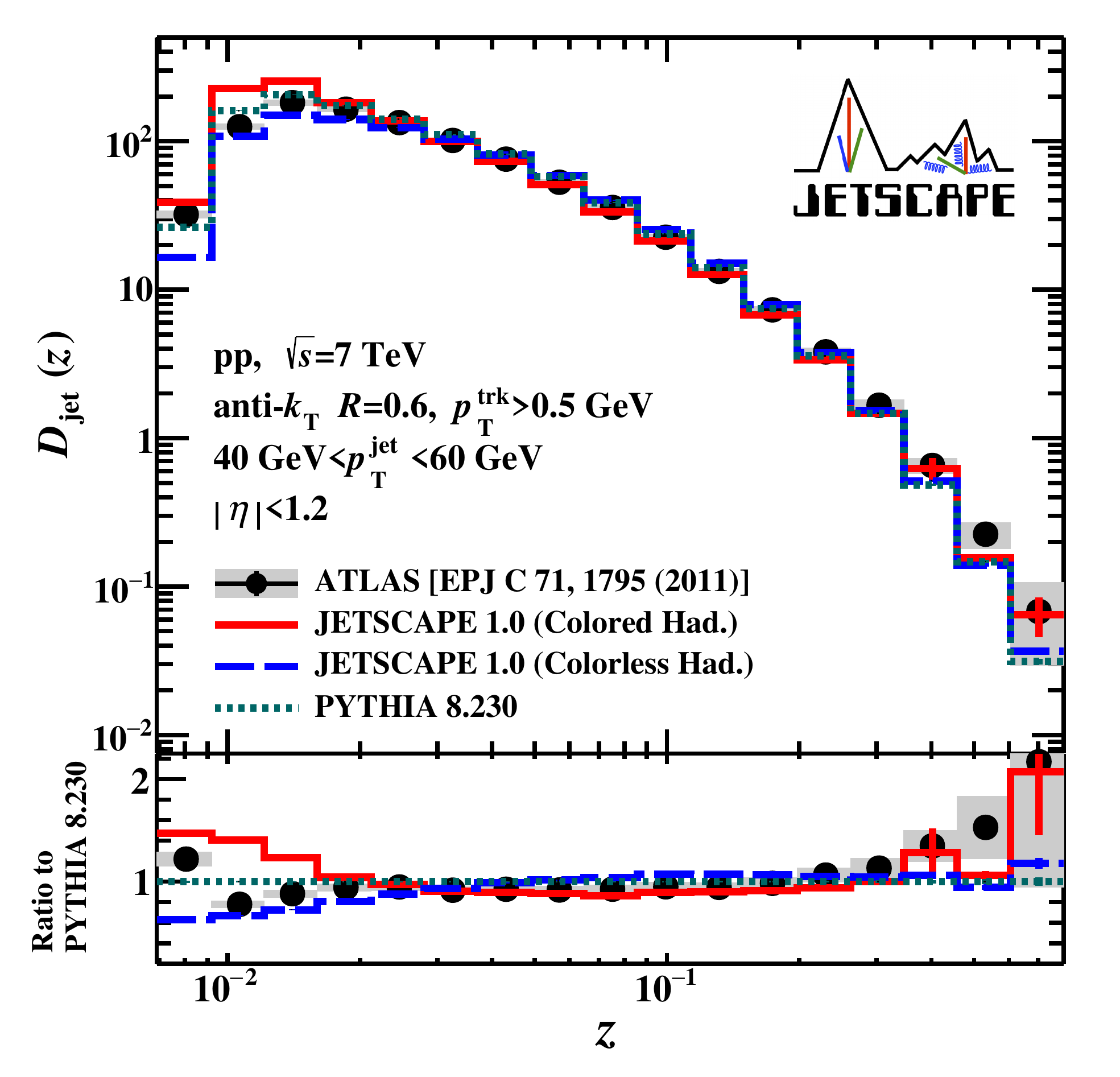}
\hspace{15pt}
  \includegraphics[width=0.85\columnwidth]{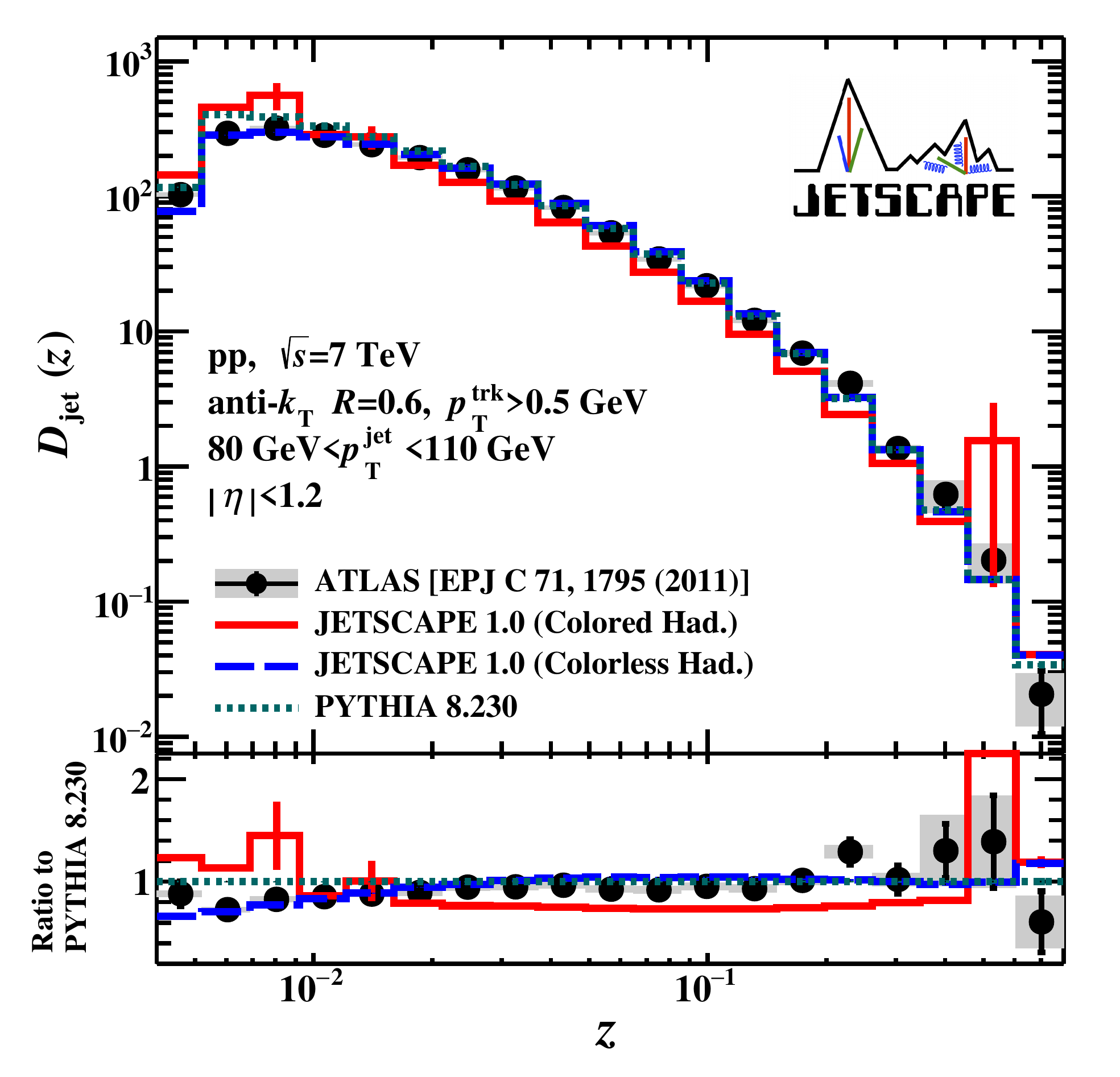}
  \\\vspace{15pt}
  	\includegraphics[width=0.85\columnwidth]{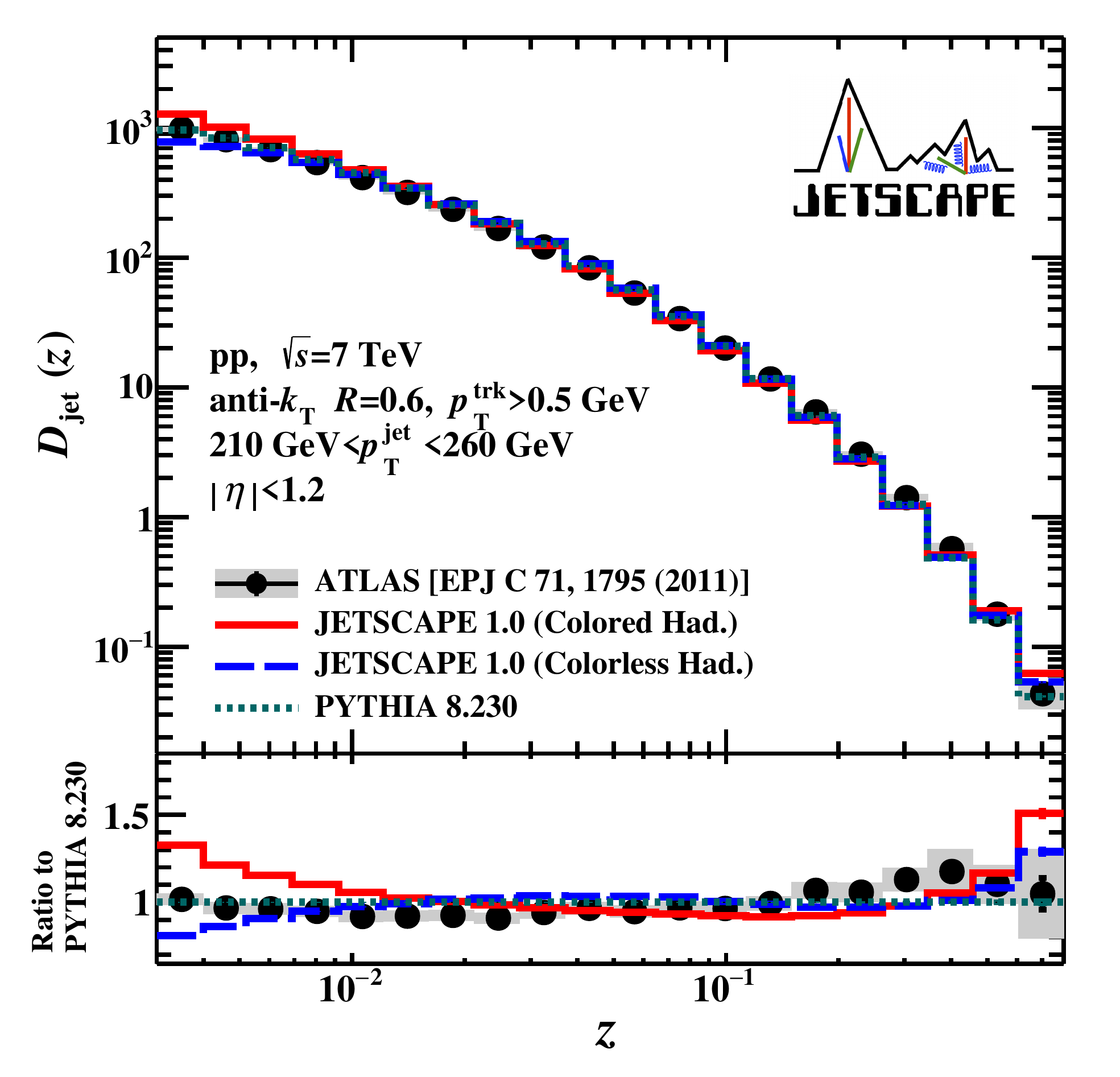}
\hspace{15pt}
  \includegraphics[width=0.85\columnwidth]{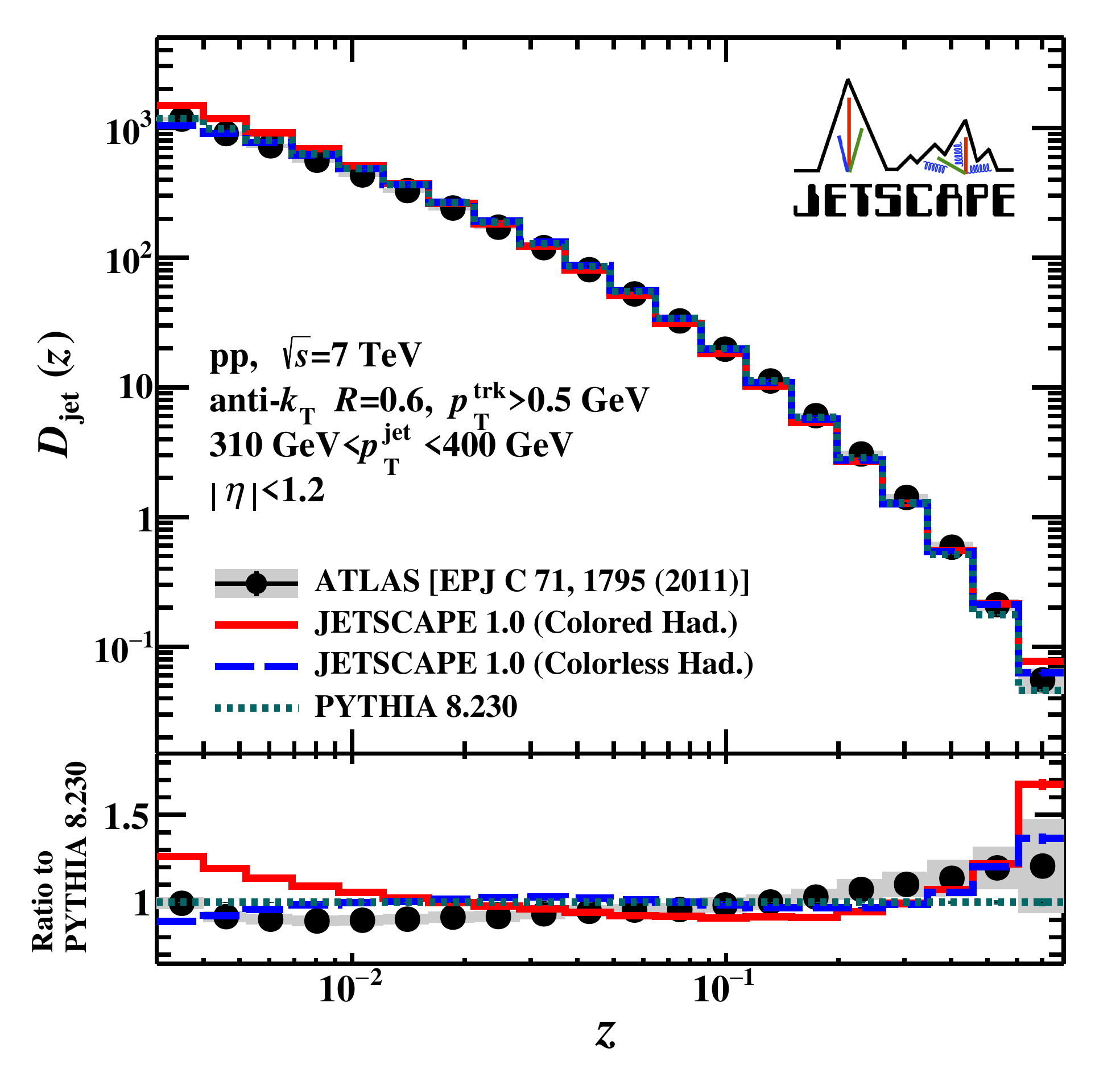}
  \caption{Jet fragmentation function $D(z)$ for charged hadrons in jets of radius $R=0.6$ with pseudorapidity $|\eta|<1.2$ in $p+p$ collisions at 7 TeV, and its ratio to the PYTHIA 8 Monte Carlo results.
Data points are from the ATLAS collaboration \cite{Aad:2011sc}
 Three different Monte Carlo calculations are presented: JETSCAPE Colored Hadronization (solid red line), 
  JETSCAPE Colorless Hadronization (dashed blue line), and PYTHIA 8 (dotted green line). Statistical errors (black error bars) and systematic errors 
  (grey bands) are plotted with the data. 
  Upper left panel: Results for jets with $p_T$ between 40 and 60 GeV/$c$.
   Upper right panel: Results for jets with $p_T$ between 80 and 110 GeV/$c$.
   Lower left panel: Results for jets with $p_T$ between 210 and 260 GeV/$c$.
   Lower right panel: Results for jets with $p_T$ between 310 and 400 GeV/$c$.   
  \label{fig:fragratio7R}}
\end{figure*}

In this subsection we discuss fragmentation functions both as functions of $z$ and $p_T$ in $p+p$ collisions at $\sqrt{s}=2.76$ TeV and 7 TeV.
At $\sqrt{s}=2.76$ TeV charged particles with transverse momentum larger 1 GeV/$c$ inside $R=0.4$ jets are used to calculate the fragmentation function.
Jets have transverse momenta between 100 and 398 GeV/$c$ to match the ATLAS experiment \cite{Aaboud:2017bzv}. 
The panels in Fig.\ \ref{fig:frag276R04had} show the fragmentation function as a function of $z$ and $p_T$ respectively, for jets around midrapidity 
($|y|<0.3$). Fig.\ \ref{fig:frag276R04hady12-21} shows the same fragmentation function around rapidity 2 with ATLAS data. Ratios of JETSCAPE results and data to PYTHIA 8
are plotted in panels below the fragmentation functions.

It is informative to calculate the same fragmentation functions defined for the final parton configuration in each Monte Carlo simulation.
We show in Fig.\ \ref{fig:frag276R04part} the results for parton jets at midrapidity. 
Fig.\ \ref{fig:frag276R04part} demonstrates large differences in the longitudinal structure between MATTER vacuum showers and
PYTHIA 8 showers. Final state radiation in PYTHIA 8 produces more partons at small $z$ and small transverse momentum 
compared to MATTER. On the other hand, at the highest $z$ and $p_T$ bins, the PYTHIA 8 distribution is slightly suppressed relative to MATTER. Turning back to the 
fragmentation functions for hadrons these large differences seen on the parton side tend to be washed out by hadronization, however significant differences 
between PYTHIA 8 and MATTER final state showers remain. 
We can conclude that fragmentation functions at low $z$ ($\lesssim 10^{-2}$) are very sensitive to hadronization. Moreover, measured fragmentation functions of hadrons
do not constrain parton distributions in jets well.

The sensitivity to hadronization is reflected in the differences between the JETSCAPE hadronization models. They are typically less than 15\% except for the 
highest and lowest $z$ or $p_T$ bins where the two calculations diverge noticeably. The uncertainty from the hadronization model is also larger than the 
experimental error bars for most $z$ or $p_T$ bins. PYTHIA 8 results tend to be bracketed between the two JETSCAPE results except for the 
largest $z$ or $p_T$ bins where JETSCAPE systematically predicts more hadrons than PYTHIA 8, consistent with the same observation for parton jets. 
Thus, while the large suppression of 
JETSCAPE parton showers at low $z$ and $p_T$ compared to PYTHIA 8 seems to be mitigated by hadronization, the enhancement at large $z$ or $p_T$ 
remains after hadronization. We note that overall PYTHIA 8 describes data on a level of accuracy comparable or slightly better
than JETSCAPE.

Fig.\ \ref{fig:fragratio7R} explores the dependence of fragmentation functions on the jet transverse momentum for $p+p$ collisions
at 7 TeV. We use jets with radius $R=0.6$ reconstructed in the pseudorapidity region $|\eta|<1.2$
for four jet $p_T$ bins.
Since the jet momentum bins are narrow we only show $D_\mathrm{jet}(z)$. 
The ratios to PYTHIA 8 results are provided in bottom panels. We find a consistent picture in all momentum bins. Again 
the two JETSCAPE hadronization models bracket both PYTHIA 8 and ATLAS data \cite{Aad:2011sc} except for very large $z$ where JETSCAPE overestimates
the fragmentation function. 

To summarize, the longitudinal structure of jets in vacuum is more challenging to compute in JETSCAPE than the transverse jet structure. Hadronization
helps to mitigate differences between PYTHIA and MATTER shower Monte Carlos at low $z$ but significant differences between the Monte Carlos remain
at large $z$. Uncertainties from the hadronization procedure are at least as large as current experimental uncertainties. While the overall agreement of JETSCAPE 
with data could be improved its performance is comparable to PYTHIA 8.

\subsection{Inclusive hadron production}
\label{subsec:hadxsec}

\begin{figure*}[tb]
	\includegraphics[width=0.85\columnwidth]{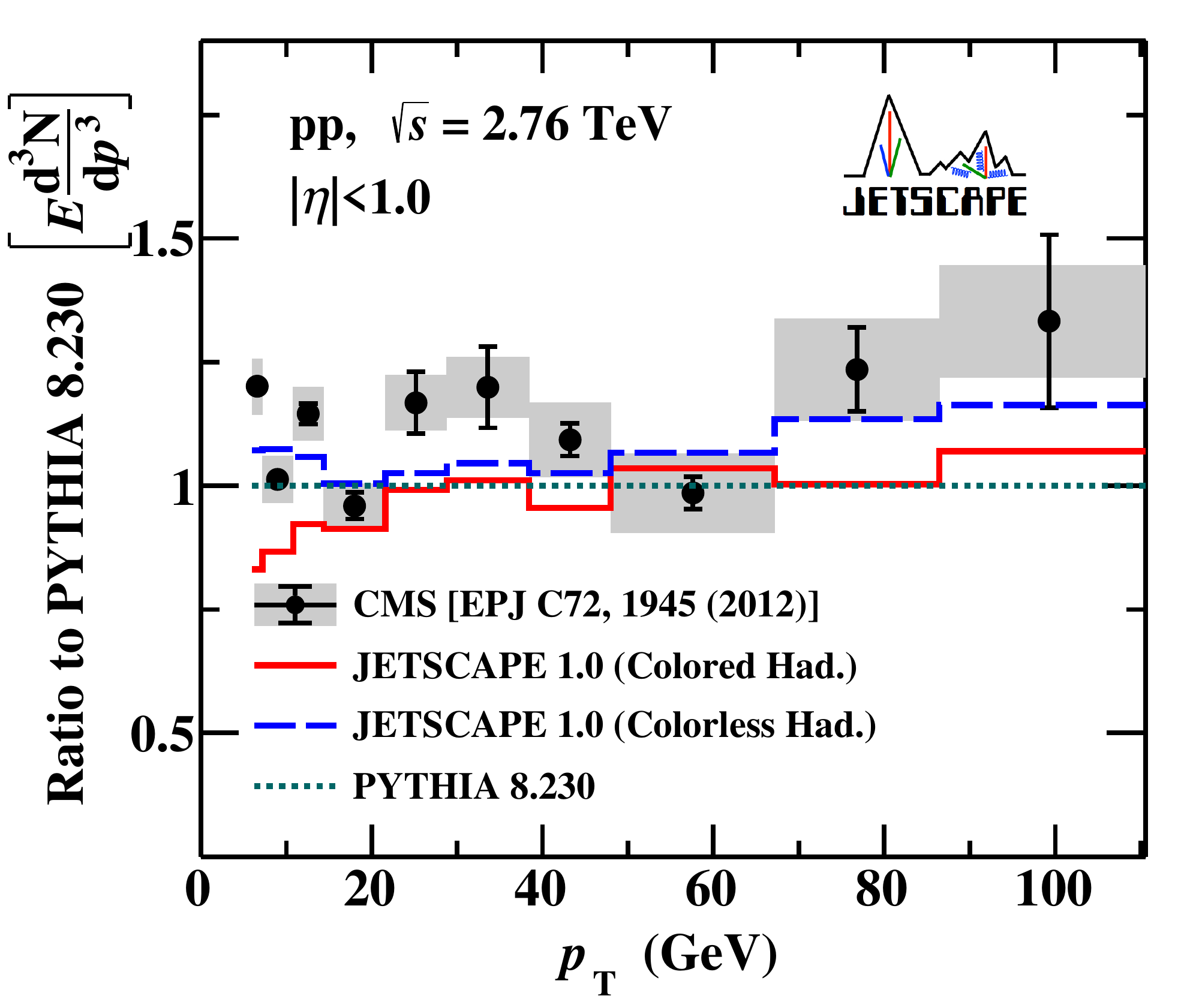}
  \includegraphics[width=0.85\columnwidth]{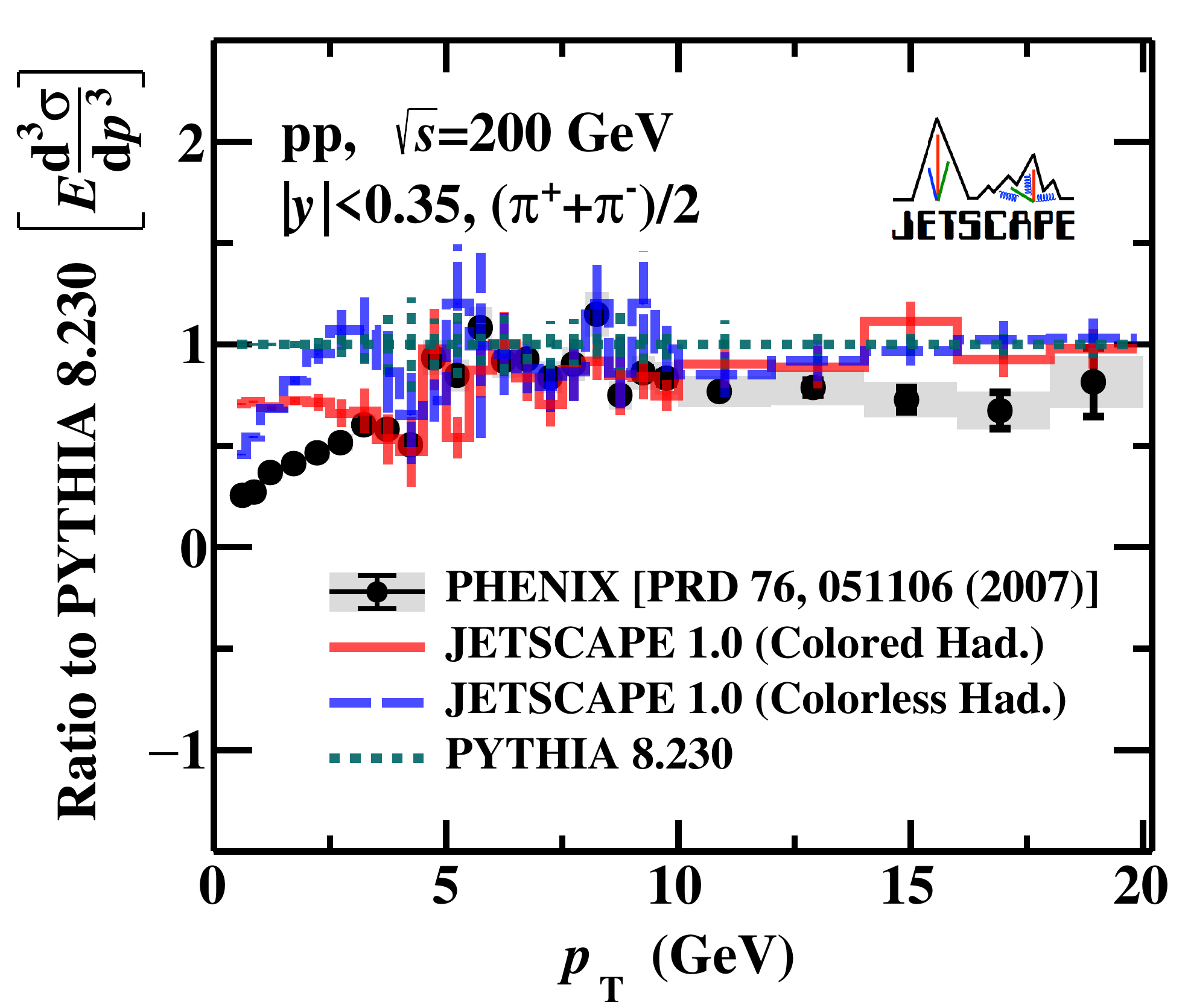}
  \caption{Ratio of inclusive hadron cross sections in $p+p$ collisions to the PYTHIA 8 Monte Carlo results. T hree different Monte Carlo calculations are presented: JETSCAPE    
  Colored Hadronization (solid red line), JETSCAPE Colorless Hadronization (dashed blue line), and PYTHIA 8 (dotted green line). 
  Statistical errors (black error bars) and systematic errors (grey bands) are plotted with the data.
  Left panel: Unidentified charge hadrons at 2.76 TeV compared to data from the CMS experiment \cite{CMS:2012aa}. 
  Hadrons are required to have rapidities $|\eta|<1.0$.
  Right panel: $(\pi^++\pi^-)/2$ at 200 GeV compared to $\pi^0$ data from the PHENIX collaboration \cite{Adare:2007dg}.
  \label{fig:chhad276}}
\end{figure*}

Now we discuss the performance of JETSCAPE for the  calculation of inclusive hadron cross sections.
The hadron decay settings in the PP19 tune are chosen for calculations of jets and unidentified charged hadron. We focus
here on charged hadrons and charged pions. The latter are compared to neutral pion measurements at RHIC energies. The breaking of
isospin symmetry at RHIC at large momentum is small enough to make this a meaningful comparison.
Fig.\ \ref{fig:chhad276} shows the ratio of the charged hadron cross sections at $\sqrt{s}=2.76$ TeV around midrapitidy for our three different Monte Carlo
calculations and data from CMS \cite{CMS:2012aa}. We also provide the comparison of $(\pi^++\pi^-)/2$ at 200 GeV compared to $\pi^0$ data from the PHENIX
experiment \cite{Adare:2007dg}.

The difference between the two JETSCAPE hadronization models is about 10\% for hadron cross sections. The PYTHIA 8 result for the hadron cross section is 
bracketed by the two JETSCAPE results for charged hadrons but the PYTHIA 8 calculation lies above JETSCAPE results for pions.
All Monte Carlo results slightly underpredict the charged hadron data with JETSCAPE Colorless Hadronization coming closest to data. For pions at RHIC
JETSCAPE describes data between 5 and 15 GeV/$c$.

\subsection{Dijet Mass}
\label{subsec:dijetmass}

\begin{figure*}[tb]
\includegraphics[width=0.85\columnwidth]{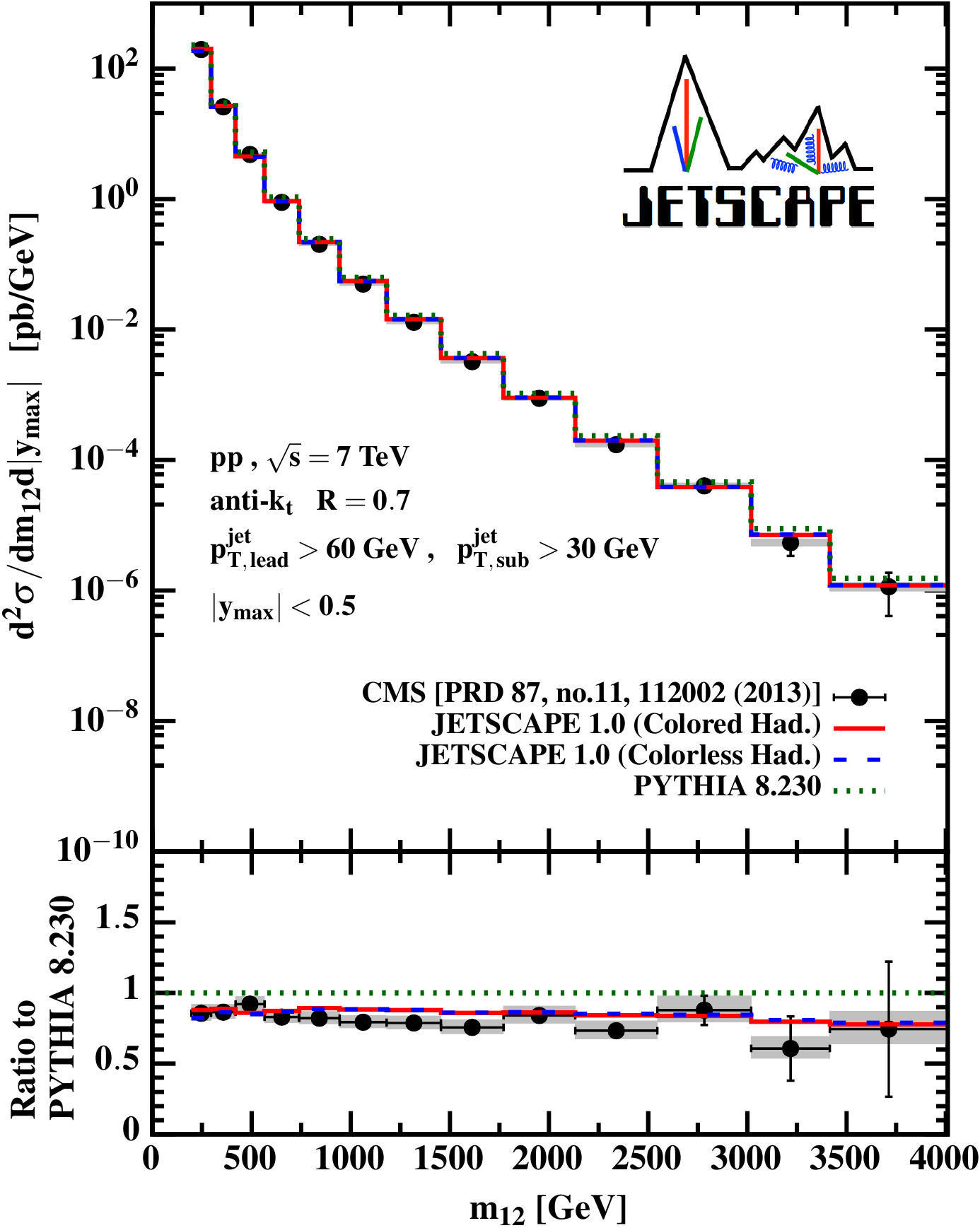}\hspace{15pt}
\includegraphics[width=0.85\columnwidth]{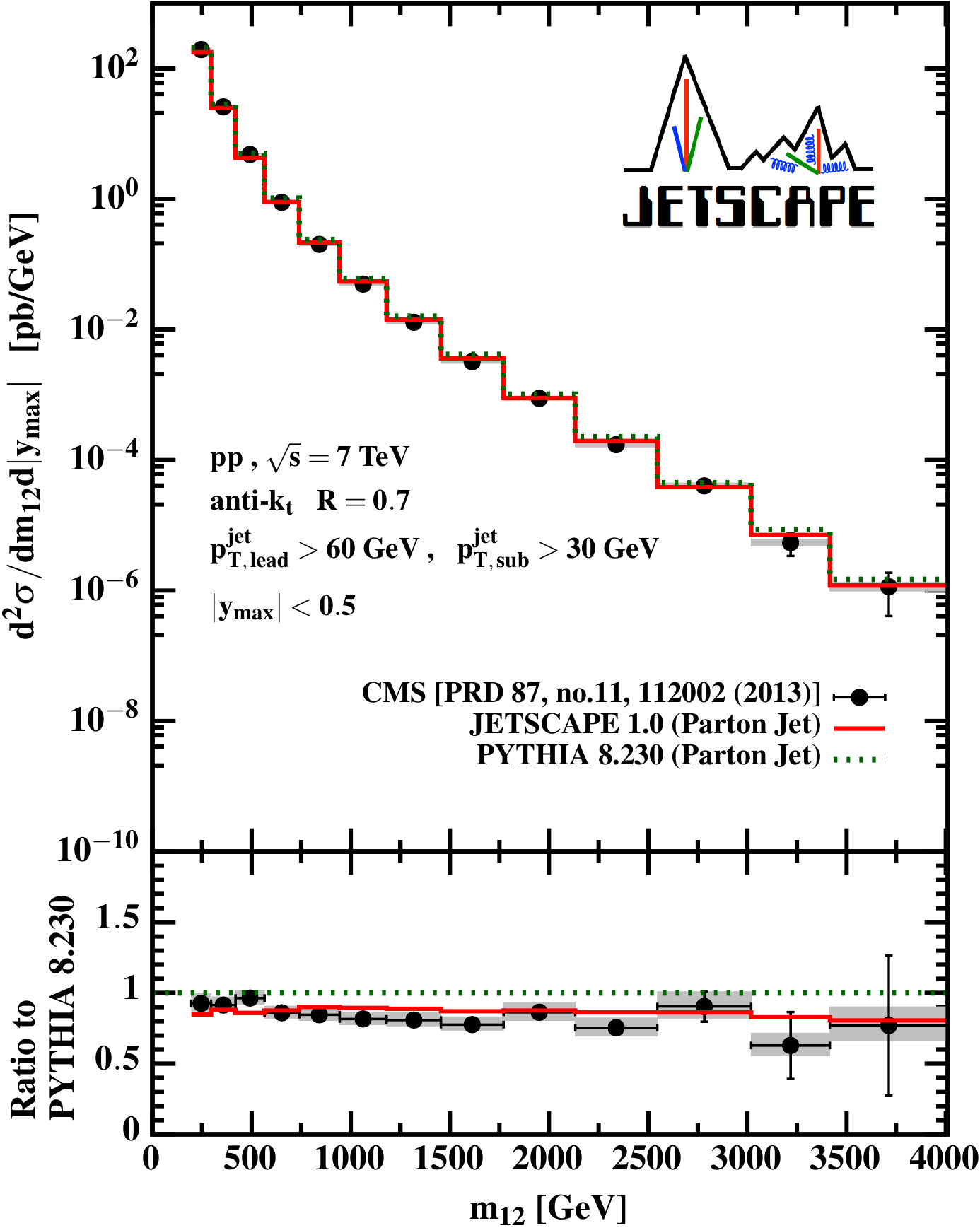}
  \caption{Dijet mass cross section $d^2\sigma/dMdy_\text{max}$ at 7 TeV for $R=0.7$ jets with $|y_\text{max}|<0.5$
  The left panel shows results from PYTHIA 8 and JETSCAPE with both colored and colorless hadronization
  together with data from CMS \cite{Chatrchyan:2012bja}. The right panel shows results for parton jets created with PYTHIA 8 and
  JETSCAPE. Bottom panels give the ratio of all results and data to PYTHIA 8.
  \label{fig:jetmasscmsy0}}
\end{figure*}

\begin{figure*}[tb]
\includegraphics[width=0.85\columnwidth]{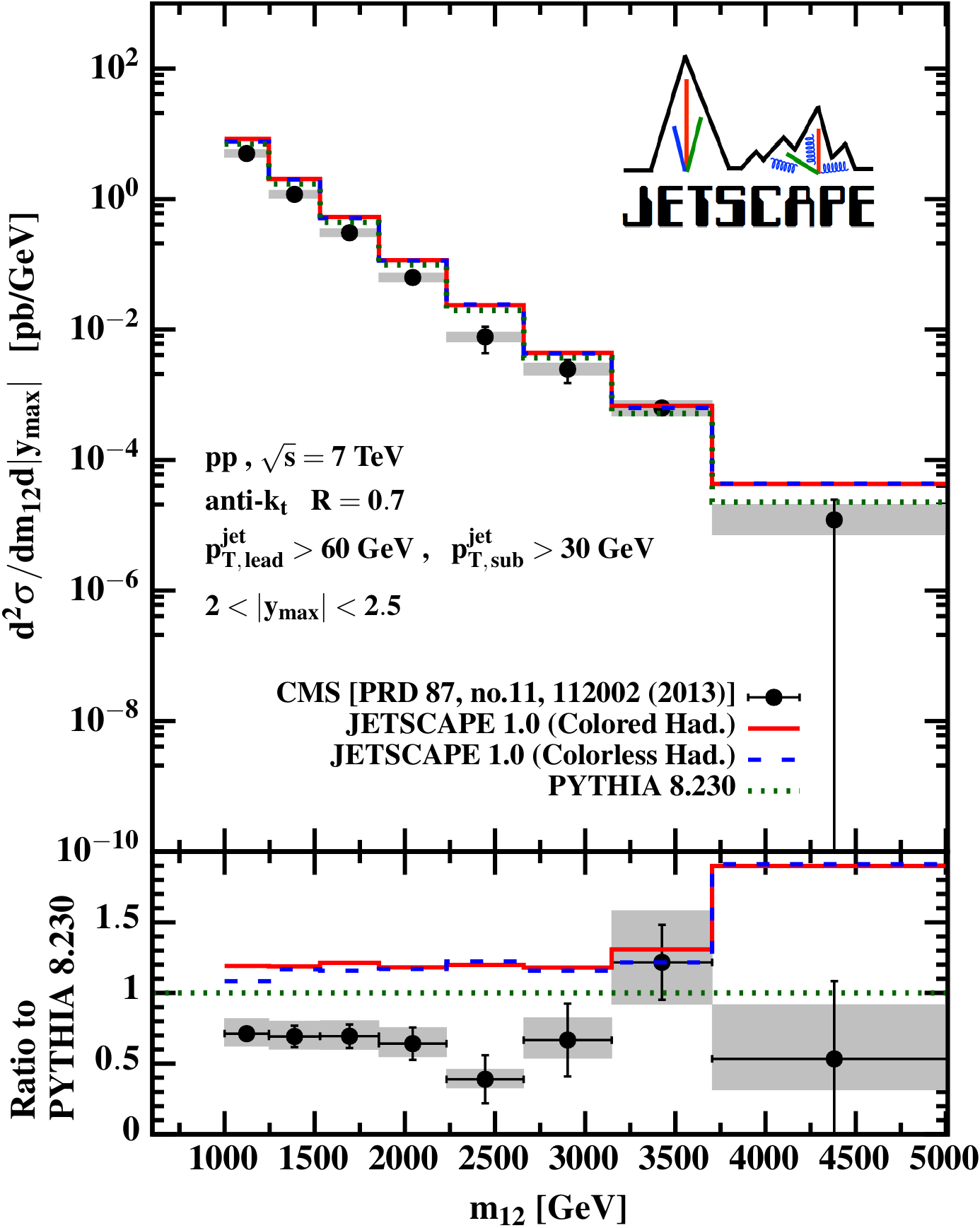}\hspace{15pt}
\includegraphics[width=0.85\columnwidth]{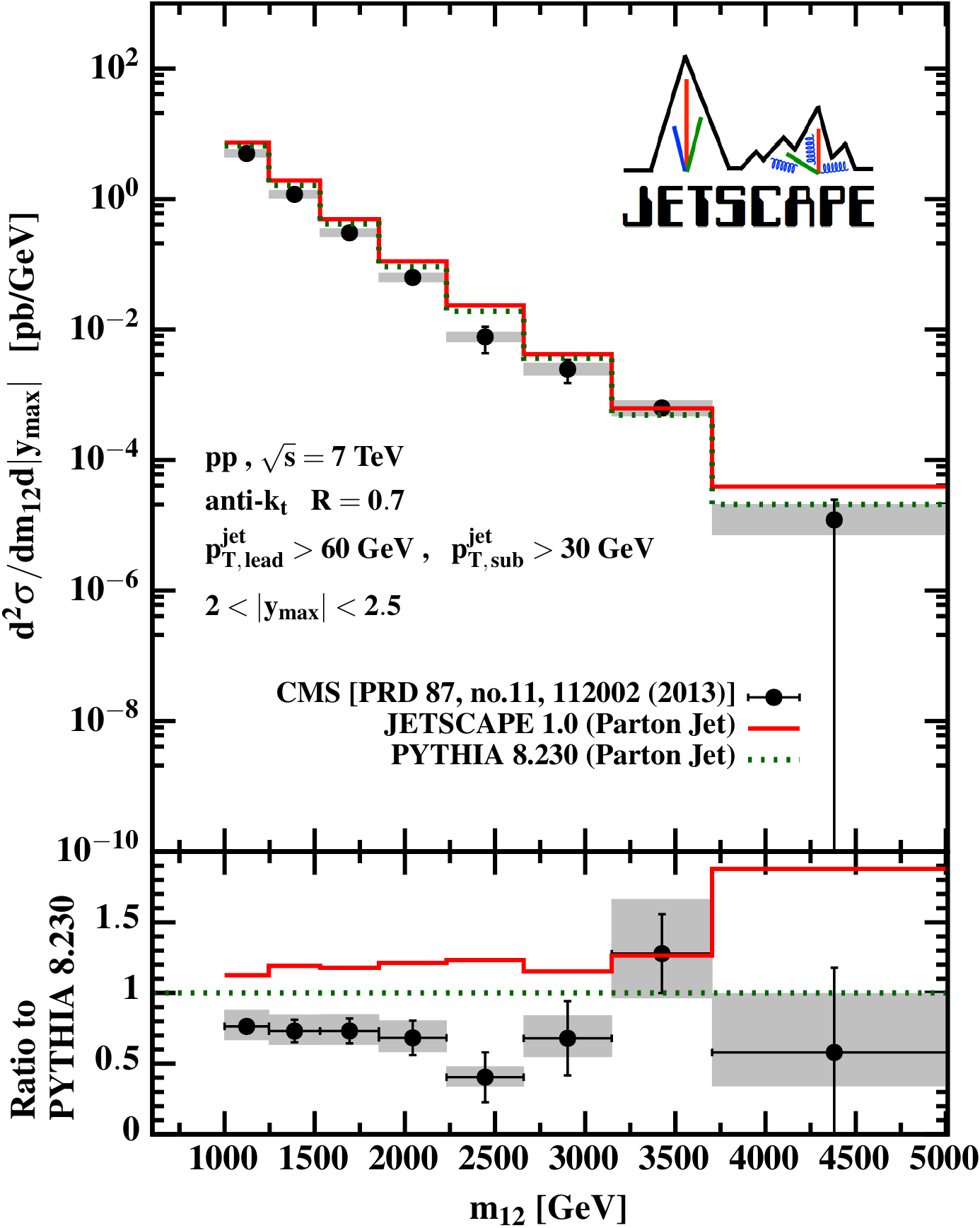}
  \caption{Same as Fig.\ \ref{fig:jetmasscmsy0} for jets with $2\leq |y_\text{max}| \leq 2.5$.
  \label{fig:jetmasscmsy2}}
\end{figure*}

\begin{figure*}[tb]
\includegraphics[width=0.85\columnwidth]{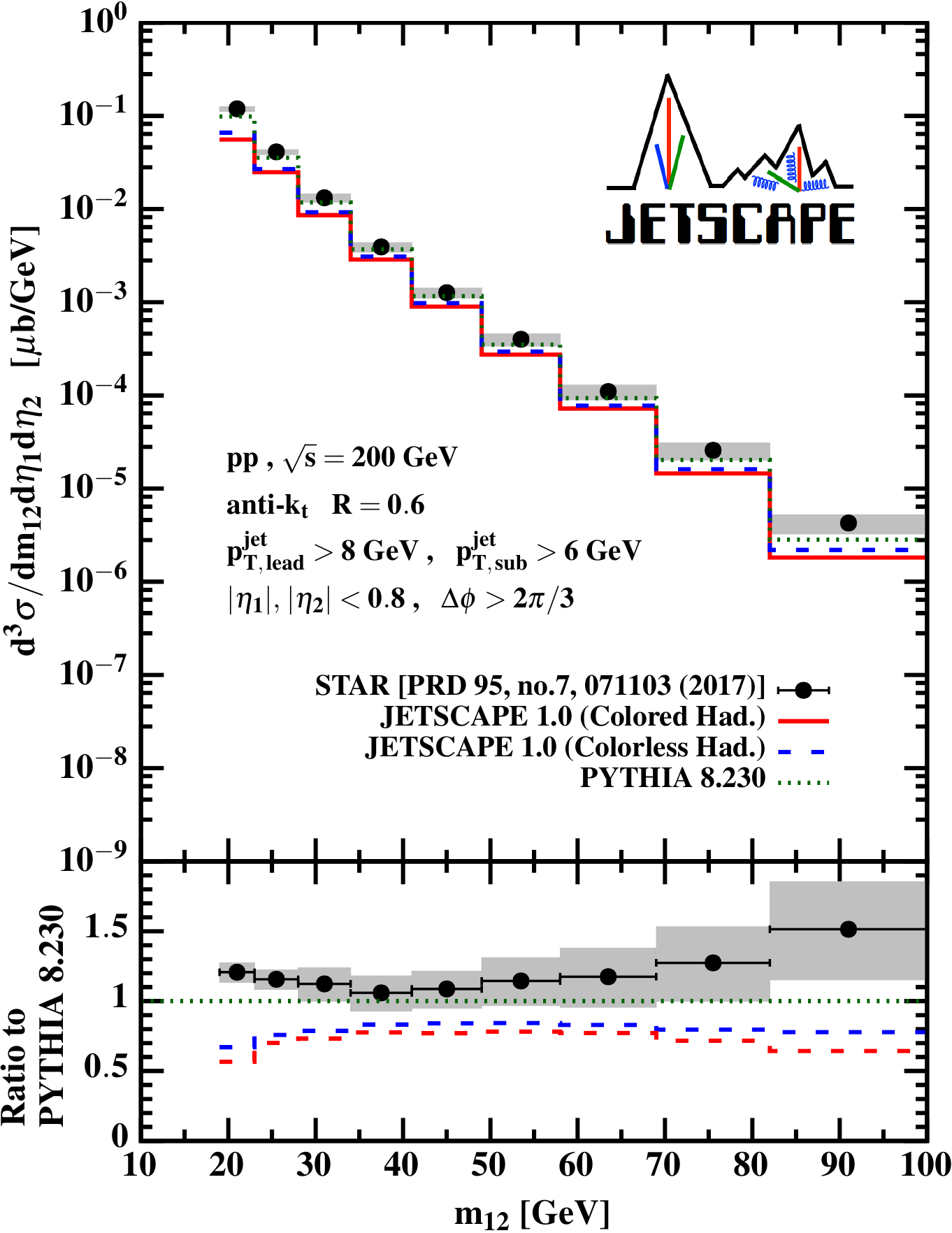}\hspace{15pt}
\includegraphics[width=0.85\columnwidth]{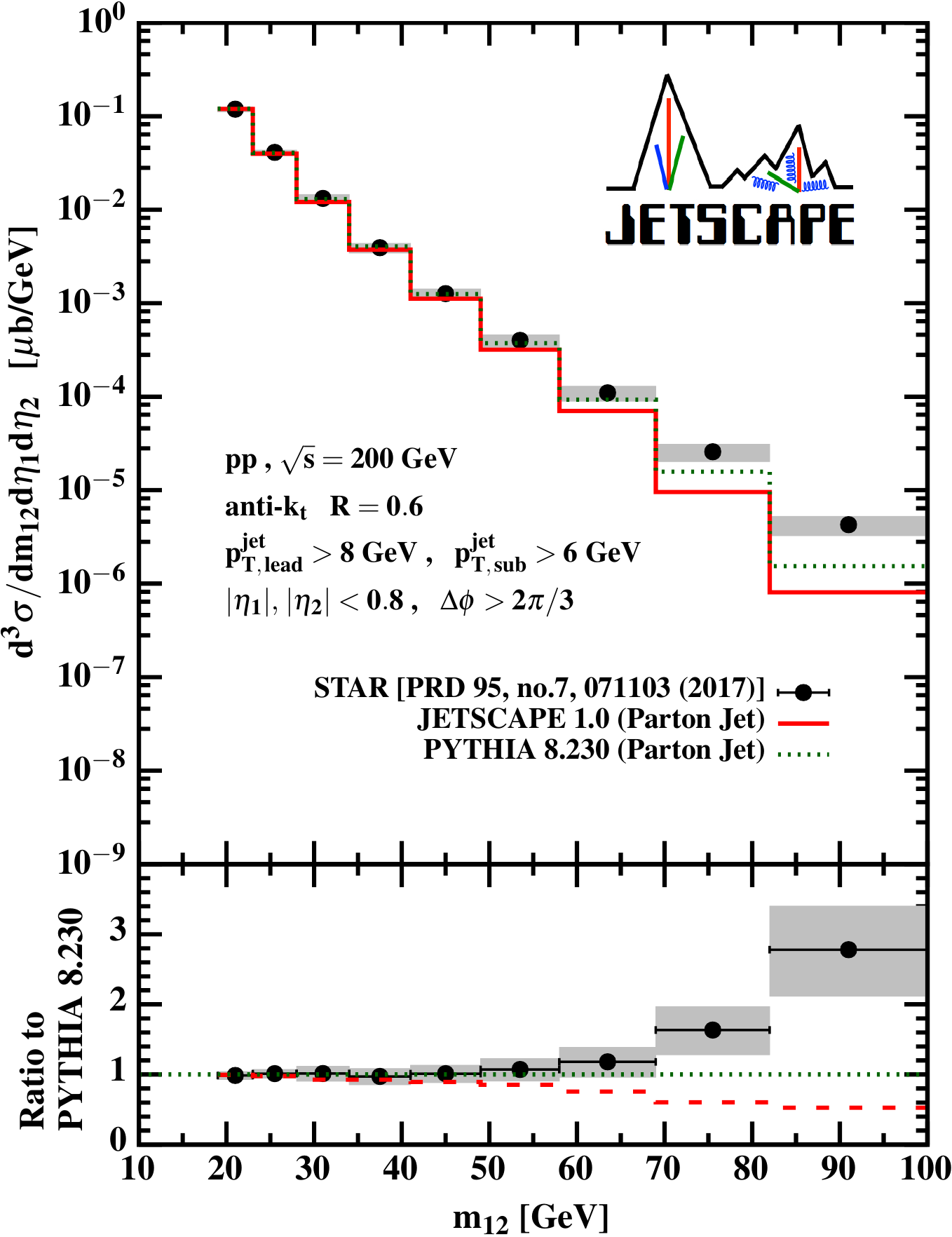}
  \caption{Dijet mass cross section $d^2\sigma/dMd\eta_1d\eta_2$ at 200 GeV for $R=0.6$ jets with $\eta<0.8$.  
  The left panel shows results from PYTHIA 8 and JETSCAPE with both colored and colorless hadronization
  together with data from STAR \cite{Adamczyk:2016okk}. The right panel shows results for parton jets created with PYTHIA 8 and
  JETSCAPE. Bottom panels give the ratio of all results and data to PYTHIA 8.
  \label{fig:jetmassstar}}
\end{figure*}

We present results for dijet mass spectra. For each event the two largest momentum jets for a given jet
radius $R$, satisfying certain cuts explained in detail for each calculation below, are chosen. If two such jets can not be found the event is discarded. 
The invariant mass of the dijet system is calculated from the four-momentum vectors of the two jets. The dijet mass observable is complementary 
to other jet measurements discussed thus far. It is likely that a dijet pair comes from the same underlying hard QCD scattering between partons in 
the beams, and it is thus sensitive to additional features of the hard QCD process and to parton distribution functions. 
We will focus our work on two cases: dijets in $p+p$ collisions at $\sqrt{s}=200$ GeV compared
to STAR data \cite{Adamczyk:2016okk} and dijets in collisions at 7 TeV compared to data measured by the CMS experiment \cite{Chatrchyan:2012bja}.

In order to compare to CMS dijet data we choose $R=0.7$ and calculate the cross section
\begin{equation}
  \frac{d^2\sigma}{dM dy_\text{max}}
\end{equation}
where $y_\text{max}$ is the larger of the two jet rapidities by magnitude. Dijet systems are accepted if the 
leading jet transverse momentum is above 60 GeV/$c$ and the sub-leading jet transverse momentum is above 30 GeV/$c$. 
Fig.\ \ref{fig:jetmasscmsy0} shows the results of JETSCAPE calculations for hadronic jets (left panel), using both hadronization options, and 
parton jets (right panel) for $ |y_\text{max}| \leq 0.5$. 
PYTHIA 8 results and CMS data are included for comparison, and the bottom panels show ratios of
JETSCAPE results and data to PYTHIA 8 results. There is little difference between the two JETSCAPE hadronization 
models and JETSCAPE 1.0 results are consistent with data within error bars. PYTHIA 8 slightly overpredicts the
dijet mass spectrum from very small to very large dijet masses. These observations are consistent between parton and hadron jets.

The picture changes when at least one jet is required to have large rapidity, see Fig.\ \ref{fig:jetmasscmsy2}. In the case of $2 \leq y_\text{max}
\leq 2.5$ the overprediction of experimental data becomes significant.
The roles are reversed with PYTHIA 8 doing better than JETSCAPE compared to data. Again parton jets show the
same behavior than hadron jets, and the two JETSCAPE hadronization models produce very similar results.

We conclude that for the large dijet masses at LHC hadronization has little bearing on dijet cross sections. Since the 
underlying hard processes for PYTHIA 8 and JETSCAPE are computed in the same manner the 
results could indicate needed improvements in cross section calculations in which one jet is at forward rapidity.
We can further confirm that the dijet mass is quite sensitive to details of the final state parton shower as shown
by the relative difference between parton results for JETSCAPE and PYTHIA 8.

At RHIC energy we calculate the triple differential cross section
\begin{equation}
  \frac{d^3\sigma}{dM d\eta_1 d\eta_2}
\end{equation}
for $R=0.6$ jets. $\eta_1$ and $\eta_2$ are the pseudorapidities of both jets. The jets were required to 
have $p_T^\mathrm{jet}  > 8$ GeV/$c$ for the leading jet and $p_T^\mathrm{jet} > 6$ GeV/$c$ for the sub-leading jet. The pseudorapidity
for both jets was constrained to satisfy $\eta \le 0.8$.
Fig.\ \ref{fig:jetmassstar} shows the results for both hadron (left panel) and parton (right panel) jets calculated with
JETSCAPE using both hadronization models. We also show PYTHIA 8 results and data from STAR.
Bottom panels once more indicate ratios with respect to PYTHIA 8.

Both hadronization models in JETSCAPE give consistent results, however a comparison of parton and hadron jets 
indicates the presence of hadronization effects in this case. Hadronization tends to push calculated dijet mass spectra lower. 
Hadron dijet mass spectra from JETSCAPE underpredict measured spectra. The deviations of PYTHIA 8 calculations from data are less 
severe and consistent with data except for very small ($<20$ GeV) and 
large masses ($>80$ GeV) available from experiment.

\section{Summary and discussion}
\label{sec:discussion}

In this work we have introduced the JETSCAPE PP19 tune based on JETSCAPE 1.0. We present the first systematic and comprehensive 
evaluation of the JETSCAPE event generator for important observables in $p+p$ collisions. We have studied JETSCAPE
at three different collision energies ($\sqrt{s}=0.2$, 2.76 and 7 TeV). Our results quantify results of the JETSCAPE framework in relation
to PYTHIA 8 and data. They also serve as a benchmark for JETSCAPE users who wish to test their setup against a comprehensive 
set of calculations.

We have calculated inclusive jet cross sections, transverse jet shapes, jet fragmentation functions, charged hadron cross sections, and dijet 
mass cross sections. The emerging picture from this body of work is that overall agreement of JETSCAPE with PYTHIA 8 and experimental data
is satisfactory, but there is room for future improvements. Inclusive jet cross sections, dijet mass cross sections and transverse jet shape observables 
at LHC energies calculated with JETSCAPE are typically compatible with data within experimental error bars for all jet radii considered, as long as 
$p_T^\mathrm{jet}$ is larger than 40 GeV/$c$. The only exception for jet cross sections
is the 2.76 TeV CMS data for jet radii between 0.2 and 0.4 which is overpredicted by both JETSCAPE and PYTHIA 8. However, uncertainties from 
hadronization for these small jet radii are appreciable. When interpreting these results it should be kept in mind that the two adjustable parameters in 
MATTER have been optimized to describe inclusive jet cross section without additional UE subtraction.
JETSCAPE calculations of the jet shape variable $\rho$ bracket PYTHIA 8 and data within the unceratainties of the data. Ratios of jet cross sections of
different jet radii are described well above $p_T^\mathrm{jet}= 40$ GeV/$c$.
The picture is different at RHIC energies. Deviations between PYTHIA 8 and JETSCAPE results for inclusive jet cross sections are as large as 50\%,
with most of the STAR data falling within the band defined by the calculations. Clearly, differences in final state parton showers and details of
the underlying event subtraction matter greatly at low jet momenta and JETSCAPE procedures need improvement in this case.

Overall, for LHC energies, around midrapidity, and for jet momenta above 40 GeV/$c$ jet cross sections, jet shapes and dijet mass spectra calculated with 
JETSCAPE are well suited as benchmarks for heavy ion collisions. At other energies and smaller jet momenta uncertainties in JETSCAPE calculations and 
deviations from data can exceed 10\% and need to be considered when A+A collisions are compared to $p+p$.

Jet fragmentation functions and charged hadron cross sections at LHC energies show two noteworthy features. JETSCAPE overpredicts fragmentation 
functions for very large $z$ or $p_T$, and large transverse jet momenta. The deviation can be 20\% to 60\% for 
$z\approx 0.7 \ldots 1$. JETSCAPE results are consistent with data on fragmentation functions within experimental errors starting from 
$z\approx 0.5$. Large-$z$ deviations are also less pronounced for smaller momentum jets. The high-$z$ excess in JETSCAPE can be traced back 
to differences in the final state showers between PYTHIA 8 and MATTER. 
Moreover, at small $z$ uncertainties from hadronization are very large compared to experimental error bars. JETSCAPE results are consistent with 
PYTHIA 8 and data within those uncertainties. 

We find no clear tendency that data would favor one JETSCAPE hadronization model over the other. Differences between calculations 
using the two models are useful to explore uncertainties from hadronization. Uncertainties are largest for jet cross sections at small momenta 
$p_T^\mathrm{jet} \lesssim 30$ GeV/$c$  (up to $\sim 30\%$) and for fragmentation functions at small $z$ (up to $\sim 50\%$).
As discussed above, both hadronization models have strength and weaknesses, with Colorless Hadronization preferable in $A+A$
collisions and Colored Hadronization in $p+p$ collisions. In absence of a clear conclusion one should understand the two results as
an uncertainty band for uncertainties in modeling hadronization.

Differences between JETSCAPE and PYTHIA 8, indicative of an important role for the final state shower Monte Carlo and the underlying event, 
can be seen most prominently for fragmentation functions for very large $z$, dijet mass cross sections, and inclusive jets cross sections for 
small radius $R$. 
Our study has produced a quantitative map of the accuracy of the JETSCAPE 1.0 event generator in $p+p$ collisions. 
Deviations from $p+p$ results seen in A+A calculations need to be evaluated in the context of the uncertainties for $p+p$ documented here.
As an example, the interpretation of medium-modified fragmentation functions needs to be discussed with the large dependence of low-$z$ fragmentation 
functions on hadronization in mind. 
Future improvements to JETSCAPE in $p+p$ would involve a proper treatment of the underlying event 
and a careful simultaneous tuning of JETSCAPE and PYTHIA 8 parameters in connection with a rigorous statistical analysis of data.


\section*{Acknowledgments} 

This  work  was  supported  in  part  by  the National Science Foundation (NSF) within the framework of the JETSCAPE collaboration, under grant numbers ACI-1550172 (G.R.), ACI-1550221 (R.J.F., M.K. and Z.Y.), ACI-1550223 (D.E., M. M. and U.H.), ACI-1550225 (S.A.B., J.C., T.D., W.F., W.K., R.W. and Y.X. ), ACI-1550228 (L.-G.P., X.-N.W., P.J.), and ACI-1550300 (S.C., L.C., K. K., A.K., M.E.K., A.M., C.N., D.P., J.P., L.S., C.Si. and R.A.S.). It was also supported in part by the NSF under grant numbers PHY-1207918 (M.K.), 1516590 and 1812431 (R.J.F., M.K. and Z.Y.), and by the US Department of Energy, Office of Science, Office of Nuclear Physics under grant numbers \rm{DE-AC02-05CH11231} (Y.H. and X.-N.W.), \rm{DE-AC52-07NA27344} (R.A.S.), \rm{DE-SC0012704} (C.S.), \rm{DE-SC0013460} (S.C., A.K., A.M., C.S. and C.Si.), \rm{DE-SC0004286} (L.D., U.H.),  \rm{DE-SC0012704} (K.K. and B.S.), \rm{DE-FG02-92ER40713} (J.P.), \rm{DE-SC0011088} (Y.C.), and \rm{DE-FG02-05ER41367} (T.D., W.K., J.-F.P., S.A.B. and Y.X.). The work was also supported in part by the National Science Foundation of China (NSFC) under grant number 11521064, Ministry of Science and Technology (MOST) of China under Projects number 2014CB845404 (Y.H., T.L. and X.-N.W), and in part by the Natural Sciences and Engineering Research Council of Canada (C.G., S.J., C.P. and G.V.), the Fonds de recherche du Qu\'ebec Nature et technologies (FRQ-NT) (G.V.) and by the Office of the Vice President for Research (OVPR) at Wayne State University (Y.T.).  This work used the Extreme Science and Engineering Discovery Environment (XSEDE), which is supported by National Science Foundation grant number ACI-1548562. Computations were made in part on the supercomputer \emph{Guillimin} from McGill University, managed by Calcul Qu\'ebec and Compute Canada. The operation of this supercomputer is funded by the Canada Foundation for Innovation (CFI), NanoQu\'ebec, R\'eseau de M\'edicine G\'en\'etique Appliqu\'ee~(RMGA) and FRQ-NT. Computations were also carried out on the Wayne State Grid funded by the Wayne State OVPR. Data storage was provided in part by the OSIRIS project supported by the National Science Foundation under grant number OAC-1541335. C.G. gratefully acknowledges support from the Canada Council for the Arts through its Killam Research Fellowship program. C.S. gratefully acknowledges a Goldhaber Distinguished Fellowship from Brookhaven Science Associates.


\bibliography{JS002_References}


\end{document}